\documentclass[largeformat]{interact}

\usepackage{braket}
\usepackage[capitalise]{cleveref}
\usepackage{chemformula}
\usepackage{bm}
\usepackage{xcolor}
\usepackage{epstopdf}
\usepackage[caption=false]{subfig}

\usepackage[numbers,sort&compress]{natbib}
\bibpunct[, ]{[}{]}{,}{n}{,}{,}
\makeatletter
\def\NAT@def@citea{\def\@citea{\NAT@separator}}
\makeatother

\theoremstyle{plain}

\theoremstyle{definition}

\theoremstyle{remark}

\usepackage{tikz}

\def\plaquette{\tikz[baseline=.05ex]{
		\fill (0,0) circle (1pt) coordinate (A);
		\fill (1.5ex,0) circle (1pt) coordinate (B);
		\fill (1.5ex,1.5ex) circle (1pt) coordinate (C);
		\fill (0,1.5ex) circle (1pt) coordinate (D);
		\draw (A)--(B);
		\draw (B)--(C);
		\draw (C)--(D);
		\draw (D)--(A);}
}

\def\plaquettev{\tikz[baseline=.1ex]{
		\fill (0,0) circle (1pt) coordinate (A);
		\fill (1.5ex,0) circle (1pt) coordinate (B);
		\fill (1.5ex,1.5ex) circle (1pt) coordinate (C);
		\fill (0,1.5ex) circle (1pt) coordinate (D);
		\draw (A)--(B);
		\draw [ultra thick] (B)--(C);
		\draw (C)--(D);
		\draw [ultra thick] (D)--(A);}
}

\def\plaquetteh{\tikz[baseline=.1ex]{
		\fill (0,0) circle (1pt) coordinate (A);
		\fill (1.5ex,0) circle (1pt) coordinate (B);
		\fill (1.5ex,1.5ex) circle (1pt) coordinate (C);
		\fill (0,1.5ex) circle (1pt) coordinate (D);
		\draw [ultra thick] (A)--(B);
		\draw (B)--(C);
		\draw [ultra thick] (C)--(D);
		\draw (D)--(A);}
}

\begin{document}


\title{Emergent particles and gauge fields in quantum matter}

\author{
\name{Ben J. Powell\thanks{Email: powell@physics.uq.edu.au}}
\affil{School of Mathematics and Physics, The University of Queensland, QLD 4072, Australia}
}

\maketitle

\begin{abstract}
I give a pedagogical introduction to some of the many particles and gauge fields that can emerge in correlated matter. The standard model of materials is built on Landau's foundational principles: adiabatic continuity and spontaneous symmetry breaking. These ideas lead to quasiparticles that inherit their quantum numbers from  fundamental particles, Nambu-Goldstone bosons, the Anderson-Higgs mechanism, and topological defects in order parameters. I then describe the modern discovery of physics beyond the standard model. Here, quantum correlations (entanglement) and topology play key roles in defining the properties of matter. This can lead to fractionalised quasiparticles that carry only a fraction of the quantum numbers that define fundamental particles. These particles can have  exotic properties: for example Majorana fermions are their own antiparticles, anyons have exchange statistics that are neither bosonic nor fermionic, and magnetic monopoles do not occur in the vacuum. Gauge fields emerge naturally in the description of highly correlated matter and can lead to  gauge bosons. Relationships to the standard model of particle physics are discussed.
\end{abstract}


\section{Introduction}

One of the major goals of elementary particle physics is to examine higher and higher energies. This is motivated, in part, by the quest for `unification'. At low-energies we observe four different forces: electromagnetism, the weak and strong nuclear forces, and gravity. These forces are described in a mathematical language known as gauge field theory. Unification is the idea that, at higher energies, one can describe two or more of these forces as different aspects of a single theory. The best developed and evidenced example of unification is electroweak theory. Above some unification energy (of order the Higgs vacuum expectation value, 246~GeV) electroweak theory predicts that electromagnetism and the weak interaction merge into a single unified force. One way to reach the unification energy scale would be to heat a system to $\sim10^{15}$~K -- a temperature that was last achieved just moments after the big-bang. Therefore, electroweak theory predicts that  a symmetry between electromagnetism and the weak interaction appears at high temperatures, and that this symmetry is spontaneously broken at `low' temperatures. 

Many low-temperature physicists are engaged in the opposite program. As we lower the temperature of materials we often find that symmetries are broken. Such spontaneously broken symmetries occur at phase transitions, for example when a liquid crystallises on cooling. A liquid looks the same under any translation, but in a crystal, only  translations commensurate with the  lattice leave the local environment unchanged. This causes dramatic changes in the low-energy physics and the macroscopic properties of the material. For example, new `emergent' particles are typically found in the low-temperature phase. In a crystal there are new massless bosons, called (acoustic) phonons, that are not present in the high-temperature disordered phase; and a crystal is rigid whereas a liquid is not. These two phenomena are deeply connected. 

Over the last few decades it has become clear that not all phase transitions correspond to a broken symmetry. In, so called, quantum materials the properties of the wavefunction describing a material can qualitatively change at a phase transition without breaking a symmetry. In such quantum matter, there are new particles and often new gauge fields. Below we will explore how concepts of quantum entanglement and topology give new insights into quantum materials. We will see that the particles in quantum materials are often  exotic: they have properties that are not found in the electrons, protons and neutrons that make up ordinary matter. Indeed, below we will meet particles with traits not observed in particles in the vacuum. 

This is not intended to be an exhaustive review. Instead, I aim to give a flavour of the exotic particles and fields that can emerge in strongly correlated matter  by picking out some examples, which have been chosen for their simplicity and the degree to which  they illustrate broadly relevant concepts. Similarly, I have selected references for their pedagogical value rather than on the basis of priority.

\section{The standard model of materials}\label{sect:SM}

Before we can discuss exotic particles and fields in quantum materials, we need to understand what the low-energy excitations of materials usually look like. The standard model of materials \cite{AndersonBasic} is built from four ingredients: quasiparticles, Nambu-Goldstone bosons,  the Anderson-Higgs mechanism, and topological defects. These ideas are of similar importance to condensed matter physics as  quarks and leptons, gauge bosons, and the Higgs boson are for particle physics.

\subsection{Adiabatic continuity and quasiparticles}

For all practical purposes, we know the theory of everything for most materials \cite{ToE}: Newton's equations for nuclei, the Schr\"odinger equation for electrons, and electromagnetic interactions between  particles. There are counterexamples to each of these restrictions, but in more materials than not, this is sufficient. But, solving these equations exactly for a macroscopic system is impossible. We are left with two possibilities: (i) make some, often very severe, approximation or (ii) study a simpler problem, which we  hope captures the essential physics.  (i) is very powerful, and these `first principles' approaches, particularly density functional theory, have made many important contributions to condensed matter physics. But, in this article, I want to focus on approach (ii).

Imagine that we have a simple model for some material that we can solve exactly or with  high accuracy. Natural questions are: which of the conclusions, drawn from this model, apply to a wide range of models and materials? and which conclusions are specific to this model? A powerful way to think about these questions is to apply perturbations to our model and see what changes. 

To get an idea of how this works, consider the one-dimensional simple harmonic oscillator with an additional Dirac delta function potential at the origin. If we choose our units so that $\hbar=m=\omega=1$ (respectively the reduced Planck constant, mass, and frequency) then the Schr\"odinger equation is
\begin{equation}
\left[-\frac{1}{2}\frac{\partial^2}{\partial x^2} + \frac12 x^2 + g\delta(x)\right]\psi_n(x)=E_n\psi_n(x),
\label{eq:sho+del}
\end{equation}
where $g$ sets the strength of the $\delta$-potential and entirely specifies the problem.

\begin{figure}
	\centering
	\includegraphics[width=\textwidth]{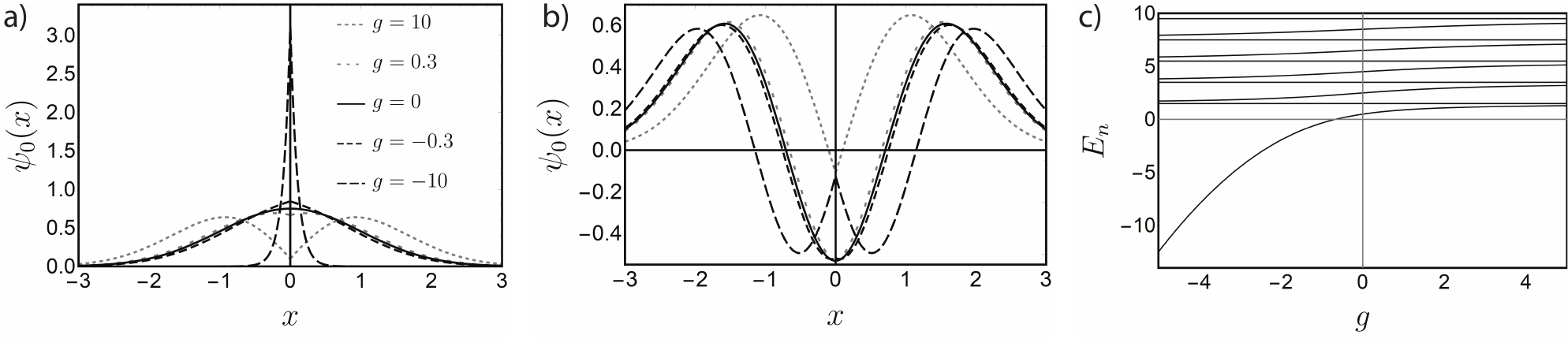}
	\caption{A simple example of adiabatic continuity is provided by the solution of the simple harmonic oscillator with a $\delta$-potential at the origin (Eq. \ref{eq:sho+del}): (a) the ground state wavefunction has no nodes for any $g$ and (b) the second excited state wavefunction has two nodes of all $g$ (solutions are shown for the same strengths of the $\delta$-potential, $g$, as in panel a); and (c) the $g$ dependence of the energies of the first ten eigenstates. They do not cross for any $g$. For details of the solutions see \cite{sho_delta}.} \label{fig:sho_delta}
\end{figure}

For $g=0$, this is just the simple harmonic oscillator. The eigenstates have energy $E_n=n+\frac12$, where $n=0, 1, 2,\dots$ The quantum number, $n$, has a simple physical interpretation: it is the number of nodes in the wavefunction, $\psi_n(x)$. If we apply a  perturbation ($g\ne0$) then the eigenstates and their energies change; just a little for small $g$, but more dramatically for $g\gg1$, Fig. \ref{fig:sho_delta}. Nevertheless, the two most important characteristics of the solution do not change regardless of the value of $g$: (i) the eigenstates are uniquely labelled by the number of nodes in the wavefunction, $n$, and (ii) $E_n<E_m$ if and only if $n<m$. That is the perturbation does not change anything really important, anything qualitative, about the solution. This is the basic idea of adiabatic continuity. If we can move from one Hamiltonian to another by slowly adding in more terms and the spectrum of solutions does not change qualitatively then the solutions are said to be adiabatically connected. More precisely, Hamiltonians are adiabatically connected if the set of quantum numbers does not change and the ordering of the energy eigenvalues does not change when one Hamiltonian is slowly transformed into the other.

However, in many interesting problems perturbations do change the essential physics. Sometimes even  arbitrarily small perturbations change everything. When we are working in the thermodynamic limit (of infinite system size and finite density)  the failure of adiabatic continuity indicates a phase transition. This is extremely helpful as phase transitions are usually easy to detect (both experimentally and theoretically). So, if we can establish that a toy model is in the same thermodynamic phase as the material or model that we really care about then adiabatic continuity tells us that the toy model and the real model will share their most important characteristics. 

Conduction electrons, even in simple elemental solids like gold or copper, interact strongly with one another. Indeed a simple estimate\footnote{p336 of \cite{Ashcroft}} suggests that in metals the kinetic energy of an electron at the Fermi energy is often smaller than the potential energy due to the Coulomb interaction with other electrons. Yet it is possible to understand the physics of these metals without a deep understanding of quantum many-body theory. In fact, most of their properties can be understood in terms of weakly interacting fermions. This should shock you. It is one of the great examples of adiabatic continuity. If one starts with a non-interacting gas of fermions  it is possible to turn on interactions so slowly that neither the labelling nor the ordering of the eigenstates change.  

\begin{figure}
	\centering
	\includegraphics[width=0.6\textwidth]{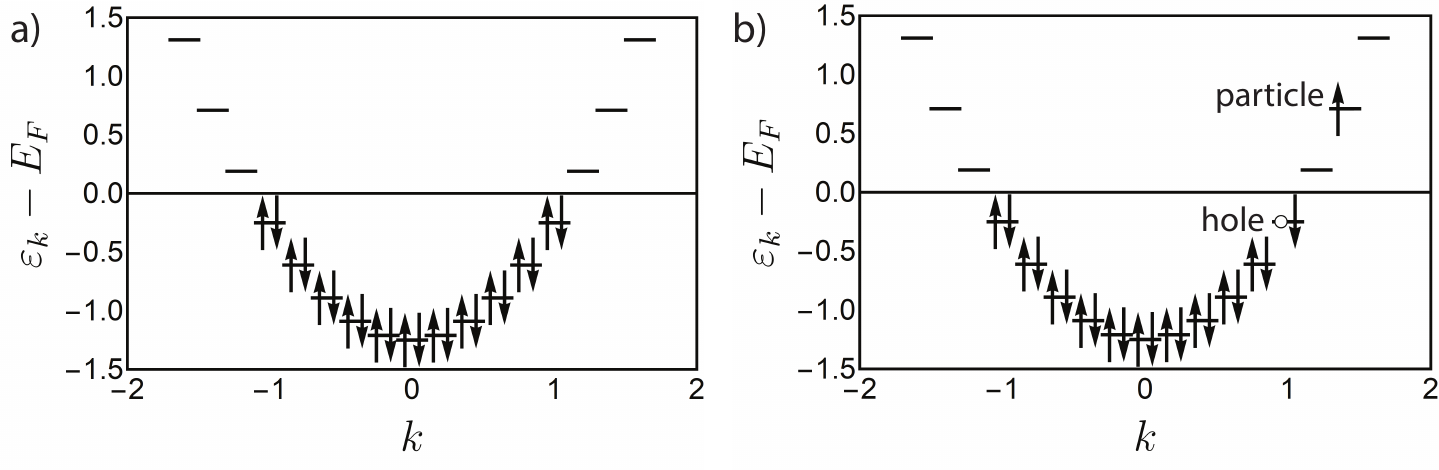}
	\caption{Sketches of low energy states in a non-interacting Fermi gas. (a) The ground or vacuum state. All states with energy $\varepsilon_k<E_F$ the Fermi energy are occupied by two fermions, one spin up, the other spin down. Note that every state is uniquely labelled by the momentum, $k$, and the spin. In a real metal there are many more states than shown here. (b) A particle-hole excitation is equivalent to moving one electron from a state below the Fermi energy to a state above it. } \label{fig:Fermi}
\end{figure}

The low-energy excited states of a non-interacting gas of fermions involve moving single particles  from below the Fermi energy to above the Fermi energy  (Fig. \ref{fig:Fermi}). A more efficient description of this is to redefine the word `vacuum' to mean the ground state and to say that the excited state contains one particle (above the Fermi energy) and one hole (empty state below the Fermi energy). This is precisely analogous to the modern treatment of positrons and electrons and the abandonment of Dirac's idea of a sea of negative energy electrons. The particles and the holes are each characterised by two quantum numbers: their spin and their momentum. 

If adiabatic continuity holds as we turn on electron-electron interactions then an interacting gas of fermions is described by Landau's ``Fermi liquid'' theory. In a Fermi liquid the lowest energy excitations are again particles and holes. However, these are no longer single particle excitations -- they involve changes in the quantum states of many electrons. Nevertheless, adiabatic continuity ensures that the key qualitative features of the  excitations are identical to those of the non-interacting system. The particles and holes are still defined by two quantum numbers: their spin and their momentum. Therefore, these excitations behave a lot like the excitations of the non-interacting Fermi gas, and are called \textit{quasiparticles}. 

However, some of the properties of the quasiparticles are  changed by the interactions. 
Most importantly, electron and hole excitations of the non-interacting Fermi gas are exact eigenstates; this means that they are stable. Quasiparticles are not exact eigenstates because they weakly interact with one another; this means that the quasiparticles have a finite lifetime. Quasiparticle interactions also impede the motion of particles -- this leads to an increase in the effective mass of the quasiparticles compared to that of the original non-interacting fermions. In some materials, known (understandably) as heavy fermion materials, the effective mass can be as much as 1000 times that of the bare electron \cite{ColemanHandbook}. The lifetime of the quasiparticles and their effective mass are actually two aspects of the same physics (they are related via a fluctuation-dissipation theorem \cite{Jacko09}). A more detailed introduction to Landau's Fermi liquid theory can be found in \cite{Schofield}.

Adiabatic continuity is not guaranteed. Not all systems of interacting fermions are Fermi liquids. Such systems can undergo various instabilities, leading to new phases of matter such as magnetic phases, (Mott and Wigner) insulators, and superconductors/superfluids. 
More subtle changes can also occur -- leading to systems of strongly interacting fermions where the low-energy excitations are not like the single particle excitations of Fermi gases. Metals without quasiparticles are found in various materials and are often called `bad' or `strange metals' \cite{KotliarVollhardt}.
In one-dimensional (1D) metals one finds a Luttinger liquid rather than a Fermi liquid. The quasiparticles in Luttinger liquids are not electron-like. Instead spin and charge are carried by  separate excitations, known as spinons and holons respectively; an effect known as spin-charge separation (a clear introduction to this is given in \cite{Schofield}, more details are given in \cite{Giamachi}). Other qualitatively different quasiparticles can also emerge, and some of which will be discussed in section \ref{sect:BSM}.

Nevertheless, in most metals under most conditions, Landau's Fermi liquid theory provides a good description of the low-energy physics. This provides the first ingredient of our standard model of materials. In metals and semiconductors we expect quasiparticles that behave qualitatively like weakly-interacting fermions.

\subsection{Spontaneously broken symmetry}

\subsubsection{Landau theory of a ferromagnet}

Ferromagnetism is a long-range order of the spins of the electrons in a material. 
A microscopic theory of ferromagnetism would seek to describe the behaviour of the spins of all of the electrons. However, even if we do not understand the microscopic mechanisms at play (which is often the case), we can make significant progress by thinking about the symmetry of the problem. The key to this  approach is a coarse-grained description of the physics based on an order parameter, which is zero in the high temperature, disordered phase and non-zero in the low temperature ordered phase. In the case of a ferromagnet an appropriate order parameter is the magnetisation, ${\bm m}$. 

If ${\bm m}$ is small, which it is just below the transition temperature, we can write the free energy difference between the ferromagnetic and paramagnetic states, $\Delta F$, as a power series in ${\bm m}$. This must respect the symmetries of the system. For example,  many materials respect \textit{spin rotation symmetry}, \textit{i.e.}, the free energy does not depend on what direction $\bm m$ points in. Therefore, we cannot include odd powers of ${\bm m}$, and  the free energy of the ferromagnet is
\begin{eqnarray}
\Delta F = \alpha|{\bm m}|^2 + \beta|{\bm m}|^4 +\dots, \label{eq:Flan}
\end{eqnarray}
where  $\alpha$ and $\beta$ are  parameters of the theory, which we do not know \emph{a priori} (we could calculate them from a microscopic theory or determine them experimentally). If we stop the expansion here, then we must have $\beta>0$, otherwise the equilibrium (lowest free energy) state would be $|{\bm m}|\rightarrow\infty$, which is unphysical.

For simplicity we will consider this theory in a small temperature range near the critical temperature, where the transition occurs $T_c$. The temperature dependence of  $\beta$ is not usually important in this range, but that of $\alpha$ is. Therefore, we will treat $\beta$ as a constant and expand $\alpha$ to linear order:  $\alpha=\alpha'(T-T_c)$, where $\alpha'=d\alpha/dT|_{T=T_c}$. Notice that this expansion assumes $\alpha=0$ at $T=T_c$; we will see below that this is really the definition of $T_c$ in this Landau theory. Hence, 
\begin{eqnarray}
\Delta F = \alpha'(T-T_c) |{\bm m}|^2 + \beta |{\bm m}|^4.  \label{eq:Flan2}
\end{eqnarray}

\begin{figure}
	\begin{center}
		\includegraphics[width=0.7\columnwidth]{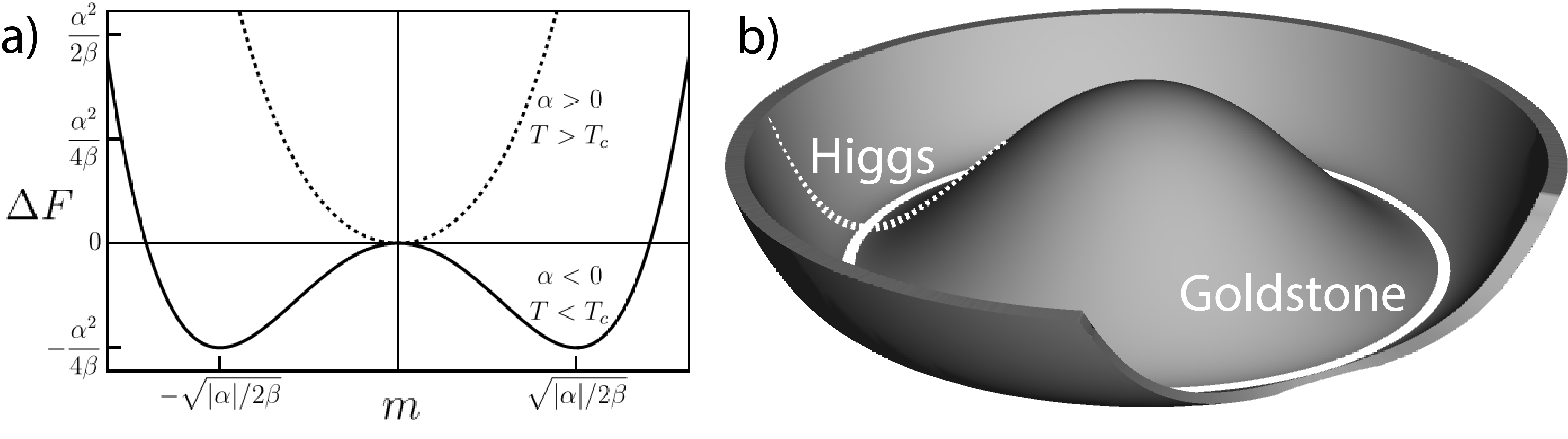}  
		\caption{Free energies in Landau theories. a) Broken discrete symmetry in a system with Ising or $Z_2$ symmetry, such as an easy axis ferromagnet. The order parameter, $m$, is a scalar and the free energy is given by Eq. \ref{eq:Flan2}. For $T<T_c$ there are two degenerate ground states: $m=\pm\sqrt{|\alpha|/2\beta}$. b) Broken continuous symmetry for a system with XY or $U(1)$ symmetry. The order parameter, $\Psi$, is complex [or equivalently a two dimension vector, say $(\text{Re}\Psi,\text{Im}\Psi)$]. The  free energy is given by Eq. \ref{eq:Flan-XY}. Phase fluctuations lead to a new massless mode -- the Nambu-Goldstone boson, in the large wavelength limit this corresponds to running around the bottom of the potential, solid white line. If the order parameter couples to a gauge field the massless Nambu-Goldstone mode and massless gauge boson are replaced by a single massive mode with both transverse and longitudinal components due to the Anderson-Higgs mechanism. This is the origin of the Meissner effect in superconductors and crucial for understanding electroweak symmetry breaking in the vacuum. The Higgs boson corresponds to amplitude fluctuations of the order parameter, dashed white line.}\label{fig:Landau}
	\end{center}
\end{figure}

For $\alpha>0$ ($T>T_c$) the free energy is minimised for ${\bm m}={\bm 0}$, Fig. \ref{fig:Landau}a. Thus, the lowest energy state is paramagnetic when $\alpha>0$  ($T>T_c$).

For $\alpha<0$  ($T<T_c$) there are minima   at $|\bm m|= m_0 = \sqrt{{|\alpha|}/{2\beta}}$, \cref{fig:Landau}a,
i.e., the equilibrium state is ferromagnetic. A phase transition from a paramagnet to a ferromagnet therefore occurs when $\alpha$ changes sign. 

In the paramagnetic phase there is only one solution, $\bm m=\bm 0$.
In the ferromagnetic phase there are infinitely many solutions: the condition $|\bm m|=m_0$ defines a sphere in `magnetisation space'. In order to minimise its free energy the system must `pick' one of the minima, $\bm m$ must point in some specific direction.
In practice, this `choice' might be made by the presence of a stray magnetic field, $\bm H$; despite our best efforts, such fields are always present in any experiment. The field induces a term proportional to $\bm H \cdot \bm m$ that lifts the degeneracy even for arbitrarily small $\bm H$. 

Note that  states with $\bm m\ne\bm 0$  are not  spin rotation symmetric. Therefore, these states `break' a symmetry of the underlying Hamiltonian.
 We call this a \emph{spontaneously broken symmetry} as the free energy is the same when we rotate the magnetisation, but the actual state of the system breaks this symmetry.

\subsubsection{Quantum mechanics and spontaneously broken symmetries}\label{sect:QM+SBS}

Landau theories  are essentially classical. But, in many real materials quantum mechanics is vital for understanding the macroscopic behaviour.

The magnetisation of a ferromagnet  is straightforwardly related to the microscopic behaviour of a subset, $N$, of the electronic spins: 
\begin{equation}
\bm{m} = \langle \hat{\bm{m}} \rangle = 
\frac{1}{N} \left\langle \sum_{i=1}^N \hat{\bm{S}}_i \right\rangle,
\label{eq:magnetisation}
\end{equation}
where  ${\hat {\bm S}}_i=(S^x_i,S^y_i,S^z_i)$ is the spin of the $i$th electron. 

We can define the $z$-axis to be the parallel to $\bm m$, without loss of generality.  
 For many reasonable model Hamiltonians, including the Schr\"odinger equation with Coulomb interactions between particles, the Hamiltonian commutes with all ${\hat S}^z_i$, and hence
\begin{eqnarray}
\left[ \hat{H}, {\hat m}^z\right] =0.
\label{eq:commutes}
\end{eqnarray}
An important result in quantum mechanics is that if any two operators commute then an eigenstate of one operator is also an eigenstate of the other. So, in this case energy eigenstates are also eigenstates of  ${\hat m}^z$.

A simple model where we can see this explicitly is the two site Heisenberg model:
\begin{eqnarray}
\hat H = J \hat{\bm S}_1 \cdot \hat{\bm S}_2 + h \sum_{i=1}^2  S^z_i.
\label{eq:H2Heis}
\end{eqnarray}
This is a simple model of an O$_2$ molecule, where each site represents an oxygen atom. 

Because the Hamiltonian commutes with the total spin and the $z$-component of the total spin (Eq. \ref{eq:commutes}) the eigenstates, $\ket{S_{tot},S^z_{tot}}$, can be labelled by their total spin, $S_{tot}$, given by the solution of $S_{tot}(S_{tot}+1)=\langle (\hat{\bm S}_{1}+\hat{\bm S}_{2})^2 \rangle$, and  $S^z_{tot}=S^z_{1}+S^z_{2}$.
For ferromagnetic interactions ($J<0$) and zero field ($h=0$) the ground state is three-fold degenerate, and hence known as a triplet, 
\begin{subequations}
	\begin{eqnarray}
		\ket{1,1} &=& \ket{\uparrow, \uparrow} \\
		\ket{1,0} &=& \left(\ket{\uparrow, \downarrow } + \ket{\downarrow, \uparrow }\right)/\sqrt{2} \\
		\ket{1,-1} &=& \ket{\downarrow, \downarrow },
	\end{eqnarray}
where the states on the right-hand-side are written in the site basis $\ket{S^z_1,S^z_2}$. All three of the degenerate ground states have the same $S_{tot} = 1$, but  $S_{tot}^z$ is different in each state. The O$_2$ molecule  has a spin triplet ground state \cite{valence}.

A small magnetic field will pick out a unique ground state. For example, $h>0$ makes $\ket{1,-1}$ the  ground state. Crucially, if we prepare our system in this state and then turn off the field ($h\rightarrow0$) adiabatically, the system will remain in this state because $\ket{1,-1}$ is an eigenstate for all values of $h$ and eigenstates are stationary states. Thus, the final Hamiltonian (with $h=0$) is rotationally invariant, but the prepared ground state, $\ket{1,-1}$, breaks rotational symmetry. 

For antiferromagnetic interactions ($J>0$) the ground state is the singlet
\begin{eqnarray}
\ket{0,0} &=& \left(\ket{\uparrow, \downarrow } - \ket{\downarrow, \uparrow }\right)/\sqrt{2}.
\label{eq:2estatesS}
\end{eqnarray}
\label{eq:2estates}
\end{subequations}
This state retains the full rotational symmetry of the Hamiltonian.

\begin{figure}
	\begin{center}
		\includegraphics[width=0.5\columnwidth]{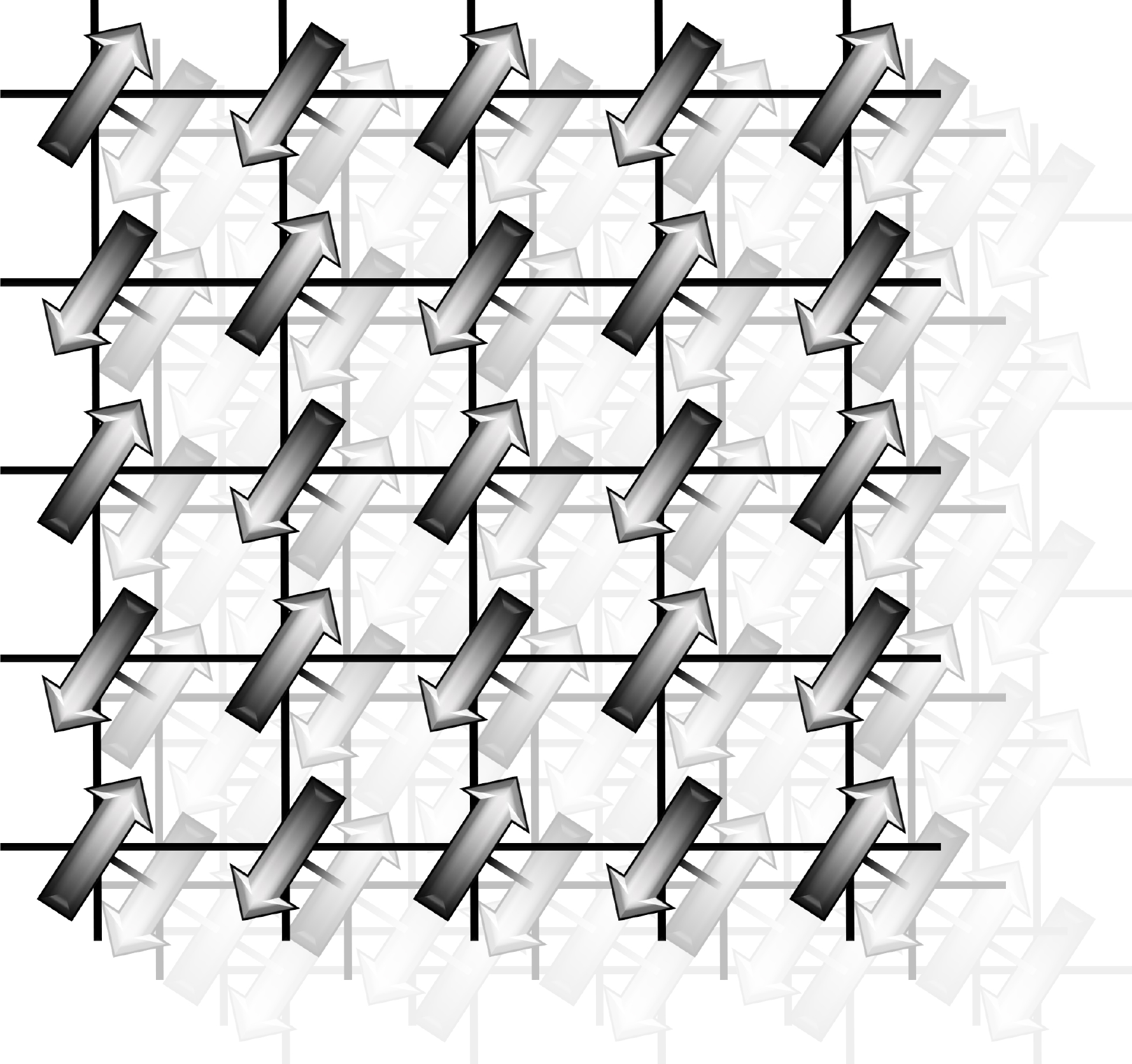}  
		\caption{Sketch of N\'eel order on a cubic lattice. The magnetic moments (spins) point in opposite directions on nearest neighbouring sites.}\label{fig:Neel}
	\end{center}
\end{figure}

The simplest type of antiferromagnetism is N\'eel order, sketched in Fig. \ref{fig:Neel}, characterised by zero magnetisation, but a non-zero `staggered magnetisation', 
\begin{eqnarray}
\bm{m}_s = \langle \hat{\bm{m} }_s\rangle = \frac{1}{N} \left\langle \sum_{i=1}^N (-1)^i \hat{\bm{S}}_i \right\rangle.
\end{eqnarray}
The  broken symmetry N\'eel states for our two-site model, for example $\ket{\uparrow, \downarrow }$ or $\ket{\downarrow, \uparrow }$, are not eigenstates (Eqs. \ref{eq:2estates} are the complete set of eigenstates). A  constant magnetic field (Eq. \ref{eq:H2Heis}) will not pick out the N\'eel state, but a staggered field, $h^s$, will give:
\begin{eqnarray}
\hat H = J \hat{\bm S}_1 \cdot \hat{\bm S}_2 + h^s \sum_{i=1}^2 (-1)^i  \hat{S}^z_i.
\label{eq:H2HeisStag}
\end{eqnarray}
In the limit $h^s/J\rightarrow\infty$ the ground state is $\ket{\uparrow,\downarrow}$. This is only an eigenstate  in this limit. Therefore, if we return $h^s\rightarrow0$ adiabatically, so that the system remains in its ground state throughout, we do not end up with a broken symmetry state. Instead the system returns to the symmetrical state $\ket{0,0}$.

Our two-site model is just a special case of a more general problem: the Heisenberg model on a bipartite lattice. A bipartite lattice is one where every site can be placed into two sets, $\mathcal{A}$ and $\mathcal{B}$, known as sublattices, such that there are no interactions between sites on the same sublattice (only interactions between sites on different sublattices are allowed). So in the two-site model we might choose $\mathcal{A}=\{1\}$ and  $\mathcal{B}=\{2\}$. In the N\'eel ordered state on a cubic lattice, Fig. \ref{fig:Neel}, we might choose that  the sites with up spins belong to sublattice $\mathcal{A}$ and  the sites with down spins belong to sublattice $\mathcal{B}$, which would make the lattice bipartite if the interactions correspond to the lines. It can be proven (Marshall's theorem) that for any bipartite lattice with equal size sublattices the ground state of the antiferromagnetic Heisenberg model is a spin singlet \cite{Auerbach}. Here a spin singlet is defined as any state with zero total spin, which therefore does not have degeneracy arising from the spin degrees of freedom. This generalises the state defined in \cref{eq:2estatesS} to any number of spins.  
Singlets are invariant under spin rotation. Therefore, the ground state must always have $ m^s =  0$. So Marshall's theorem seems to prove that we cannot break the symmetry in this case and hence that antiferromagnets cannot exist! This conflicts with the overwhelming experimental evidence for antiferromagnetism in real materials \cite{Shull,Brockhouse}.

The crucial difference from the ferromagnetic case is that the antiferromagnetic order parameter does not commute with the Hamiltonian:
\begin{eqnarray}
\left[ \hat{H}, \hat{m}_s^z \right] \ne 0.
\end{eqnarray}
Therefore, eigenstates of the staggered magnetisation are not, in general, energy eigenstates and \textit{vice versa}. Thus, eigenstates of the staggered magnetisation, such as N\'eel ordered states, are not stationary states. This means that although one could prepare a N\'eel state, say by applying a staggered field the system will evolve away from this state. The key question is then: how long does this evolution take? 

We can think of a superposition of classical states, like Eq. \ref{eq:2estatesS}, as the system quantum tunnelling between the classical states. From this perspective the question becomes: what is the characteristic time-scale for the tunnelling? Which is equivalent to asking how high the barrier between the classical states is. Intuition suggests, and  calculations confirm \cite{Coleman}, that the larger the system the higher the barrier will be. If the system is infinitely large, then the barrier is infinite; there is no tunnelling, and the broken symmetry state is a stationary state again. Mathematically, 
\begin{eqnarray}
\lim_{h^s\rightarrow0} \lim_{N\rightarrow\infty} {m}_s \ne 0.
\end{eqnarray}
Note that the order of limits is crucial here,
\begin{eqnarray}
\lim_{N\rightarrow\infty} \lim_{h^s\rightarrow0} {m}_s = 0,
\end{eqnarray}
consistent with the two site example above.
This shows us that phase transitions, such as antiferromagnetism, that involve order parameters that do not commute with the Hamiltonian can only really happen in the thermodynamic limit ($N\rightarrow\infty$).
In contrast, if the order parameter commutes with the Hamiltonian, as in ferromagnetism, then the thermodynamic limit is not essential and the order parameter can be non-zero even in very small systems, such as an oxygen molecule \cite{Reimers-chapter}, molecular magnets \cite{Schnack} or a few cobalt  atoms \cite{Brune}.

Real materials are often very large on the scale of atoms, but they are not strictly in the thermodynamic limit. So it is important to ask how much this affects the theory of spontaneous symmetry breaking. Back-of-the-envelope calculations \cite{Coleman} suggest that just a few hundred spins  are enough to give a lifetime for the N\'eel state of order years, and that in an antiferromagnet containing Avogadro's number of spins the symmetry broken state would have a lifetime that is much longer than the age of the universe. So for all practical purposes, bulk materials are in the thermodynamic limit and we should expect to see spontaneously broken symmetries even if the order parameter does not commute with the Hamiltonian.

Real magnets also often have other weaker interactions -- such as magnetic anisotropy and spin-orbit coupling -- that result in a the true Hamiltonian having a lower symmetry than that of the Heisenberg model. Such \textit{explicit} symmetry breaking terms in the Hamiltonian should not be confused with the \textit{spontaneous} symmetry breaking described above. Even when such explicit symmetry breaking terms are present, spontaneous breaking of the remaining symmetries of the Hamiltonian can still occur and, when it does, leads to a phase transition. 

Spontaneously broken symmetries are responsible for many other phase transitions. In fact spontaneously broken symmetries are at the heart of most of the phases of matter that we understand (important classes of phases that are not described by spontaneously broken symmetries will be discussed in section \ref{sect:BSM}). For example, superconductivity, superfluidity, crystallisation, and electroweak symmetry breaking all involve order parameters that do not commute with the Hamiltonian, see Table \ref{tab:SBS}.

\begin{table}
	\begin{center}
		\small
		\begin{tabular}{lccccc}
			Low-$T$ phase		& High-$T$ phase 		&  order parameter										& Nambu-Goldstone mode(s)							& $\omega({q\rightarrow0})$ \\
			\hline
			Ferromagnet			& Paramagnet			& magnetisation, $\bm m$								& magnon 									& $q^2$ \\
			Antiferromagnet		& Paramagnet			& \begin{tabular}{c}staggered\\
															magnetisation, $\bm m_s$\end{tabular}				& magnons 									& $q$ \\
			Crystal				& Liquid				& \begin{tabular}{c}density at inverse	\\
															lattice vectors, $\rho_{\bm G}$\end{tabular}		& \begin{tabular}{c} (longitudinal and\\ two transverse) \\
																													acoustic phonons\end{tabular}			& $q$ \\
			Neutral superfluid	& Fluid					& \begin{tabular}{c}macroscopic\\
															wavefunction, $\Psi$\end{tabular}					& phonon 									& $q$ \\
			Superconductor			& Metal			& \begin{tabular}{c}macroscopic\\
															wavefunction, $\Psi$\end{tabular}					& \begin{tabular}{c}none,\\Anderson-Higgs \\ 
																													(massive plasmon)\end{tabular} 			& $2\Delta$ \\
			Present universe	& Early universe		& Higgs field, $\phi$											& \begin{tabular}{c}none,\\ Anderson-Higgs \\
																													(massive weak bosons)\end{tabular} 		& $m_W\simeq m_Z$ 
		\end{tabular}
	\end{center}
\caption{Some examples of spontaneous symmetry breaking. Here $T$ is temperature, $\omega(q)$ is the dispersion relation (constants of proportionality are not included), $q$ is the momentum.}
\label{tab:SBS}
\end{table}

\subsection{Nambu-Goldstone bosons}

\subsubsection{Superfluidity}

Superfluidity is a many-particle quantum phenomena characterised by frictionless flow of the fluid. It has been observed in many systems including:
liquid helium-4, liquid helium-3, dilute atomic gases, and electron fluids (where it is known as superconductivity). Superfluidity is even believed to occur in neutron stars (pulsars).
Superconductivity was discovered in 1911; however, a microscopic theory was not developed until 1957. Nevertheless Ginzburg and Landau were able to predict many phenomena using a Landau theory. 
`All' they needed to do was to find the appropriate order parameter. Ginzburg and Landau's brilliant insight was to consider a complex scalar field, $\Psi({\bm r})$. Microscopic theory reveals that $\Psi$ is basically a macroscopic wavefunction (see \cite{Annett} for details).

The free energy is  real and analytic,  so cannot depend on odd powers of $\Psi$ or $|\Psi|$. Therefore, for small $\Psi$, the difference between the free energies of the superconducting and normal (non-superconducting) phases must be
\begin{eqnarray}
\Delta F= \alpha'(T-T_c) |\Psi|^2 + \beta |\Psi|^4. 
\label{eq:Flan-XY}
\end{eqnarray}
For $T>T_c$ the free energy is minimised when $\Psi=0$; for $T<T_c$ the free energy is minimised by
$\Psi=\Psi_0e^{i\theta}$, where $\Psi_0=\sqrt{{\alpha'(T_c-T)}/{2\beta}}$. 
Again, we have a freedom of choice -- any phase, $\theta$, leads to the same free energy.


\subsubsection{Phonons are Nambu-Goldstone bosons}

So far we have treated the order parameter as uniform. But, it can have both temporal and spatial variations. This has important consequences. To understand some of these, let us focus on a neutral superfluid such as superfluid $^4$He or a Bose-Einstein condensate in a dilute atomic gas. 

As when we developed our Landau theory, we can use symmetry to write down the lowest order allowed terms describing spatial variations. Again linear terms are forbidden by symmetry. Typically, variations in the phase of the order parameter are much lower energy than variations in its magnitude. Thus, we can write $\Psi(\bm r)=\Psi_0e^{i\theta(\bm r)}$, which yields ${\bm\nabla}\Psi(\bm r)=i\Psi_0{\bm\nabla}\theta(\bm r)$. The lowest order terms allowed in the Lagrangian are 
\begin{eqnarray}
{\cal L} = \frac{\rho_s}{2} \left[ \frac{1}{c^*}\left(\frac{{\partial}\theta}{\partial t}\right)^2 - ({\bm\nabla}\theta)^2 \right].
\label{eq:actionSuperfluid}
\end{eqnarray}
where the material specific constant, $\rho_s$, is known as the superfluid stiffness and $c^*$ is a material specific constant with dimensions of speed. The superfluid stiffness measures the rigidity of the phase of the superfluid, \textit{i.e.}, how hard it is to induce a phase difference between two spatially separated points. A similar analysis can be carried out for other broken symmetry phases and this kind of generalised rigidity consistently emerges. For example, in an antiferromagnet the equivalent constant of proportionality is known as the spinwave stiffness.

\begin{figure}
	\centering
	\includegraphics[width=0.5\textwidth]{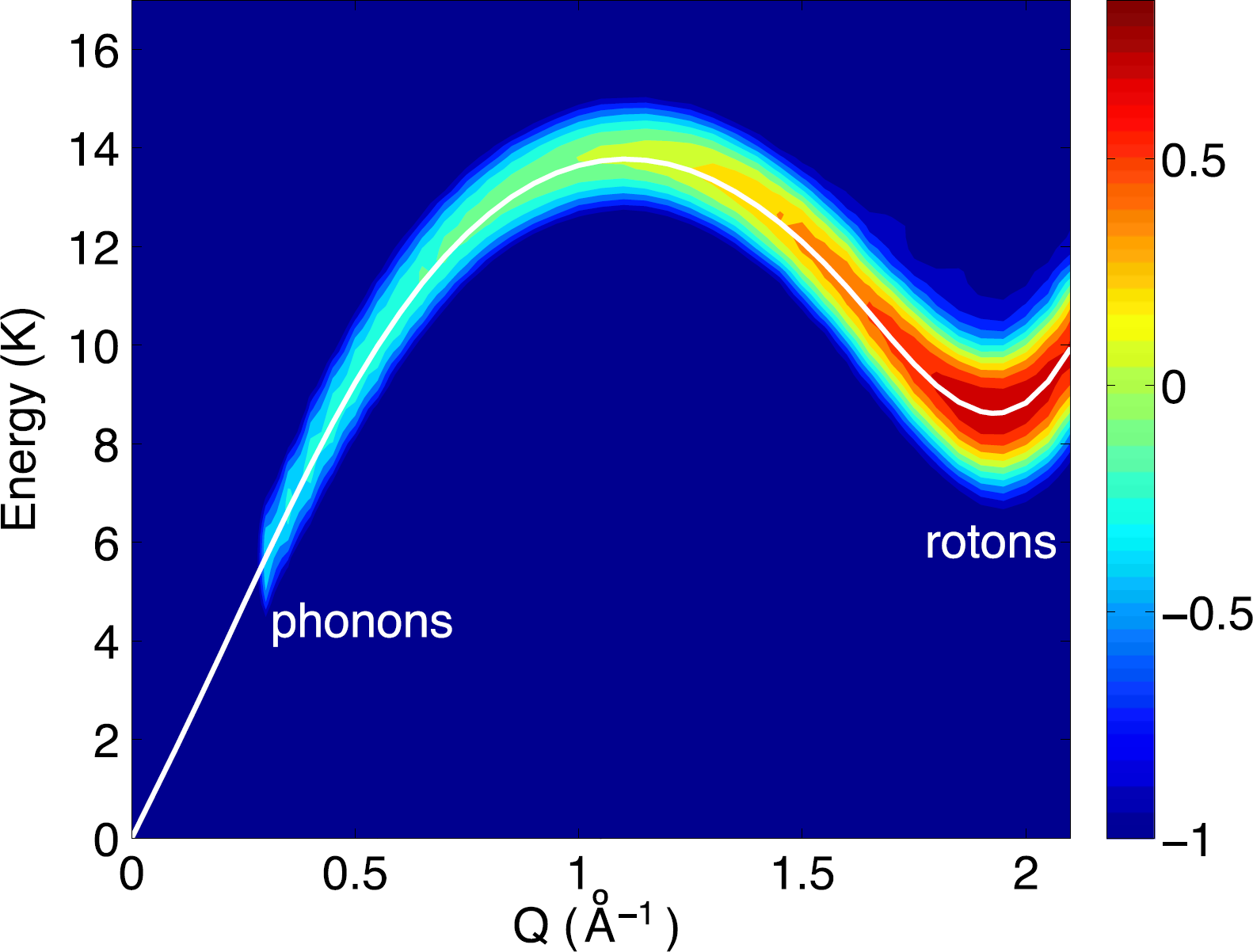}
	\caption{Dispersion relation for superfluid $^4$He measured via neutron scattering \cite{Andersen}. The massless phonon is observed at long wavelengths (small $Q$). The  minimum at higher wavenumbers is due to another emergent particle, the roton, that we will not have space to discuss here.  Reprinted  with permission from \cite{Fak} Copyright (2012) by the American Physical Society.} \label{fig:phonon}
\end{figure}

The principle of least action \cite{FeynmanHibbs} allows us to derive an equation of motion for the phase:
\begin{eqnarray}
 {\bm\nabla}^2\theta  =  \frac{1}{c^*}\frac{{\partial}^2\theta}{\partial t^2},
\end{eqnarray}
which has the solution
\begin{eqnarray}
\omega = c^* q \hspace{1cm} \textrm{or equivalently} \hspace{1cm} E = c^*p, \label{eq:Nambu-Goldstone}
\end{eqnarray}
where $E=\hbar \omega$ is the excitation energy (frequency) and $p=|\bm p|=\hbar q=\hbar|\bm q|$ is the momentum (wavenumber) of the excitation. This prediction is in excellent agreement with experiment, Fig. \ref{fig:phonon}.
The dispersion relation (Eq. \ref{eq:Nambu-Goldstone}) tells us that we have a  massless mode  due to collective phase fluctuations. The mode is called massless by comparison to the relativistic dispersion relation $E^2 = m^2c^4 + p^2c^2$, which simplifies to Eq. \ref{eq:Nambu-Goldstone} for $m=0$. Masslessness tells us that the excitation energy, $E=\hbar\omega$, goes to zero at long wavelengths, $\lambda=2\pi/q$.
The massless mode is only found below $T_c$ and is called the phonon. From \cref{eq:Nambu-Goldstone} we see that $c^*$ is the speed of sound in the superfluid.

This result is not peculiar to a superfluid. Generically, whenever a continuous symmetry is spontaneously broken there is a new massless mode in the broken symmetry phase. 
Our Ginzburg-Landau--like theory, above, was classical. A quantum calculation \cite{Fradkin} gives the same dispersion relation (Eq. \ref{eq:Nambu-Goldstone}) and also finds that the classical wave becomes a quantum particle with bosonic statistics. This is not entirely surprising, as quantising a classical mode usually results in a boson.
Thus, the massless mode is called the Nambu-Goldstone boson. Some examples of Nambu-Goldstone bosons are given in Table \ref{tab:SBS}.

The masslessness of the Nambu-Goldstone boson is a profound consequence of the spontaneously broken symmetry. To see the connection  think about the `excitations' at $q=0$. As the excitation energy is $E=0$, these are not excitations at all but degenerate ground states. The phonon in the superfluid is a long wavelength variation in the phase of the order parameter. As $\lambda\rightarrow\infty$ ($q\rightarrow0$), the  phase difference between two points at a fixed distance goes to zero. So at $q=0$ the phase is the same everywhere; we have a uniform change in phase throughout the superfluid. Thus, the degenerate states at $q=0$ are just the degenerate ground states related by the symmetry that has been spontaneously broken. This gives a simple physical picture of the Nambu-Goldstone boson: as $q\rightarrow0$  this mode corresponds to running around the bottom of the Ginzburg-Landau potential, Fig \ref{fig:Landau}b. 

The picture of the Nambu-Goldstone boson running around the bottom of the Ginzburg-Landau potential demonstrates that Nambu-Goldstone bosons only occur when the broken symmetry is continuous,  Table \ref{tab:SBS}. If the broken symmetry is discrete,\footnote{For example a mirror symmetry -- Alice can be on this side of a mirror or through the looking glass but never part way through} then the entire construction  collapses. There is no concept of moving to another degenerate ground state that is almost the same as the initial ground state (as there is in the superfluid, where this corresponds to changing the phase by an infinitesimal amount). So, if the broken symmetry is discrete, we cannot run around the bottom of the Ginzburg-Landau potential and there are no new long wavelength modes  in the broken symmetry  phase.

\subsubsection{How many magnons? Counting Nambu-Goldstone bosons}

\begin{figure}
	\centering
	\includegraphics[width=0.5\textwidth]{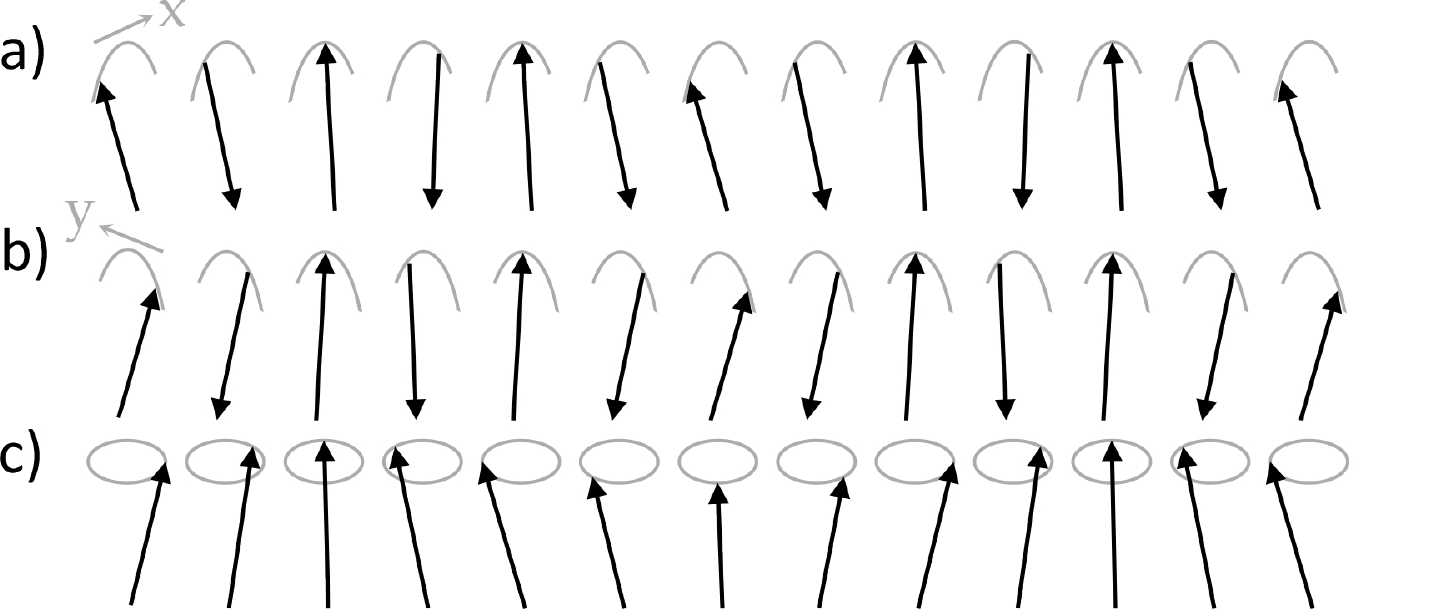}
	\caption{Sketches of the differences between magnons in antiferromagnets and ferromagnets. (a,b) There are two perpendicular polarisations of magnons in an antiferromagnet. Both have linear dispersions at long wavelengths. (c) In a ferromagnet the two polarisations are coupled by a Wess-Zumino term, resulting in a single mode with a quadratic dispersion.} \label{fig:magnon}
\end{figure}

Multiple symmetries are broken in some phase transitions; for example, in an antiferromagnet two continuous symmetries are broken\footnote{Actually, a third symmetry -- translation by a lattice vector -- is also broken. But, this is a discrete symmetry and so is not relevant to our discussion here.}: rotation about the $x$ and $y$ axes (for $\bm{m}_s\|{\bm{z}}$, which we assume below without loss of generality). Again we can write down a Lagrangian that will describe the low-energy physics by keeping only the lowest order terms consistent with the symmetry of the problem. This yields the `non-linear sigma model' \cite{Fradkin}:
\begin{equation}
{\cal L}_\text{NL$\sigma$M} = \frac{\rho_\text{AFM}}{2} \left[ \frac{\dot{\bm{n}}_s^2}{c^{*2}} - (\bm{\nabla}\bm{n}_s)^2  \right], \label{eq:Lafm}
\end{equation}
where $\rho_\text{AFM}$ is the spin wave stiffness and $m_s\bm{n}_s=\bm{m}_s$ with $|\bm{n}_s|=1$ and $m_s=|\bm{m}_s|$.  The principle of least action yields two independent equations, 
\begin{eqnarray}
{\bm\nabla}^2 n_S^x  =  \frac{1}{c^{*2}}\frac{{\partial}^2 n_S^x}{\partial t^2}, \\
{\bm\nabla}^2 n_S^y  =  \frac{1}{c^{*2}}\frac{{\partial}^2 n_S^y}{\partial t^2}.
\end{eqnarray}
Thus, we find two Nambu-Goldstone modes (magnons), both are massless and have linear dispersion relations ($\omega=c^*p$). (Figure \ref{fig:magnon}a,b)

Similarly, crystallisation breaks three symmetries: translational symmetry in the $x$, $y$ and $z$ directions.  For example, in a crystal there are three Nambu-Goldstone modes: the longitudinal acoustic phonon and two transverse acoustic phonons, all of which are massless and have linear dispersion relations.

The Lagrangians describing the low-energy dynamics of a superfluid,  Eq. (\ref{eq:actionSuperfluid}), an antiferromagnet, Eq. (\ref{eq:Lafm}), or a crystal, all have a pseudo-Lorentz invariance (they would be Lorentz invariant for $c^*=c$, the speed of light). If the low energy dynamics are governed by a (pseudo-)Lorentz invariant Lagrangian,  then one can prove (Goldstone's theorem \cite{Watanabe-review,LancasterBlundell}) that the number of Nambu-Goldstone modes is equal the number of broken symmetries. In a Lorentz invariant theory (such as the standard model of particle physics) the Lagrangian must be Lorentz invariant, and so this is all that can happen. But, the absence of Lorentz invariance in crystals allows for richer possibilities. 

In a ferromagnet, the rotational symmetries about the $x$ and $y$ axes are broken (for $\bm{m}\|{\bm{z}}$), the same continuous symmetries that are broken in an antiferromagnet. However, a detailed calculation for the ferromagnetic Heisenberg model \cite{Auerbach} finds only one massless mode with $E\propto J(1-\cos{q})\simeq Jq^2$ at long wavelengths. Clearly, something is missing from our low-energy description.

In classical mechanics the Lagrangian and the Hamiltonian are related via 
	${\cal L} = \sum_i p_i\dot{q_i} -  {\cal H}$,
where the coordinates $q_i$ and $p_i$ are the canonical momenta.
In quantum mechanics $q_i$ and $p_i$ obey a canonical commutation relations,
	$[q_i,p_i]=i$,
for $\hbar=1$. In quantum field theory one can use the same trick to write down a Lagrangian whenever one has canonically commuting variables \cite{LancasterBlundell}. For example, consider a quantum field theory of a ferromagnet. Again we will assume that $\bm{m}\|\bm{z}$ for our notational convenience. The individual spins obey the canonical commutation relation $[\hat{S}_i^x,\hat{S}_i^y]=i \hat{S}_i^z$ and therefore
$[\hat{m}^x,\hat{m}^y]=i \hat{m}^z$.
As we are dealing with a ferromagnet $m^z=\langle \hat{m}^z \rangle \ne 0$ and the $x$ and $y$ components of the magnetisation do not commute,  even in a coarse-grained theory. Thus, the low-energy collective dynamics of a quantum ferromagnet are described by the Lagrangian 
\begin{equation}
	{\cal L}_\text{FM} = m \frac{n^y\dot{n}^x - n^x\dot{n}^y}{1+n^z} + \frac{\rho_\text{FM}}{2} \left[ \frac{\dot{\bm{n}}^2}{c^{*2}} - (\bm{\nabla}\bm{n})^2  \right], \label{eq:WZHeis}
\end{equation}
where $m\bm{n}=\bm{m}$ with $|\bm{n}|=1$ and $m=|\bm{m}|$. The first term 
known as a Wess-Zumino term. The principle of least action and the assumption that $1-n^z$ is small lead to the (linearised) Landau-Lifshitz equations \cite{Fradkin}:
\begin{eqnarray}
	\dot{n}^x &=& \frac{\rho_\text{FM}}{m} \bm{\nabla}^2 n^y; \\
	\dot{n}^y &=& -\frac{\rho_\text{FM}}{m} \bm{\nabla}^2 n^x.
\end{eqnarray}
Notice that the effect of the Wess-Zumino term is to couple the $n^x$ and $n^y$ degrees of freedom. The  solution of the Landau-Lifshitz equations is 
\begin{equation}
\omega = \frac{\rho_\text{FM}}{m} p^2,
\end{equation}
consistent with the microscopic solution. (Figure 6c)

In an antiferromagnet 
\begin{equation}
[\hat{m}_s^x,\hat{m}_s^y]=i\hbar \hat{m}^z.
\end{equation}
Note that it is the magnetisation (not the staggered magnetisation) that appears on the right-hand side. However, $m^z\equiv\langle \hat{m}^z \rangle =0$, so the components of the order parameter commute in our effective theory. This means that there is no Wess-Zumino term and the non-linear sigma model is the correct description.

This result can be made general \cite{Watanabe}. Whenever the effective Lagrangian has a pair of variables that obey a canonical commutation relation  then there is a topological term in the Lagrangian involving both variables. This `interaction' means that the na\"ively expected pair of linearly-dispersing Nambu-Goldstone bosons are replaced by a single quadratically-dispersing Nambu-Goldstone boson. Examples of broken symmetry phases with quadratically-dispersing Nambu-Goldstone bosons and a correspondingly `reduced' number of Nambu-Goldstone bosons include the phonons in crystals of skyrmions,  Wigner crystals  in a magnetic field,  spinor Bose-Einstein condensates,  and kaon condensates with a non-zero chemical potential \cite{Watanabe-review}.

Spontaneously broken symmetries and Nambu-Goldstone bosons are the second ingredient of the standard model of materials. If a continuous symmetry is broken we expect to have massless bosonic modes. But, there are important exceptions to this, which we will explore  next.

\subsection{How the photon got its mass}\label{sect:Higgs}

But, you might object, photons are massless. This is true in the vacuum. But not, we shall see, in superconductors. The most famous property of superconductors is that their electrical resistance is exactly zero. However, that would also be true of a perfect conductor -- a metal with no imperfections cooled to absolute zero. The defining property of a superconductor is perfect diamagnetism, the complete exclusion of magnetic field from the bulk, which is known as the Meissner effect. Zero resistance is a consequence of the Meissner effect \cite{Annett}.  

\subsubsection{Gauge invariance in electromagnetism}

In classical electromagnetism the physical observables are the electric and magnetic fields, $\bm E$ and $\bm B$. However, it is often convenient to rewrite the theory in terms of vector and scalar potentials, $\bm A$ and $\phi$, that are not themselves observable \cite{Jackson},
\begin{subequations}
	\begin{eqnarray}
\bm B&=&\bm\nabla \times \bm A \\
{\bm E}&=&-{\bm \nabla}\phi -\frac{\partial \bm A}{\partial t}.
\label{eq:EBgauge}
\end{eqnarray}
\end{subequations}
All physical observables (including, $\bm E$, $\bm B$, and the Aharonov-Bohm and Berry phases \cite{Bernevig}) are invariant under a gauge transformation defined by 
\begin{subequations}
\begin{eqnarray}
	\psi&\rightarrow&\psi e^{i\Lambda(\bm r,t) q/\hbar}, \\
	\phi&\rightarrow&\phi-\frac{\partial\Lambda(\bm r,t)}{\partial t} \\
	{\bm A}&\rightarrow&{\bm A}+{\bm \nabla}\Lambda(\bm r,t), \label{eq:gaugetransA}
\end{eqnarray}
	\label{eq:gaugetrans}
\end{subequations}
for any scalar field $\Lambda(\bm r,t)$

Sometimes the invariance of physical observables under gauge transformations is referred to as `gauge symmetry'. This is misleading. Usually,  when we talk of a symmetry we mean that there are two or more states with the same physical properties. The `generator' of the symmetry maps us from one state to another with the same properties. A gauge transformation is not a generator -- it maps us from one state to \textit{the same state}. All that changes is our mathematical description of the state. For example, in classical electromagnetism $\bm E$ and $\bm B$ are unchanged; only $\bm A$, $\phi$ and the phase of the wavefunction, $\psi$, which are unobservable, have changed.

It is obvious from this that `gauge symmetry' cannot be broken -- it is not a physical symmetry, just a redundancy in our mathematical description of the physics. Nevertheless, you can read statements in the literature that gauge symmetry is broken in superconductors, because the spontaneously broken symmetry picks out a phase for the macroscopic wavefunction. Clearly, something more subtle must be happening in a superconductor than a broken gauge symmetry. It turns out that this has physical consequences for the Nambu-Goldstone mode, the photon and leads directly to the Meissner effect.

\subsubsection{The Anderson-Higgs mechanism}

The essential difference between a superfluid and a superconductor is that the superconductor carries an electrical charge, and hence the dissipationless flow becomes a current that flows without resistance. To understand gauge transformations in a superconductor we need to include the `minimal coupling' between a charged particle and electromagnetic fields in our Lagrangian. Thus, the Lagrangian of a neutral superfluid, Eq. \ref{eq:actionSuperfluid}, is replaced by a `gauged' contribution of the condensate to the Lagrangian, ${\cal L}_\Psi$. As the photon is massless, we also need to add in the electromagnetic Lagrangian, ${\cal L}_\text{EM}$ \cite{FeynmanHibbs}, in order to properly describe the low energy dynamics. Hence, ${\cal L}={\cal L}_\Psi+{\cal L}_\text{EM}$, where
\begin{eqnarray}
{\cal L}_\Psi = \frac{\rho_s}2 \left[ \left(\frac{\dot\theta}{c^*}+\frac{e^*}{\hbar}\phi\right)^2 - \left(\bm\nabla\theta-\frac{e^*}{\hbar}\bm A\right)^2 \right],
\end{eqnarray}
$e^*$ is the effective charge and
\begin{eqnarray}
{\cal L}_\text{EM} =  \left(\frac{\bm E}c \right)^2- \bm B^2. 
\end{eqnarray}

As the scalar and vector potentials always occur in the same, gauge invariant, combination with the gradients of the phase, we can simply redefine the electromagnetic fields to absorb the gradients of the phase:
\begin{subequations}
	\begin{eqnarray}
	\bm A &\rightarrow& \bm A+\frac\hbar{e^*}\bm\nabla\theta, \\
	\phi &\rightarrow& \phi-\frac\hbar{e^*c^*}\dot\theta.
	\end{eqnarray}
	\label{eq:trans}
\end{subequations}
This gauge transformation leaves $\bm E$ and $\bm B$  unchanged  (compare Eqs. \ref{eq:trans} with Eqs. \ref{eq:gaugetrans}). However, this transformation couples the vector potential, which is associated with transverse modes, with the gradients of the phase, which are longitudinal -- thus, the new field has both transverse and longitudinal components.
In terms of the new fields, the Anderson-Higgs Lagrangian is
\begin{eqnarray}
{\cal L} =   \frac1{2\mu_0\lambda_L^2}\left[ \left(\frac\phi{c^*}\right)^2 - \bm A^2 \right] + \frac1{2\mu_0}\left[ \left(\frac{\bm E}c\right)^2-\bm B^2\right],
\end{eqnarray}
where $\lambda_L\equiv\sqrt{\hbar^2/\rho_s \mu_0 e^{*2}}$ is known as the London penetration depth. Thus, we have a purely electromagnetic action, but there is a new quadratic term in the Lagrangian of the electromagnetic field (the term proportional to $1/\lambda_L^2$). The effect of `integrating out' the matter terms from our effective low-energy action is to create a mass in our field.

In the non-superconducting case ($\rho_s=0$ or equivalently $\lambda_L=\infty$), taking variations in  with respect to $\phi$ and $\bm A$ allows us to derive Maxwell's equations. If we do this for the full Anderson-Higgs Lagrangian with $\rho_s\ne0$ we find that Gauss' law is 
$\varepsilon_0\bm\nabla\cdot\bm E-\rho=0$,
where $\rho = -\frac1{\mu_0c^{*2}\lambda_L^2}\phi$ and $\varepsilon_0=\frac1{\mu_0c^2}$; and Ampere's law is
$\frac1{\mu_0}\left(\frac1{c^2}\dot{\bm E}-\bm\nabla\times\bm B\right)+\bm j=0$,
where $\bm j = -\frac1{\mu_0\lambda_L^2} \bm A$.
These can be combined via standard vector identities  
to give
\begin{eqnarray}
\left[ \bm\nabla^2-\frac1{c^2}\frac{\partial^2}{\partial t^2}-\frac1{\lambda_L^2}  \right]\bm A= \left[1 - \left(\frac{c^*}{c}\right)^2\right]\bm\nabla(\bm\nabla\cdot{\bm A}).
\end{eqnarray}

If we substitute $\bm A=A_0e^{i(\bm p\cdot\bm x-E_{\bm p}t)/\hbar}\hat{\bm e}$, where $\hat{\bm e}$ is a unit vector, we find that the dispersion relation for longitudinal waves ($\hat{\bm e}\|\bm p$) is
\begin{subequations}
	\begin{eqnarray}
	E_{\bm p}=\sqrt{\left(m_Ac^2\right)^2+(\bm pc^*)^2},
	\end{eqnarray}
	whereas  the dispersion relation for transverse waves ($\hat{\bm e}\perp\bm p$) is
	\begin{eqnarray}
	E_{\bm p}=\sqrt{\left(m_Ac^2\right)^2+(\bm pc)^2},
	\end{eqnarray}
	\label{eq:masses}
\end{subequations}
where 
$m_A=\hbar/{\lambda_L c}$.
These are just the usual relativistic dispersion relations for particles of mass $m_A$. The photons have developed a mass due to the Anderson-Higgs mechanism!

If there is no coupling between the matter and light fields (as in a neutral superfluid) then we have massless photons with only transverse polarisations that travel at the speed of light, $c$, and massless Nambu-Goldstone modes (phonons) with only  longitudinal polarisations that travel at the speed of sound, $c^*$. The coupling between the fields changes this dramatically. In a superconductor there is only one mode, which is usually called a plasmon, that is massive and has fast transverse components (travelling at $c$) and a slow longitudinal component (propagating at $c^*$). This is the Anderson-Higgs effect.

The mass of the photon inside a superconductor has important experimental consequences. In quantum field theories forces are mediated by the exchange of virtual gauge bosons. The range of the force is determined by the mass of the force carrier.
For example, the long-range electromagnetic force is carried by massless virtual photons whereas the short range of the weak nuclear force is due to the large masses of the weak bosons.
We can motivate this via the following  argument, which is far from rigorous.
The uncertainty principle allows quantum fluctuations to `borrow' some energy, $\Delta E$, from the vacuum for a  time $\Delta t\sim \hbar/\Delta E$. Therefore a long-wavelength force carrier with mass $m_A$ can travel a distance
\begin{eqnarray}
\Delta x\sim c\Delta t\sim \frac{\hbar}{m_A c}=\lambda_L.
\end{eqnarray}
In the vacuum, $m_A=0$ ($\lambda_L=\infty$) and the electromagnetic interaction is long-ranged as we expect. But, inside a superconductor $m_A\ne0$ ($\lambda_L$ is finite) and the magnetic field can only penetrate a depth $\sim \lambda_L$, which is typically tens or hundreds of nanometers. This short penetration depth for magnetic fields means that inside the bulk of a superconductor the magnetic field vanishes. This is  the Meissner effect, the defining property of a superconductor.

In particle physics the Anderson-Higgs mechanism is associated with electroweak symmetry breaking. Direct experimental evidence for this comes from the observation of the (massive) mode associated with amplitude oscillations of the condensate, Fig. \ref{fig:Landau}b, which, in a quantum theory, is known as the Higgs boson. This mode is less important in superconductors because we already have direct evidence for the Anderson-Higgs mechanism from the Meissner effect, and because we are generally most interested in the low-energy excitations of a material. The Higgs mode is challenging to observe in superconductors because it does not carry an electric charge, a  dipole, or a magnetic moment (in other words, it is a scalar boson). Nevertheless, it is possible to observe the Higgs boson in non-linear optical experiments such as third harmonic generation and pump-probe spectroscopy as it has a second order coupling to the electric field \cite{Shimano}. Similar collective modes resulting from  fluctuations in the  amplitude of the order parameter are  observed for several other spontaneously broken symmetries \cite{Varma}.

The Anderson-Higgs mechanism is the third ingredient of the standard model of materials. If the matter field couples to a gauge field there is no Nambu-Goldstone mode. Instead the gauge field `swallows the Nambu-Goldstone boson' and we are left with a single massive boson (the plasmon).

\subsection{Topological defects in the order parameter}

\begin{figure}
	\centering
	\includegraphics[width=\textwidth]{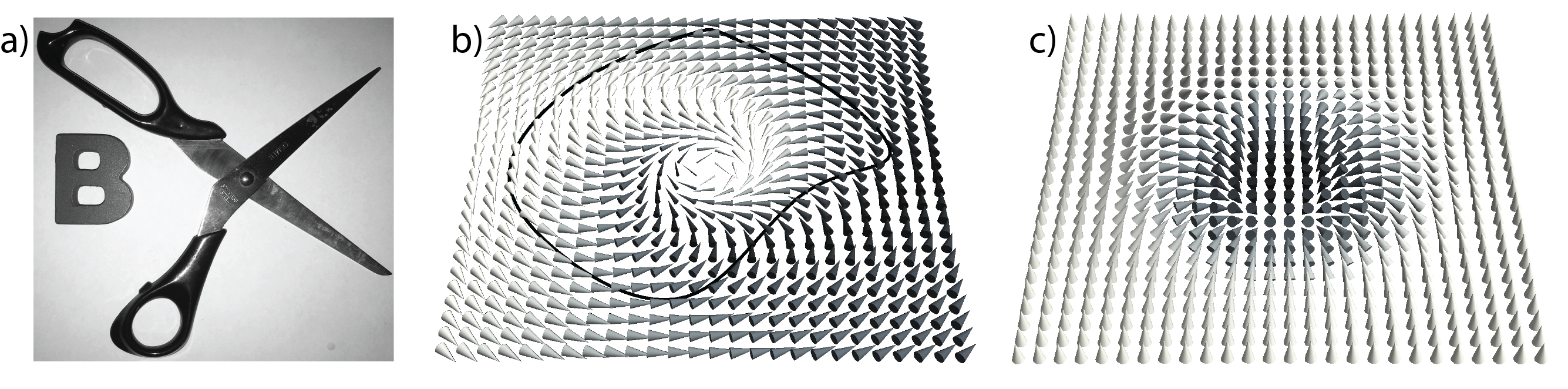}
	\caption{(a) Spot the difference. (b,c) Topological defects.  The direction and shading of the cones  indicate the direction of the order parameter at that location; their size indicates the magnitude of the order parameter. (b) Near a vortex the order parameter winds around the core. Notice that as the phase is not defined at the centre of the vortex  the magnitude must vanish. This is the same physics as the calm at the eye of a hurricane. The black line is an example of a closed loop enclosing the vortex core. (c) In a skyrmion the local order parameter reverses direction as we move from the centre to the outside. There is no continuous function that can be added to either of these vector fields that removes the defect (i.e., returns the field to the ground state where all the arrows are parallel). Thus, these are topological defects.
} \label{fig:vortex}
\end{figure}

We have seen above that the phase of a superconductor can vary in space. However, there is a constraint on this variation: the order parameter, $\Psi({\bm r})=|\Psi({\bm r})|e^{i\theta{\bm r}}$, and hence its phase and magnitude are   continuous functions of $\bm r$. Therefore, the order parameter  must return to its initial value if we traverse any closed path. We can parametrise $\bm r$ by the distance we have travelled along the path, $d$, so that $d=0$ is the beginning of the path and $d=1$ is the end. As the path is closed, $d=0$ and $d=1$ are the same point so $\Psi(d=0)=\Psi(d=1)$, Fig. \ref{fig:vortex}b.

The continuity of the order parameter imposes a weaker constraint on the phase of the order parameter, $\theta(d=0)=\theta(d=1)+2n\pi$.
The winding number, $n\in\mathbb{Z}$, counts the number of times $\theta$ winds  around itself along the closed path,  Fig. \ref{fig:vortex}b. Smooth local variations in $\theta({\bm r})$, i.e., gauge transformations (Eq. \ref{eq:gaugetrans}), cannot change $n$.
Configurations with $n=1$ ($-1$) are known as (anti)vortices.

If we imagine shrinking the path so that it  passes very close to the centre of the vortex, then something strange must happen. The problem is analogous to defining time at the south pole. If I travel due south from Brisbane (GMT+10) and my friend travels due south from Lagos (GMT+1) neither of us changes time zone and so we do not need to adjust our watches. Nevertheless, we will disagree about the time if we both travel all the way to the south pole. In a superconductor the phase is undefined at the centre of a vortex -- this is only possible if the magnitude of the order parameter, $|\Psi({\bm r})|$, vanishes at the centre of the vortex. 

These problems are related via a branch of mathematics known as topology. Topology is most simply introduced via the study of geometric objects under continuous deformations, \textit{e.g.}, twisting and stretching, but not discontinuous deformations like tearing and gluing. In a topological sense, a pair of scissors is the same as the letter B as both are tori with two holes (Fig. \ref{fig:vortex}a). 
The number of holes in a torus is a topological invariant -- it does not change under any of the allowed transformations.

Topology can be used to describe more abstract objects. For example, there is no way to continuously deform the surface of the earth to remove the problem of defining local time at the poles. Similarly, the winding number around a closed path in a superconductor is topologically invariant because it cannot be changed by smooth local deformations in the underlying field, $\theta(\bm r)$. Vortices are found in many classes of materials, such as superconductors, superfluids, ferromagnets and antiferromagnets \cite{Chaikin}. Indeed, in two dimensions vortices dominate the low-energy physics of these systems. This gives rise to a special kind of phase transition, known as the Kosterlitz–Thouless transition and leads to low-temperature phases with quasi-long-range order \cite{Chaikin}.

The phase of a superconducting order parameter is  the mathematically simplest kind of object that can host a topological defect.\footnote{A phase has a $U(1)$ symmetry, whereas a vector field in three dimensions (3D) has $SO(3)$ symmetry.} 
If the order parameter is a 3D vector (rather than a 2D vector or a phase)  more complicated topological defects can emerge, such as the skyrmions found in ferromagnets (Fig. \ref{fig:vortex}c) and antiferromagnets \cite{Lancaster}. Topological defects also play important roles in the physics of liquid crystals and determining the mechanical properties of crystals \cite{Chaikin}.

Topological defects have all the properties we usually associate with particles. They can move around, they have long range interactions, and crystals of topological defects have been observed in many contexts -- most famously the Abrikosov vortex lattice in superconductors \cite{Annett}. Topological defects are typically massive because of the energy cost of suppressing the order parameter in the core of the defect.

Topological defects are the final ingredient in the standard model of materials. Provided the order parameter has a continuous symmetry, we should expect them to occur.

\section{Beyond the standard model: entanglement and topology}\label{sect:BSM}

The standard model of materials is based on two concepts: 
\begin{enumerate}
	\item Adiabatic continuity and quasiparticles; and
	\item The order parameter.
\end{enumerate}
So, if we are looking for physics beyond the standard model of materials, a good starting point is to look for materials that are not described by these paradigms.

\begin{figure}
	\centering
	\includegraphics[width=3cm]{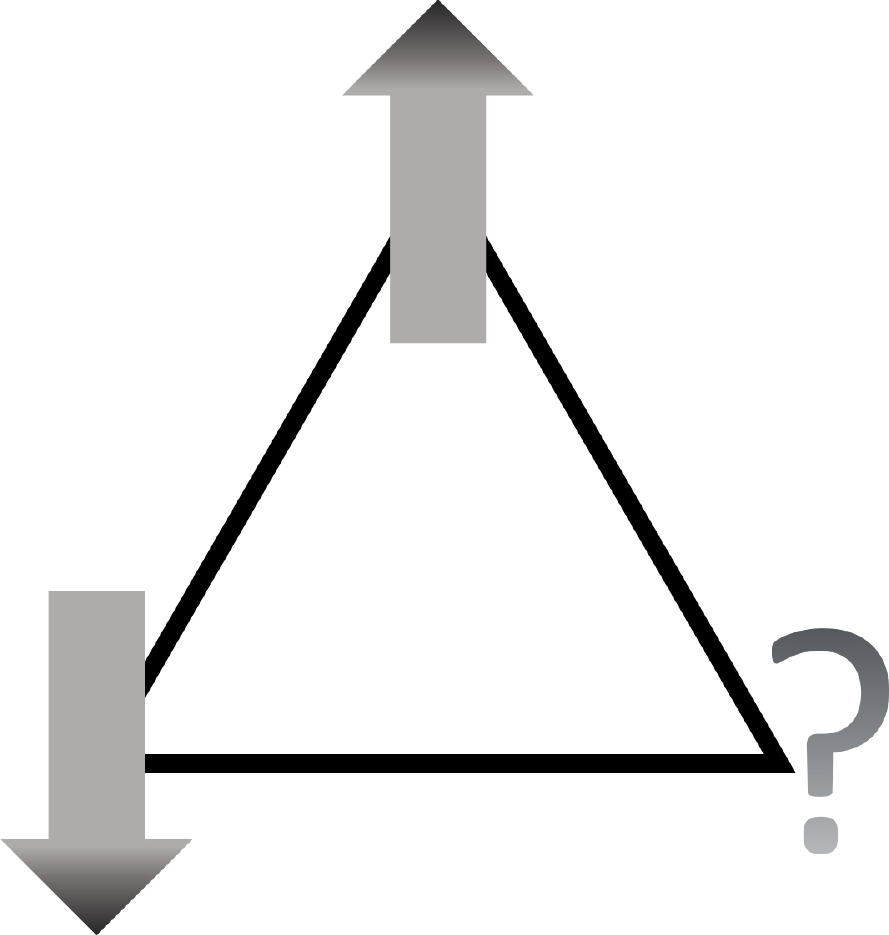}
	\caption{Geometric frustration. There is no way, classically or quantum mechanically, to arrange three spins at the corners of a triangle so that every pair of spins points in opposite directions. Here we cannot chose any orientation for the right-hand spin that is antiparallel to both the top and left-hand spins. More generally, we cannot have all nearest neighbours antiparallel on any polygon with an odd number of vertices. This means that we cannot find a state that minimises every term in, say, the Heisenberg, $H=J\sum_{ij} \hat{\bm S}_i \cdot \hat{\bm S}_j$, or Ising, \cref{eq:Ising}, models. Something more interesting must happen!  Generically, frustration leads to many competing low-energy states.} \label{fig:tri}
\end{figure}

Long-range order implies a non-zero local order parameter.
Therefore, systems with no long-range order  are  promising places to look for exotic particles.  Long-range order is largely absent in 1D  \cite{Halperin} and is suppressed in higher dimensions when the interactions are frustrated and compete in just the right way so that none of them win out. A simple example of this is three spins on a triangle with antiferromagnetic interactions: there is no way to simultaneously minimise all three interactions (Fig. \ref{fig:tri}).  
A third place we might look for new physics is when the inter-particle interactions are so strong  that adiabatic continuity to the non-interacting system fails.

In this section, we will explore examples of  these frontiers. Several leitmotifs  emerge:
\begin{itemize}
	\item Collective excitations  that carry only part of the `charge' of the elementary building blocks of the system. This can mean electrical charge, or it can  be generalised to  other properties, such as  spin or  magnetic dipole moment. When the `charge' of the emergent particles in a material is less than the charges of the basic building blocks the excitations are said to be fractionalised.
	\item Frustration leads to degeneracy in classical systems, \cref{fig:tri}. Often the degenerate states are related by a conservation law. The classical ground states can then be described by an emergent gauge field. Quantum mechanical terms couple the classically degenerate ground states. An emergent gauge field often remains in the quantum description, but  not necessarily the same gauge field as found classically. If the gauge symmetry is continuous we expect to find a gauge boson in the system.
	\item Quantum matter is often topologically non-trivial. One signature of this is that the ground state degeneracy depends on the topology of the lattice. This may enable practical applications of topological materials and metamaterials for fault tolerant quantum information processing.
	\item Topological states in the bulk of a material  typically have consequences at the  boundary between the material and the vacuum (or a topologically trivial material). This  is known as the bulk-boundary correspondence and can result in fractionalised particles on the surface.
\end{itemize}

\subsection{Correlations and entanglement}\label{sect:entag&correl}

When we say that two particles are correlated, we mean that if we know something about the behaviour of one particle, we can make a (statistical) prediction about the behaviour of the other.
For example, the spin-spin correlation function, $C(r_{ij})=\langle \bm S_i \cdot \bm S_j \rangle$, where $r_{ij}$ is the distance between sites $i$ and $j$, measures   the correlations between spatially separate spins. If there is long-range ferromagnetic order then $\lim_{r\rightarrow\infty}C(r)=m^2$, cf. \cref{eq:magnetisation}. But, in the high-temperature disordered phase $\lim_{r\rightarrow\infty}C(r)=0$ .

Correlations between matter can be classical or quantum. The traditional quantum definition of electronic correlation compares  the wavefunction with the Hartree-Fock state, which can be defined as the lowest energy state that can be written as an (antisymmetrised) product of single particle wavefunctions  (\textit{i.e.}, as a single Slater determinant) \cite{Rossler}. The correlations are the multi-determinant character neglected in the Hartree-Fock approximation. This already clearly has a quantum character and antisymmetrisation excludes the trivial correlations that are always present for fermions. Notice that this definition of correlations involves the Hamiltonian as it involves the minimum energy state.

Entanglement is a measure of  quantum mechanical correlations that  arose from the study  and quantum information and can be defined as the correlations that allow the violation of a Bell inequality \cite{Barrett}. Typically, entanglement measures, like Bell inequalities,  invoke a concept of locality. This is not always natural in a material -- \textit{e.g.}, for itinerant electrons it is  natural to describe the system in terms of delocalised (Bloch) states -- which leads to difficulties defining entanglement for itinerant systems. The singlet $\ket{S}\equiv\ket{\uparrow}_i\ket{\downarrow}_j - \ket{\downarrow}_i\ket{\uparrow}_j$,  is the prototypical entangled state:  we cannot describe it as a product of a state, $\ket{\psi}_i$, at location $i$ and another state, $\ket{\phi}_j$, in location $j$. That is $\ket{S}\ne\ket{\psi}_i\ket{\phi}_j$ for any choice of $\ket{\psi}_i$ and $\ket{\phi}_j$. In words, $\ket{S}$ is not a separable state. Note that the definition of entanglement does not involve a Hamiltonian, only the state.

The above two definitions are related but importantly different. If a state is correlated one cannot write down a wavefunction for any particle without specifying what the other particles are doing. If a state is entangled one cannot write down a wavefunction anywhere without specifying what is happening elsewhere.

To make this more concrete, consider a toy model with two electrons in two orbitals:
\begin{equation}
H=-t\sum_\sigma\left( \hat{c}^\dagger_{1\sigma} \hat{c}_{2\sigma} + H.c. \right)
+ U\sum_i  \hat{n}_{i\uparrow}  \hat{n}_{i\downarrow}
- \Delta \sum_\sigma \left(  \hat{n}_{1\sigma} - \hat{n}_{2\sigma} \right).
\end{equation} 
Where $\hat{c}^{(\dagger)}_{i\sigma}$ annihilates (creates) an electron with spin $\sigma$ on site $i$ and the number operator is $ \hat{n}_{i\sigma}=\hat{c}^{\dagger}_{i\sigma} \hat{c}_{i\sigma}$ \cite{Rossler,Reimers-chapter}.
The ground states that  occur in different parameter regimes \cite{Reimers-chapter} include:
\begin{subequations}
\begin{eqnarray}
\ket{E} &=& \frac12	\left( \hat{c}^\dagger_{1\uparrow} - \hat{c}^\dagger_{2\uparrow} \right) \left( \hat{c}^\dagger_{1\downarrow} - \hat{c}^\dagger_{2\downarrow} \right) \ket{0}
= \frac12 \left[\hat{c}^\dagger_{1\uparrow} \hat{c}^\dagger_{1\downarrow}  + \hat{c}^\dagger_{2\uparrow} \hat{c}^\dagger_{2\downarrow} - \left( \hat{c}^\dagger_{1\uparrow} \hat{c}^\dagger_{2\downarrow} - \hat{c}^\dagger_{1\downarrow} \hat{c}^\dagger_{2\uparrow} \right) \right] \ket{0}	
\\
\ket{B} &=&	\frac1{\sqrt{2}} \left(\hat{c}^\dagger_{1\uparrow} \hat{c}^\dagger_{2\downarrow} - \hat{c}^\dagger_{1\downarrow} \hat{c}^\dagger_{2\uparrow} \right) \ket{0}	\\
\ket{N} &=&		\hat{c}^\dagger_{1\uparrow} \hat{c}^\dagger_{1\downarrow}  \ket{0}	
\end{eqnarray}
\end{subequations}
 $\ket{E}$, the ground state for $t=1$ and $U=\Delta=0$, is entangled, but not correlated;
$\ket{B}$, the ground state for $U/t\rightarrow\infty$ and $\Delta=0$, is both entangled and correlated; 
and $\ket{N}$,  the ground state for $U=t=0$ and $\Delta=1$, is neither entangled nor correlated. It is not possible to write down a ground  state, for this model, that is not entangled but is correlated, whenever the ground state is separable it can always be written as a single Slater determinant.

The use of entanglement as a quantitative tool for studying materials is most developed for spin systems, where the concept of locality has a clear, intuitive meaning. Here, specifying a location also specifies the particle (spin), so entanglement and correlations are actually trying to measure the same thing!

\subsection{Confined and deconfined; bound and free}

We do not see individual quarks at low energies. Instead, quarks always arrive in colourless combinations -- for example, three quarks  in a proton or a quark  and an antiquark in a pion \cite{Griffiths}. This is an example of confinement. Confining interactions are a little like  elastic bands. At short distances (high energies) the  particles are free, but at long distances (low energies) the  force grows. Importantly, it costs an infinite amount of energy to pull a pair of confined particles infinitely far apart. 

This  contrasts with the finite energy required to separate particles in a bound state. For example, an electron is bound to a proton in a hydrogen atom. It costs about 13.6~eV to pull the electron infinitely far from the proton, \textit{i.e.}, to ionise the atom.

Quarks are always confined at low energies. This is a fact of the strong interaction that we cannot change. But strongly correlated  matter offers  richer possibilities. Many phases of matter have  been discovered where otherwise confined particles become ``deconfined" and move freely through the material.

\subsection{Magnetic monopoles in spin ices}\label{sect:IceMonopoles}

One of the most spectacular examples of deconfinement is the presence of magnetic monopoles in spin ices - a class of frustrated magnets. Magnetic monopoles have not been observed in the vacuum \cite{monopoles}, whereas electric monopoles (charged particles) are commonplace. Magnetic monopoles in spin ice is also one of the simplest examples of deconfinement to understand. This is because the emergence of magnetic monopoles in spin ices is essentially a classical phenomena. Nevertheless, it contains many of the most important ingredients of deconfinement.

\begin{figure}
	\centering
	\includegraphics[width=0.5\textwidth]{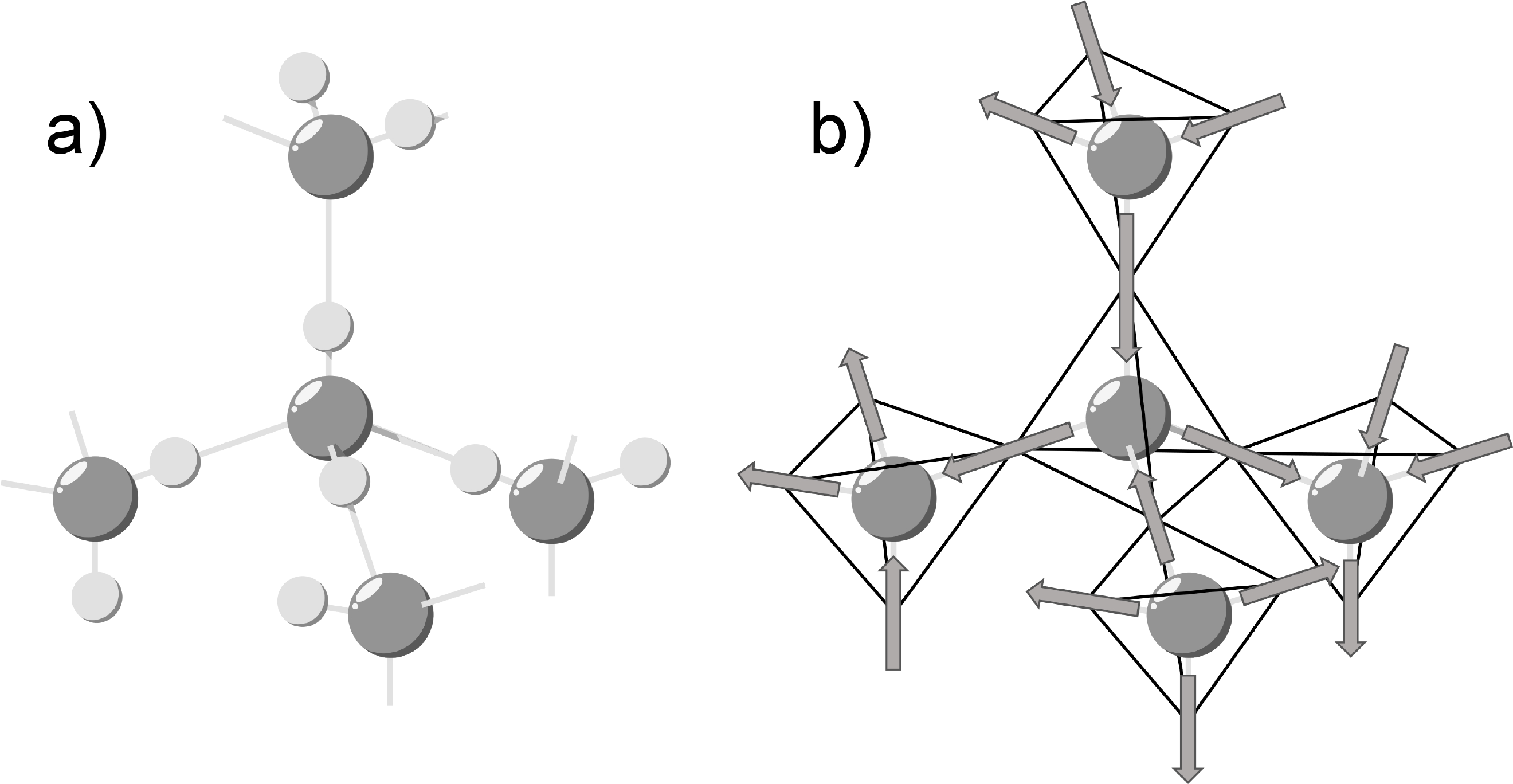}
	\caption{Ice rules for (a) water ice and (b) spin ice. In (a) the large dark grey spheres represent oxygen atoms and the light grey spheres represent hydrogen atoms. The ice rules insist that each oxygen forms covalent (short) bonds with two hydrogen atoms and hydrogen (long) bonds with another two. The spin ice materials have the generic formula $A_2B_2$O$_7$. In (b)  the large dark grey spheres represent oxygen atoms at the centres of the tetrahedra and the arrows, at the corners of the tetrahedra, represent the spins of the rare earth metals at the $B$ sites. This sublattice is therefore composed of corner sharing tetrahedra. (The $A$ sites and 6/7 of the oxygens form an interpenetrating lattice of corner sharing octahedra, which is not shown for simplicity.) The ice rule demands that two spins point in, and two spins point out of, each tetrahedron.} \label{fig:ice}
\end{figure}

Before we think about spin ices, its helpful to think a little about water ice. Ice is a remarkable material. There are at least seventeen different crystalline phases of \ch{H2O}; however, almost all of the ice in the biosphere is in the I$_\text{h}$ phase, which we will discuss here. In this phase each oxygen atom forms (short) covalent bonds with two hydrogen atoms and (long) hydrogen bonds with two other hydrogen atoms (Fig. \ref{fig:ice}a). So long as these `ice rules' are obeyed the energy is minimised. However, Pauling \cite{Pauling} showed that there is a macroscopically large number of ways to obey the ice rules. This means that the hydrogen atom positions are disordered. Hence there is a `residual' entropy in ice which Pauling estimated to be $S=(k_BN_\text{H}/2)\ln(3/2)$, where $N_\text{H}$ is the number of hydrogens. The residual entropy of water ice should remain even if the temperature were at absolute zero, in apparent violation of the third law of thermodynamics. Nature avoids this problem  because I$_\text{h}$ is not the equilibrium phase at $T=0$.

Spin ices are magnetic materials that obey ice rules. The magnetic moments in spin ices live on the `pyrochlore' lattice, these positions are essentially equivalent to the mid-points between the oxygen atoms in water ice  (Fig. \ref{fig:ice}b). There is a strong crystal field interaction (electrostatic interaction with the non-magnetic atoms) so, at low energies, the magnetic moments can only point in two directions: in  or out of the tetrahedron. The interactions between the moments are minimised when each tetrahedron has two spins pointing in  and two spins pointing out. A spin pointing out is equivalent to a (long) hydrogen bond, and  a spin pointing in is equivalent to a (short) covalent bond. Thus, Pauling's argument carries across and predicts a residual entropy of $S=(k_BN_s/2)\ln(3/2)$, where $N_s$ is the number of spins.

To make these ideas quantitative, it is helpful to introduce an Ising model with antiferromagnetic interactions ($J>0$) between nearest neighbours,
\begin{equation}
	H=J\sum_{\langle ij \rangle}\sigma_i\sigma_j, 
	\label{eq:Ising}
\end{equation}
where $\langle ij \rangle$ indicates that the sum runs over spins, $\sigma_i=\pm1$, on the same tetrahedron. First, let us consider a single tetrahedron. It is impossible to minimise all six interactions between the four spins simultaneously. The best one can do is to minimise four of the interactions, which is achieved for any 2-in/2-out state. As the interactions of the full Hamiltonian (Eq. \ref{eq:Ising}) are all within individual tetrahedra, any state that obeys the ice rules will minimise the energy of each tetrahedron individually, and thus, the total energy of the model. Thus, the pyrochlore lattice  is geometrically frustrated, cf. \cref{fig:tri}. 

On a single tetrahedron there are 6 states that obey the ice rules out of $2^4=16$ total possible states. If we ignore any short range correlations and surface effects, then $N_s$ is half the number of tetrahedra (as each spin is shared by two tetrahedra). Thus, arrive at Pauling's estimate for the total number of ice states on a lattice:  
	$N_\text{ice}=2^{N_s}\left({6}/{16}\right)^{N_s/2}$.
Hence, $S=k_B\ln N_\text{ice}=(k_BN_s/2)\ln(3/2)$.

\begin{figure}
	\centering
	\includegraphics[width=0.5\textwidth]{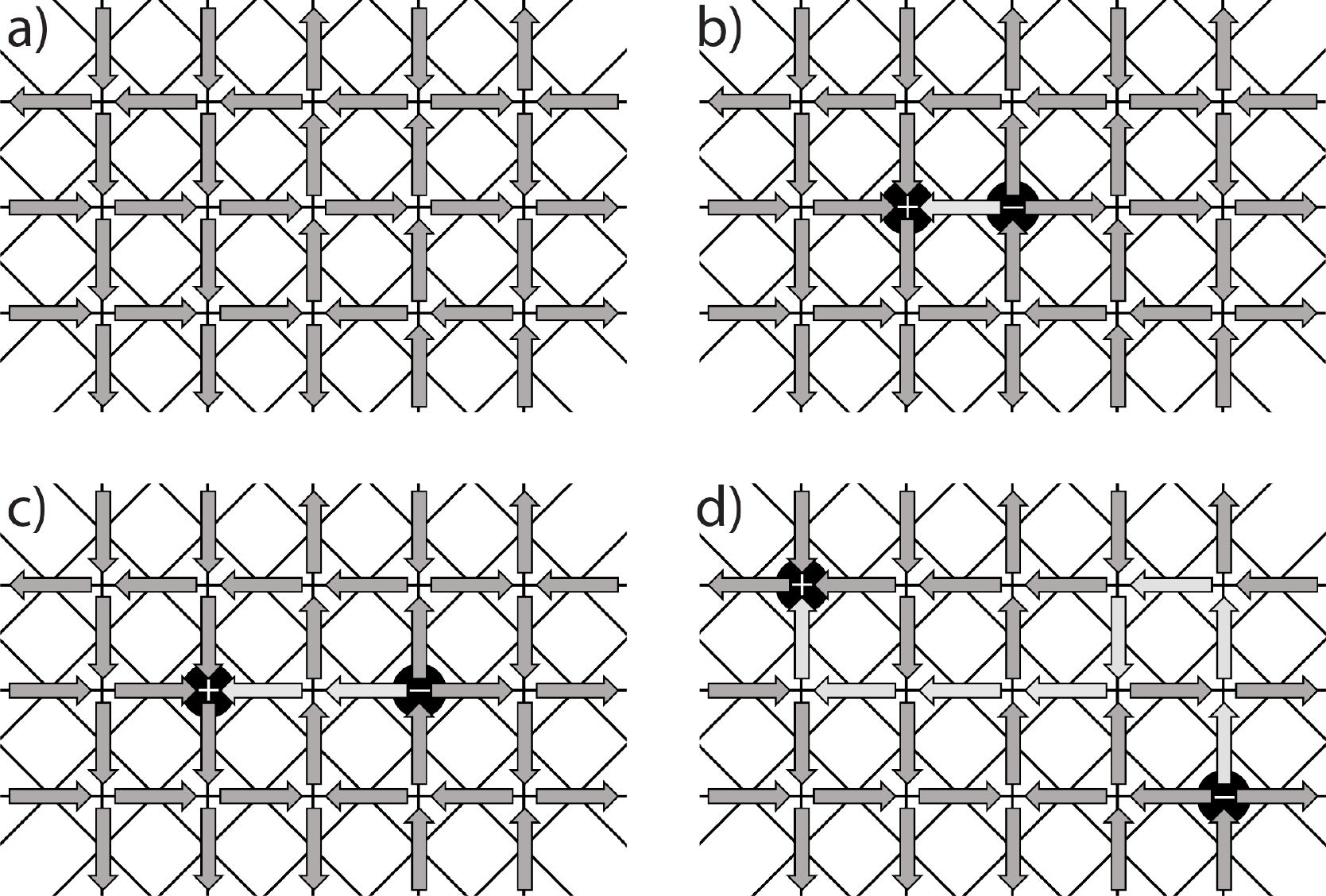}
	\caption{(a) An ice state of the checkerboard lattice.  Each of the  hatched squares, \rotatebox[origin=c]{45}{$\boxtimes$}, obeys the 2-in/2-out ice rule. The  \rotatebox[origin=c]{45}{$\boxtimes$} are  the 2D projection of the tetrahedra in the pyrochlore lattice (Fig. \ref{fig:ice}). (b) Flipping one spin (highlighted via a lighter colour) creates two defects (monopoles), labelled $+$ (3-in/1-out) and $-$ (1-in/3-out). (c) Flipping another spin (also highlighted) on one of the \rotatebox[origin=c]{45}{$\boxtimes$}  that violates the ice rule, so as to restore the ice rule on that \rotatebox[origin=c]{45}{$\boxtimes$}  moves the monopole to the other \rotatebox[origin=c]{45}{$\boxtimes$}  that hosts the flipped spin. (d) Continuing this process allows the two monopoles to move independently.} \label{fig:squareIce}
\end{figure}

We want to understand the new types of particles that can exist in  spin ice. 
As pictures are easier to draw in 2D than 3D we will consider the 2D projection of the pyrochlore lattice: the checkerboard lattice (Fig. \ref{fig:squareIce}). Spins on the checkerboard lattice with the same interaction strength between vertical, horizontal and diagonal neighbours also obey the 2-in/2-out ice rule (Fig. \ref{fig:squareIce}a) and display much of the same physics. 	
Remarkably, this physics can be realised in 2D arrays  of nanomagnets known as artificial spin ices.\footnote{In these metamaterials it is conventional to think of the spins as living on the bonds (rather than the vertices) of a lattice, so the checkerboard lattice is usually called the square lattice. Furthermore, in actual metamaterials, the interactions are slightly more complicated than the simple model discussed here \cite{Stamps}.}  

The fundamental local excitation is to flip a single spin (Fig. \ref{fig:squareIce}b). This creates defects (violations of the ice rules) on the two tetrahedra/checked squares (\rotatebox[origin=c]{45}{$\boxtimes$}) that contain the flipped spin. One expects this to be a relatively high energy excitation ($\sim4J$ in our nearest neighbour model), indicating that there is a gap to the excitations (i.e., that we are dealing with massive particles). We can restore the ice rule on one of the \rotatebox[origin=c]{45}{$\boxtimes$} by flipping another of the spins; however, this creates  another defect on the \rotatebox[origin=c]{45}{$\boxtimes$} that shares this spin (Fig. \ref{fig:squareIce}c). We expect this `move' to cost much less energy (in our nearest neighbour model this excitation is still at $\sim4J$). Another way to view our second move is that one defect hopped to another \rotatebox[origin=c]{45}{$\boxtimes$}. We can repeat this process, flipping spins on the \rotatebox[origin=c]{45}{$\boxtimes$} that contain defects and watching the defects hop around the lattice (Fig. \ref{fig:squareIce}d). The key point is that these moves are all very low  energy. So the defects move independently -- the two defects are  free particles. However, creating more defects remains energetically expensive due to their mass ($\sim 2J$). More sophisticated calculations including the long range dipolar interactions present in real spin ices confirm these conclusions \cite{CastelnovoARCMP}.

Now we would like to understand the properties of the defects. In the ground states where  the ice rules are obeyed  the net magnetic moment of the system vanishes. Flipping a spin, in violation of the ice rule, creates a local magnetic dipole with one pole at each of the two defects created. As the defects hop around, the two poles of the dipole move with them. Therefore, as the defects behave as free, independent particles, so must the two poles of the dipole -- the elementary excitations of spin ice are magnetic monopoles! In the vacuum monopoles are confined within dipoles (or higher order multipoles). In spin ice the dipoles are fractionalised into monopoles, which are deconfined.

Ices have subsequently been proposed and/or observed in various other classes of materials with a local binary (Ising) degree of freedom. In each case, the relevant degrees of freedom, for example charge \cite{Zinkin,Fulde} or spin \cite{Jace}, is fractionalised and deconfined. These ice states are very unusual as they give rise to fractionalisation in classical systems. The other examples of fractionalisation we will encounter below require quantum mechanics in an essential way.

\subsection{The integer quantum Hall effect (IQHE)}\label{sect:IQHE}

The next idea I want to introduce is the bulk-boundary correspondence. One of the simplest examples of this is the in the IQHE. 

\begin{figure}
	\centering
	\includegraphics[width=0.45\textwidth]{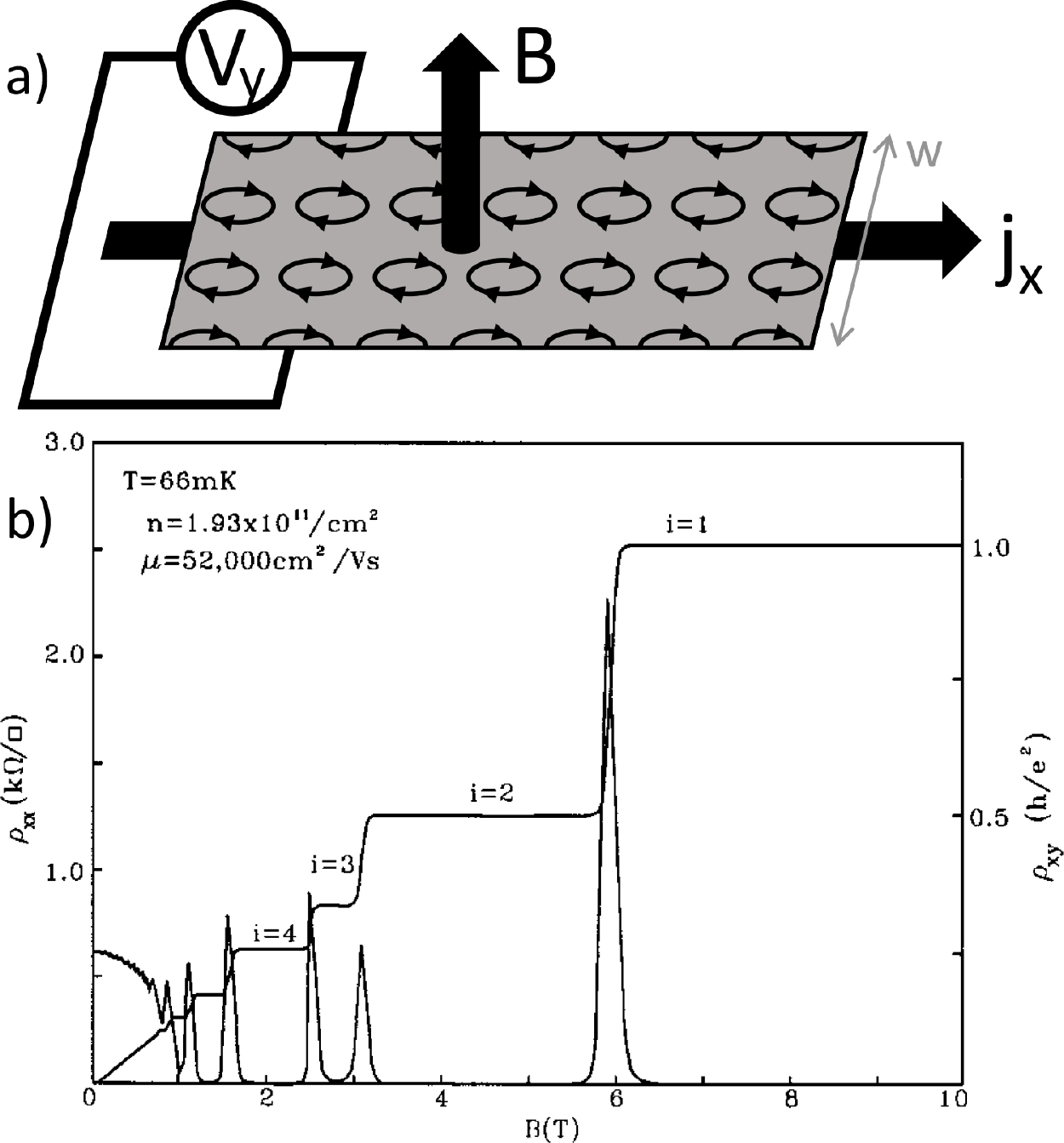}
	\caption{a) The basic set-up for a Hall measurement. The classical cyclotron orbits and skipping edge modes are sketched. b) The integer quantum Hall effect. The numbers labelled $i$ here are called $\nu$ in the main text. Reprinted with permission from \cite{Tsui} Copyright (1999) by the American Physical Society. 
	} \label{fig:Hall}
\end{figure}

The classical Hall effect, \cref{fig:Hall}a, is that a current flowing in the $x$ direction, with current density $j_x$  in a  magnetic field in the $z$ direction, with magnitude $B$, induces an electric field in the $y$ direction, $E_y=-V_y/w$, where $w$ is the width of the sample. This results from a build up of charge on the edges of the sample due to the Lorentz force. There are two properties of particular interest: the transverse  conductivity $\sigma_{xy}=-j_x/E_y$ and the Hall coefficient $R_H=E_y/j_xB=1/ne$, where  $n$ is the electron density. Importantly, $R_H$ allows for the experimental determination of the charge of the current carrier, $e$, which the longitudinal  conductivity, $\sigma_{xx}$, does not. Therefore, the classical Hall effect provides direct evidence for holes, \textit{i.e.}, positive charge carriers in metals and semiconductors.

The IQHE occurs in two dimensional electron gases (2DEGs). These can be formed at the interfaces between two different materials, \textit{e.g.}, in semiconductor heterostructures. Although disorder will play an important role below, devices need to be very clean in order to see the IQHE. We need some disorder, but not too much. 

The most spectacular  experimental manifestations of the IQHE are plateaus in the transverse resistivity at  $\rho_{xy}=\frac{h}{e^2}\frac{1}{\nu}$ for integer  $\nu$ as the magnetic field is varied, \cref{fig:Hall}b. This quantisation is so precise that  the von Klitzing constant, $\frac{h}{e^2}$, can be measured more accurately than either the Plank constant, $h$, or the fundamental charge, $e$ \cite{CODATA}. The plateaus are centred at magnetic fields $B=\frac{n}{\nu}\Phi_0$, where $\Phi_0=h/e$ is  the flux quantum.

An electron moving in a magnetic field is described by the Hamiltonian,
\begin{equation}
	H=\frac{1}{2m}\left( \bm{p}+e\bm{A} \right)^2
	= \frac{p^2}{2m} + \frac{1}{2}m\omega_c^2|z|^2 
,\label{eq:LandauLevelH}
\end{equation} 
where we have chosen to work in the `symmetric gauge', ${\bm A}=\frac{B}{2}(-y,x)$, introduced a complex coordinate $z=x+iy$, and  $\omega_c=eB/m$ is  the cyclotron frequency.  
This is just a harmonic oscillator 
which has eigenenergies
$E_{n} = \hbar \omega_c \left(n+\frac12\right)$. 
The corresponding wavefunctions are
\begin{equation}
	\psi_{n}(z)= e^{-|z|^2/4\ell_B^2} \left( \frac{\partial}{\partial z} - \frac{{z}^*}{2\ell_B^2} \right)^n f(z),  \label{eq:LandauLevelWF}
\end{equation}
where $f(z)$ is an analytic function of $z$ \cite{Wen,Tong} and $\ell_B=\sqrt{\hbar/eB}$.
These states are known as Landau levels.

\begin{figure}
	\centering
	\includegraphics[width=0.95\textwidth]{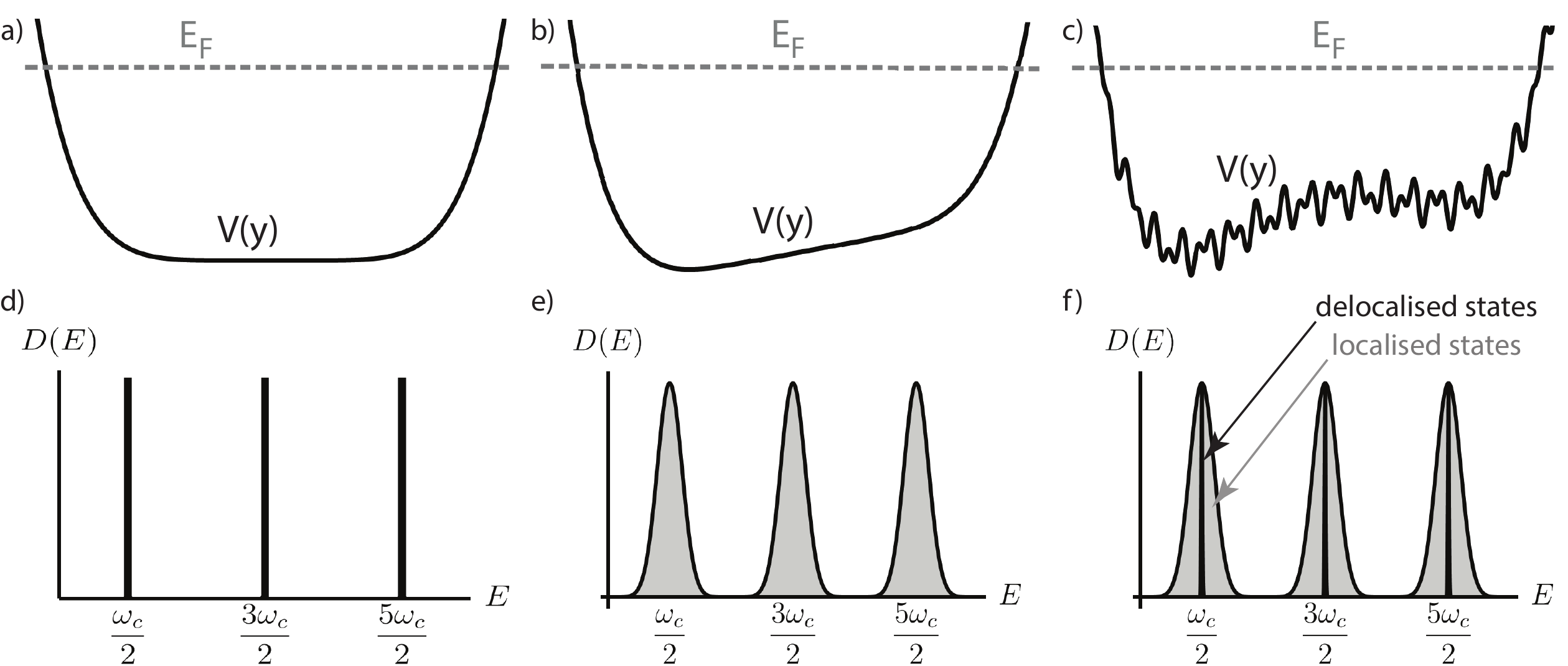}
	\caption{a) A confining potential, $V(y)$ in a Hall device. b) The confining potential plus the Hall voltage. c) The confining potential plus the Hall voltage and a disorder potential. d) The density of states due to Landau levels in a clean system. e) This is broadened by disorder. f) Disorder changes the energies of the localised states in the bulk, but has little effect on the extended states at the edges, cf. b,c.
	} \label{fig:IQHE}
\end{figure}

The filling fraction $\nu$ is  the number of electrons per flux quantum. Therefore, for $\nu=1$ every state in the zeroth Landau level is filled by an electron and there is a gap  ($\hbar\omega_c$) to the next available energy level in the bulk, \cref{fig:IQHE}d. The combination of the gap and the Pauli exclusion principle mean that the electron fluid is incompressible -- it cannot respond linearly to perturbations.

Classically, electrons in a magnetic field undertake circular orbits. This suggests that something interesting is going to happen at the edges of a quantum Hall fluid (Fig. \ref{fig:Hall}a). Classically, the cyclotron orbits are replaced by skipping modes when the orbits collide with the edge of a device. The skipping modes only allow motion of electrons in one direction; the particles  are said to be \textit{chiral}. Charged chiral particles are not allowed in one-dimension, but the edge modes are not really one dimensional; they live on the surface of a two dimensional object. The particles on opposite sides of the 2DEG have opposite chiralities. Therefore, in the absence of an electric field the net current vanishes.

To understand the quantum mechanical edge states we need to add a confining potential, $V(y)$,  (Fig. \ref{fig:IQHE}a) to our Hamiltonian (Eq. \ref{eq:LandauLevelH}). The eigenstates for $V=0$ (Eq. \ref{eq:LandauLevelWF}) are Gaussians of width $\ell_B$. Provided the potential is smooth on the scale of $\ell_B$ around the centre of the wavefunction $Y$, it is reasonable to Taylor expand the potential to linear order $V(y)=V(Y)+(\partial V/\partial y)(y-Y)$. The constant term can safely be ignored. The linear term induces a drift velocity in the $x$-direction:
	$v_x=-\frac{1}{eB}\frac{\partial V}{\partial y}$,
which has opposite signs on  opposite edges (Fig. \ref{fig:IQHE}a). Therefore, the quantum treatment also predicts chiral edge modes.

The Hall voltage (Fig. \ref{fig:Hall}a) is a potential difference between the two sides of the 2DEG. This tilts the potential (Fig. \ref{fig:IQHE}b). To calculate the current we need to sum over all states, which is equivalent to integrating over $y$:
\begin{equation}
	I_x 
	= -\frac{e}{2\pi\ell_B^2}\int dy \frac{1}{eB}\frac{\partial V}{\partial y}  = \frac{e}{2\pi\hbar}\Delta\mu = \frac{e}{2\pi\hbar}eV_y,
\end{equation}
where we have used that fact the the potential difference between the two edges is proportional to the Hall voltage, $\Delta\mu=eV_y$. 
Rearranging, we find that
\begin{equation}
\sigma_{xy} \equiv \frac{I_x}{V_y} =  \frac{e^2}{h},
\end{equation}
which is the experimentally observed Hall conductivity at the first plateau. Repeating this calculation for $\nu$ filled Landau levels yields $\sigma_{xy} = \nu e^2/h$, reproducing the experimentally observed values of the Hall plateaus.

The argument above is limited to filled Landau levels. However, the key experimental result is that even when one Landau level is partially filled we still observe a quantised Hall conductivity. Disorder is responsible for this. It causes a randomness in the potential (Fig. \ref{fig:IQHE}c). This will broaden the Landau levels and lift their degeneracy (Fig. \ref{fig:IQHE}d-e), but provided the randomness in the potential is small compared to $\hbar\omega_c$, a description in terms of quantised Landau levels is still reasonable. Even though disorder is important, if there is too much disorder, the quantum effect is washed out and only the classical Hall physics remains. 

Disorder localises the Landau levels in the bulk, as they traverse equipotentials in the $x$-$y$ plane. Only equipotentials that span the sample in the $y$-direction can contribute to the Hall current. But, these will be vanishingly rare in the bulk of a macroscopic device. However, the potential inside the material is much lower than the vacuum potential, so there must be equipotentials that span the the device \textit{near} the edges, regardless of the details of the disorder potential. The energies of extended states are not significantly changed by disorder but the energies of the localised states are (Fig. \ref{fig:IQHE}f). If we decrease $B$ with a fixed electron density, each Landau level can hold fewer electrons, so the Fermi energy increases. However, rather than populating an extended state, the electrons begin to populate the localised states. Therefore, sweeping the field does not change the conductivity until we hit a new set of extended states. Thus, disorder is vital for the jumps between Hall plateaus observed experimentally.

\subsection{Topological order, symmetry protected topological phases and the bulk-boundary correspondence}

Edge states are crucial for understanding the IQHE. The physics of the edge states is a direct consequence of the bulk properties of the system. We can see this already in our classical picture, where the skipping modes along the edges are just interrupted cyclotron orbits. In the quantum theory the edge states arise from the interplay of the bulk physics (Landau levels) with the  change in the potentials the electrons experience at the boundary between the material and the vacuum. This as an example of a general phenomenon called the bulk-boundary correspondence. 

In our discussion of the IQHE the details of the disorder potential were not very important. Whatever the potential is there will still be extended states somewhere near the edge and localised states in the bulk. The robustness to changing the details and the importance of the edge states suggest that there is a topological explanation of the IQHE.
Indeed, the argument that the edge states in the IQHE must span the system regardless of the disorder potential already has a topological character. We did not  carefully check whether the localisation of the bulk states by disorder  changes the Hall conductivity -- one might reasonably expect it to! However, Thouless, Kohomoto, Nightingale and den Nijs showed that disorder does not change the Hall conductivity because it is a topological invariant  \cite{Tong}. The disorder profile cannot change the value of the Hall conductivity just as stretching or bending cannot change the number of holes in a manifold, \cref{fig:vortex}a. The topological argument also explains why the quantisation of the conductivity is so precise. This is important because the precision of the quantisation cannot be explained by any approximate theory that does not make a topological argument and we have no hope of performing an exact calculation for a system as complicated as a semiconductor heterostructure.

There is a deep connection between topology and the bulk-boundary correspondence. The vacuum is topologically trivial. So if any material is topologically non-trivial then something interesting must happen at the boundary between the material and the vacuum. 

The most interesting way for a material to be topologically non-trivial is called topological order. Topologically ordered states do not break any symmetries of the Hamiltonian, but there is no local unitary transformation that maps them to a trivial product state. Recall that the time evolution of a Hamiltonian, $e^{iHt/\hbar}$, is a unitary operator. So  no perturbation can take us between a topologically ordered phase and a topologically trivial phase without causing a phase transition (which we would hope to  notice). The only way to achieve topological order is if there is \textit{long range} entanglement in these phases \cite{Chen,Wen}. This means we should only expect to find topological ordered materials where the interactions between the components dominate the physics.

As no symmetries are broken in a topologically ordered state they cannot be described by a local order parameter. As no perturbation connects them to the non-interacting problem these two Hamiltonians are not adiabatically connected. Therefore, the two paradigms of the standard model of materials do not help us understand topological order; we should expect new physics  beyond the standard model of materials. 

The absence of a broken symmetry, and hence an order parameter, means that many experimental probes of matter are of limited use for characterising topological ordered phases. Therefore, it is important to understand what the bulk-boundary correspondence tells us about the states that live on the boundaries, this can give vital information about the physics of the bulk. 

A second class of topologically  non-trivial states are found in symmetry protected topological (SPT) materials. The word `protected' makes it sound like SPT states are more robust than topological order; but in fact the reverse is true. SPT states cannot be transformed into trivial states by any unitary transformation that respects the `protective' symmetry. However, like topologically ordered phases, SPT phases cannot be described by a local order parameter. Entanglement is not required for SPT phases, although often short range entanglement plays an important role in the physics of SPT materials. Therefore, SPT can occur even in materials where the interactions between the components are weak -- the IQHE is an example of SPT and we were able to describe it in a theory that neglected electron-electron interactions.

There many other SPT materials where the interactions are not important. These include topological insulators \cite{Kane}, and Weyl and Dirac semimetals \cite{Yan}. The bulk-boundary correspondence leads to exotic effects on the surfaces of these materials, which have proved crucial for their experimental detection.  For example, massless chiral fermions, known as Weyl fermions, live on the surfaces of Weyl semimetals. There are no Weyl fermions in the standard model. So just like magnetic monopoles this is an example of quasi-particles that are qualitatively different from the particles observed at high energies.

\subsection{The Haldane chain}\label{sect:Haldane}

So far, we have seen fractionalisation in spin ices and the bulk-boundary correspondence in the IQHE. Now we are going to examine a SPT system that has both: the spin-one Heisenberg chain (also known as the Haldane chain). Here short-range entanglement will  provide important insights into the physics. 

Haldane employed a topological argument  \cite{Fradkin} (focusing on the Wess-Zumino term in the  non-linear sigma model, cf. Eq. \ref{eq:WZHeis}) to show that for any integer spin the 1D Heisenberg model is gapped (all excitations are massive). In contrast for half-odd-integer spin the 1D Heisenberg model is gapless (there are massless excitations).\footnote{See \cite{Schofield} for a clear introduction and \cite{Giamachi} for more details} This is sometimes known as the Haldane conjecture.
Later  Affleck, Lieb, Kennedy and Tasaki (AKLT) provided a beautifully simple picture for why the Haldane chain is gapped. 

In most cases it is impossible to calculate the exact ground state of a quantum many-body Hamiltonian. However,   $\ket{\Psi_0}$ is,  trivially, the exact ground state of $H=-\ket{\Psi_0}\bra{\Psi_0}$. Usually this is not a very helpful statement. But for the AKLT state, defined below, $-J\ket{\text{AKLT}}\bra{ \text{AKLT}}=H_\text{bb}(1/3)$, where
\begin{equation}
	H_\text{bb}(\beta)=J\sum_i \left[ \hat{\bm S}_i \cdot  \hat{\bm S}_{i+1} + \beta \left( \hat{\bm S}_i \cdot  \hat{\bm S}_{i+1} \right)^2 \right]
\end{equation}
defines the bilinear-biquadratic model and the $\hat{\bm S}_i$ describe spin-one local moments.
Although we know that $\beta=1/3$ in the AKLT model it is helpful to think about the more general bilinear-biquadratic model because it can be shown \cite{Mikeska} that there is no phase transition in the region $1/3\geq\beta\geq0$. Therefore, adiabatic continuity assures us that the key qualitative conclusions from the AKLT model will hold for the spin-one Heisenberg model, which is  the bilinear-biquadratic model at $\beta=0$.

\begin{figure}
	\centering
	\includegraphics[width=0.5\textwidth]{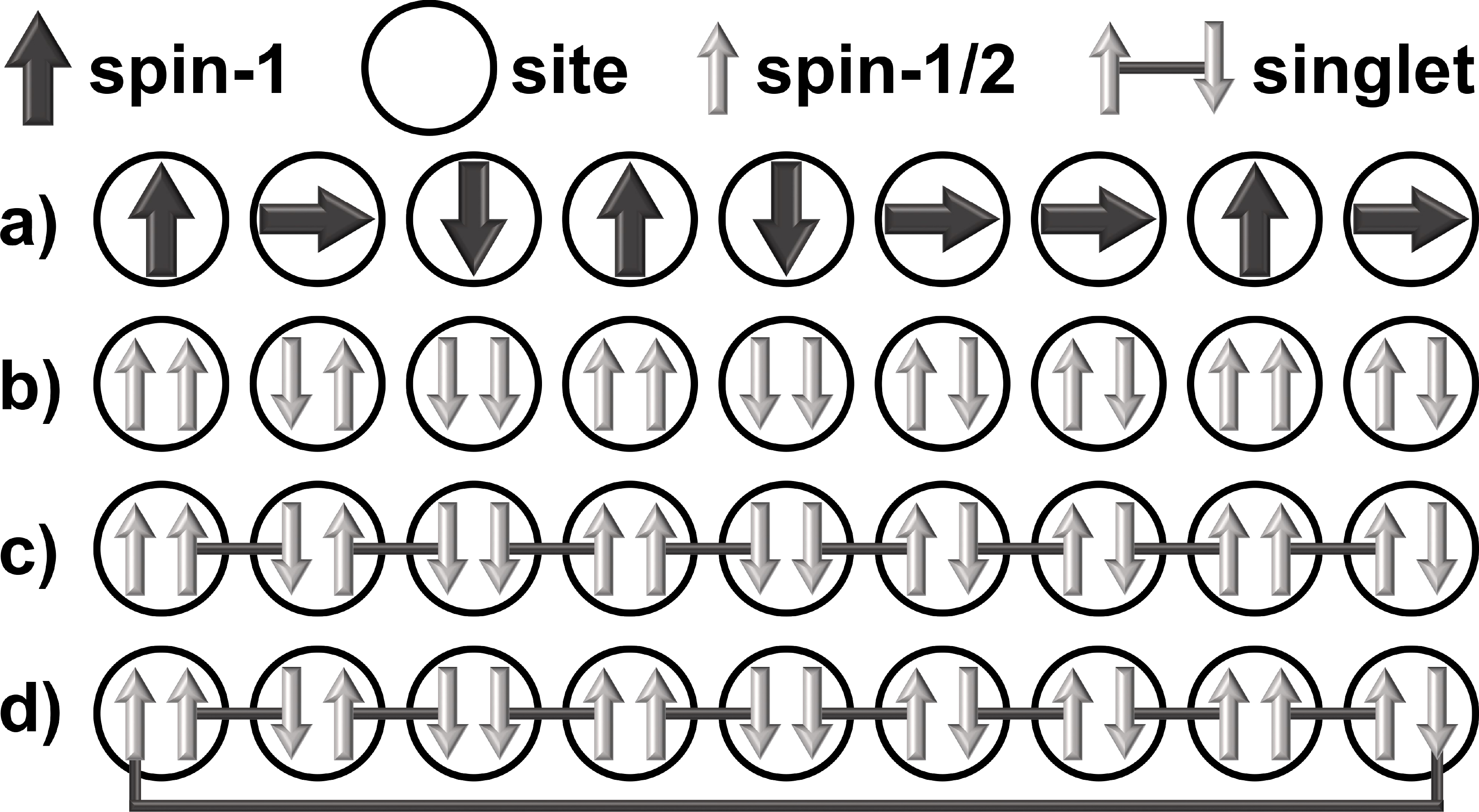}
	\caption{Cartoon of the AKLT state. (a) A spin-one chain can be represented by (b) a chain with two spin-1/2s per site provided they each form a triplet. (c) In the AKLT state one projects onto a state where each spin forms a singlet with a spin on the neighbouring site. For open boundary conditions this leaves a spin-1/2 at each end of the chain unpaired. The states of these spins do not change the energy of the system so every state, including the ground state, must be at least four fold degenerate. Excitations require breaking a singlet ``bond'', so there is a gap in the bulk. (d) For periodic boundary conditions (a ring rather than a chain) there are no unpaired spins and so the ground state is non-degenerate.} \label{fig:Haldane}
\end{figure}

There are  three steps to construct the AKLT state for a chain of length $N$ (Fig. \ref{fig:Haldane}a):
\begin{enumerate}
	\item Represent each spin-one in the chain by two spin-$1/2$s (Fig. \ref{fig:Haldane}b).
	\item \label{bullet:vb} Take the right-hand spin $\ket{\sigma}_{i,r}$ on each site, $i$, and form a singlet with the left-hand spin, $\ket{\sigma}_{i+1,l}$ on the site to its right, $i+1$  (Fig. \ref{fig:Haldane}c). That is construct the valence bond wavefunction
	$\ket{\text{VB}} =  \prod_{i=1}^{N-1} \frac{1}{\sqrt{2}} (\ket{\uparrow}_{i,r} \ket{\downarrow}_{i+1,l}  - \ket{\downarrow}_{i,r} \ket{\uparrow}_{i+1,l} )$.
	\item Project $\ket{\text{VB}}$ onto a state where there is a spin-one (triplet state, \cref{eq:2estates}a-c) on  every site,
	\begin{equation}
	\ket{\text{AKLT}} = \sum_{i=1}^N\left[
	\ket{ 1 }_i \bra{ \uparrow}_{i,l} \bra{ \uparrow}_{i,r}  +
	\ket{ 0}_i  \frac{\bra{ \uparrow}_{i,l} \bra{\downarrow}_{i,r}  + \bra{ \downarrow}_{i,l} \bra{\uparrow}_{i,r}}{\sqrt{2}} +
	\ket{-1}_i \bra{ \downarrow}_{i,l} \bra{\downarrow}_{i,r} 
	\right] \ket{\text{VB}},
	\end{equation}
	where $\ket{S^z}_i$ are  basis states of the spin-one model with $S^z_i=-1$, 0, or 1.
\end{enumerate}
The AKLT state is gapped because making an excitation requires us to break a valence bond, which costs an energy $\sim J$.

Notice that we have not specified the states of the leftmost and rightmost spin-$1/2$s, $\ket{\sigma}_{1,l}$ and $\ket{\sigma}_{N,r}$. Regardless of what state these take, $\ket{\text{AKLT}}$ is the ground state of $H_{bb}(1/3)$. Therefore, the AKLT state has a free spin-$1/2$ at each end. This is an example of a bulk-boundary correspondence. The emergent spin-$1/2$s are a consequence of the valence bonds in the bulk. It is also an example of fractionalisation. The underlying Hamiltonian only contains spin-ones, but the emergent degrees of freedom are spin-one-half. In the pure Heisenberg model ($\beta=0$), the spin-$1/2$s are not found exactly on the leftmost and rightmost sites -- but their amplitude decays rapidly as we move into the bulk.  Emergent spin-$1/2$s have been observed experimentally in magnetic resonance experiments on spin-one chains \cite{Hagiwara}.

We can  see the topological nature of the AKLT state  directly. 
As each spin-$1/2$ has two orthogonal states (say $\ket{\uparrow}$ and $\ket{\downarrow}$),  the ground state is four-fold degenerate.\footnote{In the pure Heisenberg model the finite extent of the edge states allows for an  exponentially small (in the chain length) interaction between the edge states and thus an exponentially small lifting of the degeneracy of the ground state.} 
Topology is quite limited in 1D, but we can make a topologically distinct manifold by fusing the two ends of the chain. Now we can also make a singlet from the two edge states (Fig. \ref{fig:Haldane}d). That is, we replace $\ket{\text{VB}}$ with $\ket{\text{VB}'} = \frac{1}{\sqrt{2}} (\ket{\uparrow}_{N,r} \ket{\downarrow}_{1,l}  - \ket{\downarrow}_{N,r} \ket{\uparrow}_{1,l} )\ket{\text{VB}}$ in step (\ref{bullet:vb}) of constructing the AKLT state. Now, all of the spin-$1/2$s are paired and the ground state is non-degenerate.
Thus, we see that the ground state degeneracy depends of the topology (open or closed boundary conditions) of the lattice. This dependence of the ground state degeneracy on the topology of the material is typical of topological matter.

Entanglement is central to the physics of the Haldane chain. In the AKLT state neighbouring sites are entangled: they form singlets, the quintessential entangled state. This is  typical. Entanglement is often an important ingredient for fractionalisation and topological matter. However, as we saw in the examples of spin ice and the IQHE entanglement is not necessary for either phenomenon. 

The AKLT only has entanglement between nearest neighbours ($i$ and $i+1$), so it must be an SPT state and not topological order, which requires long-range entanglement. As the Haldane state is in the same phase, it follows that the ground state of spin-one Heisenberg model is also an SPT state\footnote{Protected by inversion, time reversal and $\pi$ rotations \cite{Pollmann}.}. More generally, only short range entanglement is possible in 1D and so topological order can only occur in higher dimensions \cite{Eisert}.

\subsection{Majorana fermions}\label{sect:Majoranas}

Spin chains are among the simplest problems in many-body physics. Natural questions are: Can we find quantum phases in materials with itinerant electrons? Do such materials support fractionalised excitations? Is there a bulk-boundary correspondence? In this section we explore an example that shows that the answer to all three of these questions is yes.

In 1937 Ettore Majorana proposed a new type of particle: a fermion that is its own antiparticle. These `Majorana fermions' must be massless. It was proposed that neutrinos could be Majoranas. But,  neutrino flavour oscillations make this idea difficult, but perhaps not impossible \cite{Wilczek}, to reconcile with experiment. Nevertheless, there is growing evidence that Majoranas can emerge in condensed matter, for example, in superconductor-semiconductor heterostructures \cite{Lutchyn} and in the vortices of triplet superconductors \cite{Alicea}. To understand the basic physics of these systems, we review a toy model of  a spinless p-wave superconductor in 1D, introduced by Kitaev \cite{KitaevChain} and now generally known as the Kitaev chain. It consists of a chain of $N$ sites described by the Hamiltonian
\begin{equation}
	H=-\mu\sum_i^N \hat{c}_i^\dagger \hat{c}_i -\frac12 \sum_i^{N-1} \left( t\hat{c}_i^\dagger \hat{c}_{i+1} + \Delta e^{i\theta} \hat{c}_i \hat{c}_{i+1}  + H.c. \right), \label{eq:KitaeChainFermion}
\end{equation}
where $\hat{c}_i^\dagger$ creates a spinless fermion on site $i$, $\mu$ is the chemical potential, the hopping integral, $t$, is the amplitude for a fermion to hop between neighbouring sites, $H.c.$ stands for Hermitian conjugate, and the microscopic superconducting parameter, $\Delta e^{i\theta}\propto\Psi$, the Ginzburg-Landau order parameter discussed in section \ref{sect:Higgs}.

We define a set of Majorana operators, $\hat\gamma_{\alpha,i}$, by
	$\hat{c}_i = e^{-i\theta/2} ( \hat\gamma_{r,i} + i \hat\gamma_{l,i} )/2$.
Provided these operators obey  canonical fermion anticommutation relations,
\begin{equation}
	\left\{ \hat\gamma_{\alpha,i} , \hat\gamma_{\beta,j} \right\} \equiv 
	\hat\gamma_{\alpha,i} \hat\gamma_{\beta,j} +  \hat\gamma_{\beta,j} \hat\gamma_{\alpha,i} =
	\delta_{i,j}\delta_{\alpha,\beta},
	\label{eq:MajoranaCom}
\end{equation}
and 
$\hat\gamma_{\alpha,i} = \hat\gamma_{\alpha,i}^\dagger$, then this is a valid representation because it preserves the anticommutation relation of the underlying spinless fermions, $\left\{ \hat{c}_i , \hat{c}_j^\dagger \right\}  = \delta_{i,j}$. Notice however, that  creating a Majorana is the same as annihilating one; as trailed above,  Majoranas are their own antiparticles. This means that we cannot talk about the occupation number of a Majorana mode in the usual sense.
In the Majorana basis the Hamiltonian is 
\begin{equation}
H=-\frac{\mu}{2}\sum_i^N \left( 1 + i \hat\gamma_{r,i} \hat\gamma_{l,i} \right) -\frac{i}{4} \sum_i^{N-1} \left[ (\Delta - t) \hat\gamma_{l,i} \hat\gamma_{r,i+1}  + (\Delta + t) \hat\gamma_{r,i} \hat\gamma_{l,i+1} \right]. \label{eq:KitaeChainMajonana}
\end{equation}
This model has two phases, both of which are adiabatically connected to limits that are  trivially solved. 

\begin{figure}
	\centering
	\includegraphics[width=0.5\textwidth]{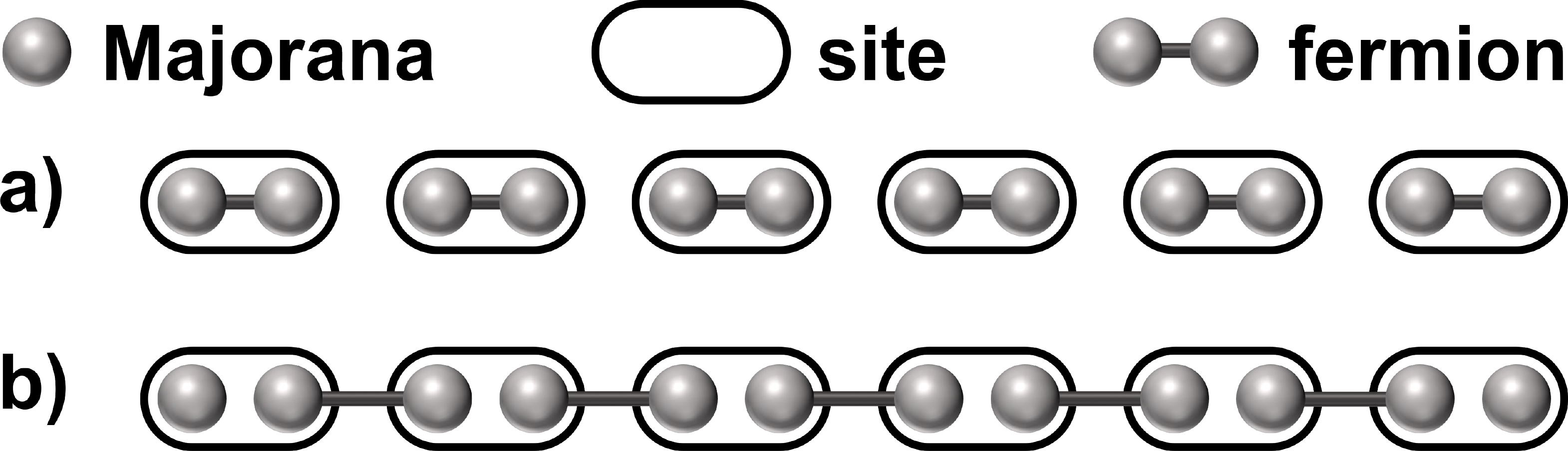}
	\caption{The fractionalisation of electrons into Majoranas in the Kitaev chain, \cref{eq:KitaeChainMajonana}. (a) In the trivial phase (which includes $\Delta=t=0$, $\mu<0$) the Majoranas on the same site pair up to form fermions. The ends of the chain are not special. (b) In the topological phase (which includes $\mu=0$, $\Delta=t\ne0$) the Majoranas on neighbouring sites pair up to form fermions. At each end a single Majorana is left unpaired. We can construct a non-local fermion at zero energy from these two Majoranas. This gives rises to a zero energy Majorana mode and a two fold degeneracy for all of the eigenstates of the chain.} \label{fig:Majoranas}
\end{figure}

For $\Delta=t=0$ and $\mu<0$ only the first term of Eq. \ref{eq:KitaeChainMajonana} remains. This term pairs Majoranas on the same site, Fig. \ref{fig:Majoranas}a. Comparing to the original Hamiltonian (Eq. \ref{eq:KitaeChainFermion}) we see that this corresponds to the vacuum state for the spinless fermions. Adding a fermion costs a finite energy, $|\mu|$. This ground state is a product state and has no entanglement.

For $\mu=0$ and $\Delta=t\ne0$ only the last term of  Eq. \ref{eq:KitaeChainMajonana} remains. This term pairs Majoranas on neighbouring  sites, Fig. \ref{fig:Majoranas}b. We can write the Hamiltonian in terms of a new set of fermions $\hat{d}_i =  ( \hat\gamma_{r,i+1} + i \hat\gamma_{l,i} )/2$:
\begin{equation}
H=t\sum_{i}^{N-1} \left(\hat{d}_i^\dagger \hat{d}_i - \frac12 \right), \label{eq:KitaeChainNewFermions}
\end{equation}
so again there is a bulk gap, here $|t|$. Each site is entangled with its two nearest neighbours. 

There are two unpaired Majoranas at the ends of the chain, $\hat\gamma_{l,1}$ and $\hat\gamma_{r,N}$. These make a new fermion, $f=(\hat\gamma_{l,1}+ i \hat\gamma_{r,N})$. As $f$ does not appear in the Hamiltonian (Eq. \ref{eq:KitaeChainNewFermions}),  occupying this state costs zero energy. This introduces a two-fold degeneracy into the model. Importantly, if $|0\rangle$ is a ground state satisfying $f|0\rangle=0$, then $|1\rangle\equiv f^\dagger|0\rangle$ is also a ground state. The pair of ground states, $|0\rangle$ and $|1\rangle$, have  potential for use as the two logical states of qubits in a quantum computer. $f$ is a highly non-local object; the two Majoranas are separated by the length of the chain, which can be arbitrarily long. Most sources of noise, which are highly detrimental to quantum computers, act locally. Therefore, as well as our fundamental interest,  Majoranas provide a promising route to quantum computing.

Notice the  similarities of the construction here to the AKLT chain,  \cref{sect:Haldane}; particularly, the similar way fractionalisation arose from local entanglement and the similar natures of the bulk-boundary correspondences.

In the Kitaev chain, it is straightforward to write down a topological invariant that classifies the two phases. We begin by writing the Hamiltonian in the `Bogoliubov-de Gennes' form
\begin{equation}
H=\frac12\sum_k \hat{C}_k^\dagger {\bm h}(k)\cdot{\bm \sigma} \hat{C}_k , \label{eq:KitaeChainBdG}
\end{equation}
where $h_x(k)=\text{Re}[\Delta e^{i\theta}]\sin k$, $h_y(k)=\text{Im}[\Delta e^{i\theta}]\sin k$, 
$h_z(k)=-t\cos k-\mu$, 
${\bm \sigma}$ is the vector of Pauli matrices,  
$\hat C_k=(\hat c_k^\dagger ~ \hat c_{-k})$, 
and $\hat c_k$ is the Fourier transform of $\hat c_i$. The sinusoidal $k$-dependence implies that  ${\bm h}(0)/|{\bm h}(0)|=\pm\hat{\bm z}$ and ${\bm h}(\pi)/|{\bm h}(\pi)|=\pm\hat{\bm z}$, therefore  
\begin{equation}
	\nu = \frac{{\bm h}(0)}{|{\bm h}(0)|} \cdot \frac{{\bm h}(\pi)}{|{\bm h}(\pi)|} =\pm1.
\end{equation}
$\nu$ is known as a $Z_2$ topological invariant (because, as a binary variable, $\nu$ is a representation of the group of integers modulo 2, typically denoted  $Z_2$ \cite{ZeeGroup}). For $|\mu|>t$ one finds \cite{Alicea} that $\nu=1$ and the model is in the trivial phase (adiabatically connected to the $\Delta=t=0$ and $\mu\ne0$ limit). For $|\mu|<t$ one finds \cite{Alicea} that $\nu=-1$ and the model is in an SPT phase (adiabatically connected to the $\Delta=t\ne0$ and $\mu=0$ limit).

\subsection{Anyons}\label{sect:anyons}

Majoranas in 2D can behave even more strangely than the fractionalised particles we found in 1D. To describe this, we first need to review the statistics of quantum particles. Consider a state, $|{\bm r}_1,{\bm r}_2\rangle$, of two identical particles at positions ${\bm r}_1$ and ${\bm r}_2$. If we exchange the positions of the particles, then they will be in the state  $|{\bm r}_2,{\bm r}_1\rangle$. Normalisation of the states requires that 
\begin{equation}
	|\langle {\bm r}_1,{\bm r}_2 | {\bm r}_1,{\bm r}_2 \rangle|^2 = |\langle {\bm r}_2,{\bm r}_1 | {\bm r}_2,{\bm r}_1 \rangle|^2.
\end{equation}
In 3D we are used to two solutions, either $|{\bm r}_2,{\bm r}_1\rangle = |{\bm r}_1,{\bm r}_2\rangle$, in which case we call the particles bosons, or $|{\bm r}_2,{\bm r}_1\rangle =- |{\bm r}_1,{\bm r}_2\rangle$, in which case we call the particles fermions. But in 2D there are two more families of solutions \cite{Alicea}:

\begin{itemize}
	\item Abelian anyons where $|{\bm r}_2,{\bm r}_1\rangle = e^{i\theta} |{\bm r}_1,{\bm r}_2\rangle$ with  $\theta\ne n\pi$ for any integer $n$.
	\item Non-Abelian anyons, where $|{\bm r}_1,{\bm r}_2\rangle$ is a fundamentally different state from $|{\bm r}_2,{\bm r}_1\rangle$. 
\end{itemize}

The 2D generalisation of the Kitaev chain is a spinless `$p+ip$' superconductor. In such a system Majoranas are bound to vortices. In a system with $2N$ Majoranas bound to vortices we can arbitrarily pair up the Majoranas to make $N$ fermions, $f_i=(\hat\gamma_{2i-1}+i\hat\gamma_{2i})/2$. Each fermion has zero energy and can either be occupied or unoccupied. Thus, the Majoranas cause a $2^N$-fold degeneracy  of the ground state.
If one now adiabatically exchanges two Majoranas, this will, in general, change the state to another of the  $2^N$ ground states. But,  eigenstates are orthogonal and must differ by more than a phase. Therefore, Majoranas in $p+ip$ superconductors are expected to be non-Abelian anyons  \cite{Stern,Alicea}.

The idea of composite particles -- particles bound to vortices or magnetic fluxes -- will be important below. So it is worth understanding a toy version of this problem 
in the context of a Bose-Einstein condensate formed from $n$-particle bound states of charge-$e$ bosons in 2D.\footnote{This is only a toy problem, so we need not worry about whether such a thing can really exist.} A vortex, \cref{fig:vortex}b, will carry  a flux $\phi=2\pi\hbar/ne$ \cite{Annett}. Firstly, consider a boson moving around  a single vortex at the origin. This can be described by the Hamiltonian
\begin{equation}
H = -\frac{\hbar^2}{2m} \left( \bm\nabla - i\frac{e}{\hbar}\bm A \right)^2,
\end{equation}
where $\bm A = \frac{\phi}{2\pi r^2 }(y,-x)$. If the boson moves on a closed loop that does not enclose the vortex then we find that $B=(\bm\nabla\times\bm A)_z=0$. But, if the loop encloses the origin then we pick up an (Aharonov-Bohm) phase\footnote{This is easily confirmed for a circular path, $\bm r=r(\cos\theta,\sin\theta)$.} of $\frac{e}{\hbar}\oint \bm A\cdot d\bm r=2\pi/n$.

\begin{figure}
	\centering
	\includegraphics[width=0.5\textwidth]{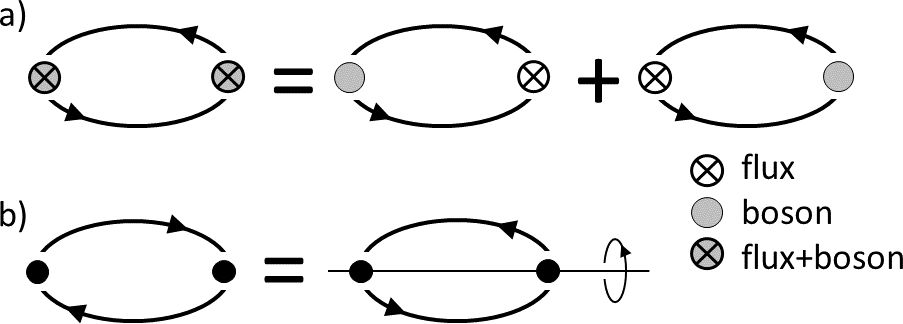}
	\caption{Exchanging anyons. (a) Exchanging two bound states of a flux/vortex and a boson requires moving each boson half way around the flux it is not bound to. (b) In 2D clockwise and anticlockwise are fundamentally different. But, in 3D they are related by a rotation, e.g., about the axis shown. This geometrical fact means that anyons can only exist in 2D.} \label{fig:anyons}
\end{figure}

Now, we want to consider two composite particles which consist of a vortex bound to a boson. What happens to the wavefunction if we exchange the two composite particles? The vortices obey Bose statistics, so exchanging  the vortices does not change the wavefunction; nor does swapping the  bosons. However, exchanging the two bound states requires each boson to move half way around the vortex it is not bound to, \cref{fig:anyons}a. Therefore, each composite particle picks up an Aharonov-Bohm phase of $\frac12\frac{2\pi}{n}$ and the total state picks up a phase of $\theta=2\pi/n$. That is, 
\begin{equation}
|{\bm r}_2,{\bm r}_1\rangle = e^{i\theta} |{\bm r}_1,{\bm r}_2\rangle = e^{i2\pi/n} |{\bm r}_1,{\bm r}_2\rangle. 
\end{equation}
Hence, for
\begin{itemize}
	\item $n=1$ the composite particles are bosons; for
	\item $n=2$ the composite particles are fermions; and for
	\item $n>2$ the composite particles are (Abelian) anyons!
\end{itemize}
For a \textit{conventional} superconductor $n=2$, but the quasiparticles are fermions, so the exchange of these gives us a second minus sign, and the composite particles are bosons.

This works in 2D because clockwise and anticlockwise rotations are fundamentally different. In 3D they are basically the same thing (they can be continuously mapped to one another by rotating the entire system, \cref{fig:anyons}b). This requires that $e^{i\theta} =e^{-i\theta}$. Hence, $\theta=N\pi$, with integer $N$, and all particles must be either bosons or fermions.


\subsection{The emergent gauge field in spin ices and the classical Coulomb phase}\label{sect:iceColomb}

The last two examples, the Haldane and Kitaev chains, are both 1D. Fractionalisation is very common in 1D systems, see  \cite{Schofield}. Fractionalisation in higher dimensions is not only more exotic, but is also accompanied by addition phenomenology. Typically, the fractionalised particles are accompanied by an emergent gauge field, just as the electromagnetic gauge field accompanies charged particles. Classical spin ices provide a simple example of an emergent gauge field \cite{CastelnovoARCMP,Henley,MoessnerPhilTrans}. In these systems the gauge theory arises in a coarse-grained theory as a consequence  the large classical degeneracy.

Consider the nearest neighbour \textit{classical} Heisenberg model on the pyrochlore lattice, \cref{fig:ice}. It is convenient to define the spin flux at a tetrahedron, ${\bm L}_\alpha = \sum_{i\in\boxtimes_\alpha} {\bm S}_i$, where $\alpha$ labels the tetrahedra ($\boxtimes$). I will refer to the lattice of tetrahedra (indexed by $\alpha$) as the dual lattice. The dual lattice for the pyrochlore lattice is the diamond lattice, which is bipartite. 

Up to a constant
\begin{equation}
H=J\sum_\alpha \sum_{i\ne j\in\boxtimes_\alpha} {\bm S}_i \cdot {\bm S}_j =\frac{J}{2}\sum_\alpha {\bm L}_\alpha^2,
\end{equation}
which is easy to confirm by expanding the right-hand-side. For antiferromagnetic coupling ($J>0$) the energy $E\geq0$. Therefore, the ground states satisfy  the flux constraint, ${\bm L}_\alpha=0$, for all $\alpha$. As we found in the Ising model on the same lattice,   \cref{sect:IceMonopoles}, we have many possible ground states, all of which obey a local constraint. The local constraint can be enforced via a gauge field theory.
To construct this, we define $\eta(\alpha)=1$ ($-1$) of the A (B) sublattice.  We now coarse grain our theory by mapping the spins onto a field $\bm B$ via $\sum_{i\in V(\bm r)} {\bm S}_i \rightarrow {\bm B}({\bm r}) V(\bm r)$, where $V(\bm r)$ is the volume characteristic of our coarse-grained description (large compared to a tetrahedron, but small compared to the system) centred at $\bm r$.

\begin{figure}
	\centering
	\includegraphics[width=0.5\textwidth]{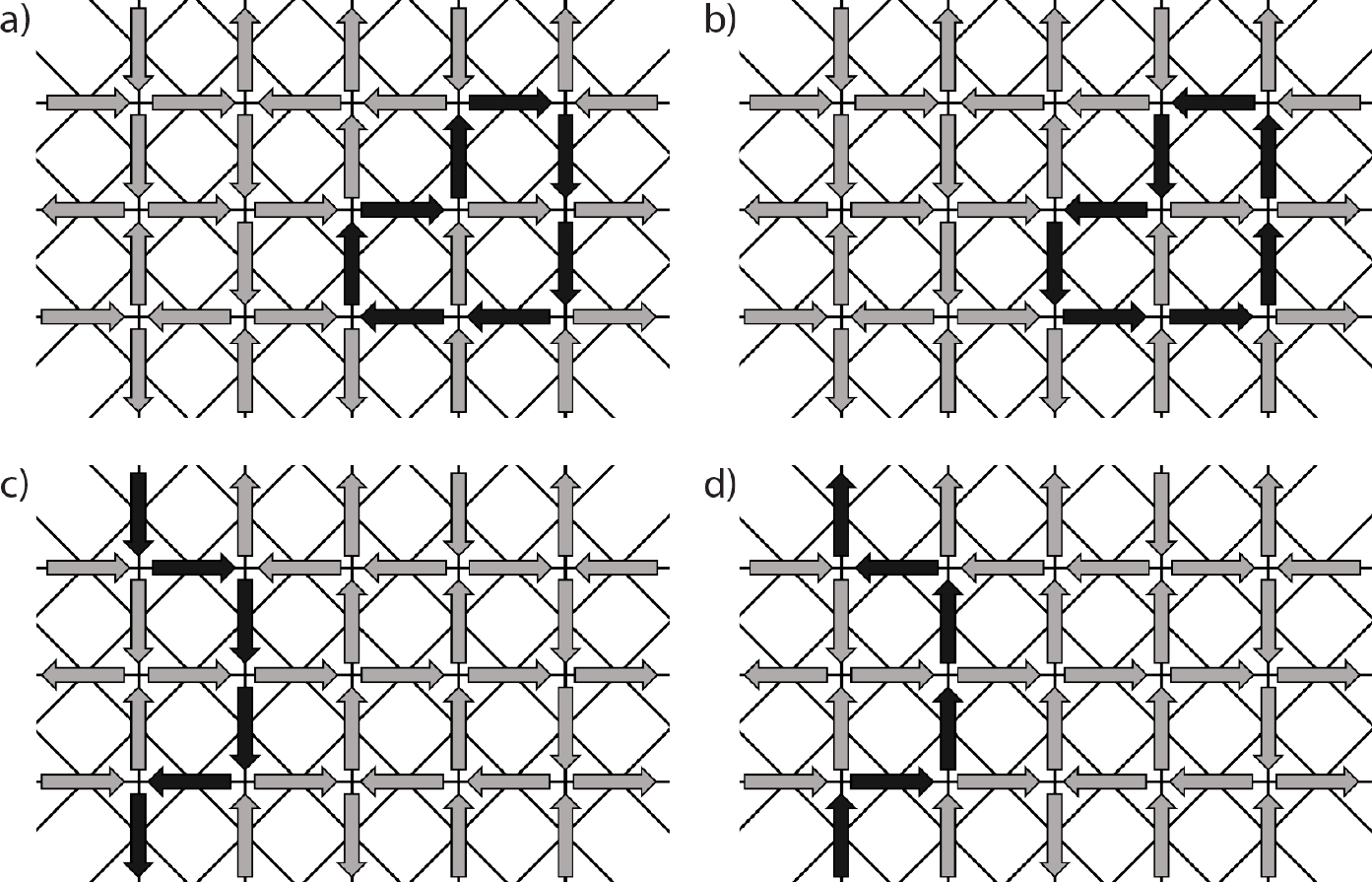}
	\caption{(a,b) Small regions of two states for the Ising model on the checkerboard lattice that both obey the (2-in/2-out) ice rule. They are identical except for flipping all of the spins in the loop highlighted black. As the loop does not span the lattice, neither the total spin nor the coarse-grained $\bm B$ for this small region are changed. (c,d) Another two states  that obey the ice rule. They are identical except for flipping all of the spins  highlighted black. As the loop spans the lattice, both the total spin and the coarse-grained $\bm B$ for this small region are changed. Note that 2D lattices are not in the Coulomb phase -- and are shown here to illustrate the physics of the pyrochlore lattice. Comparing to \cref{fig:ice}, we see that these loop moves can occur by creating a pair of monopoles, the monopoles diffusing through the lattice and then coming back together and annihilating. The flipped loop is then the path the monopoles took.} \label{fig:IceLoops}
\end{figure}

Consider one of the coarse-grained volumes with the spins in an arbitrary state that obeys the flux constraint, ${\bm L}_\alpha=0$. If a spin is flipped, then to maintain the flux constraint one must also flip another spin, with the opposite value, on the same tetrahedron. But, each spin lives at the corner of two tetrahedra. So one can only maintain the flux constraint by flipping the spins around a closed loop that does not extend over the entire system, \cref{fig:IceLoops}a,b.   More generally, flipping spins over a loop that spans the system changes $\bm B$, whereas flipping the spins on a loop that does not span the lattice moves the system between different states with the same $\bm B$.
The number of flippable loops, ${\cal N}(\bm B)$, is maximal for $\bm B=\bm 0$ but becomes increasingly small as $\bm B$ approaches its maximum magnitude (saturation). As the system becomes polarised, most of the spins start to point in the same direction. But, a closed loop requires half of its fluxes to point in the reverse direction, therefore closed loops become increasingly unlikely as $\bm B$ increases. Thus, there are many states with $\bm B=\bm 0$ but few at saturation. Thus, the specific entropy, $s(\bm B)=\lim_{V\rightarrow\infty}\ln[{\cal N}(\bm B)]/V$, is maximal when $\bm B=\bm 0$.

If we have many coarse-grained volumes, and if these volumes are large enough that the local $\bm B$-fields are not too strongly correlated, then, according to the central limit theorem, ${\cal N}(\bm B)\propto\exp(-KV|\bm B|^2/2)$. 
Therefore, $s(\bm B)=s_0-\frac12K|\bm B|^2$ and the free energy is
\begin{equation}
\frac{F(\{\bm B(\bm r)\})}{T} = S_0 + \frac{K}2 \int d^3\bm r |\bm B(\bm r)|^2,
\end{equation}
where $S_0$ is a constant.
The second term in the free energy is entirely entropic. Nevertheless, the free energy has the same form as the energy of a magnetic field (with permeability $1/K$). 

Similarly, when we coarse grain the flux constraint it becomes 
\begin{equation}
\bm\nabla\cdot\bm B(\bm r)=0, \label{eq:Guass}
\end{equation}
which is just Gauss's law in the absence of monopoles. 
Excited states contain monopoles (\cref{sect:IceMonopoles}) which cause the right-hand side to be non-zero. If one has two monopoles, or more generally, two charges magnetic, $Q_{1(2)}$,  at positions $\bm r_{1(2)}$, and no other defects are present then it can be shown  \cite{CastelnovoARCMP,Henley,MoessnerPhilTrans} that the energy increases an amount,
\begin{equation}
{\cal V} = \frac{K}{4\pi}\frac{Q_1Q_2}{|\bm r_1-\bm r_2|}.
\end{equation}
Thus, we have a theory that is  analogous to electrostatics  (or magnetostatics). This is often called the Coulomb phase. Notice that  we are free to define $\bm B(\bm r)=\bm\nabla\times\bm A$, whence our theory is invariant under the gauge transformation defined by \cref{eq:gaugetransA}. Thus, we have an emergent classical gauge field theory of the low energy states of spin ice.

The emergent gauge theory is   only valid at low energies. For example, \cref{eq:Guass} is only true for the ground states and therefore only (strictly) true at $T=0$, although the gap to the monopoles mean that this it is a  good approximation at low temperatures. In general, one does  not expect an emergent gauge transformation   to be respected by all physically allowed states. This becomes important  at temperatures that are high compared to the gap to the excited states, where we expect the gauge field to disappear.

\begin{figure}
	\centering
	\includegraphics[width=0.3\textwidth]{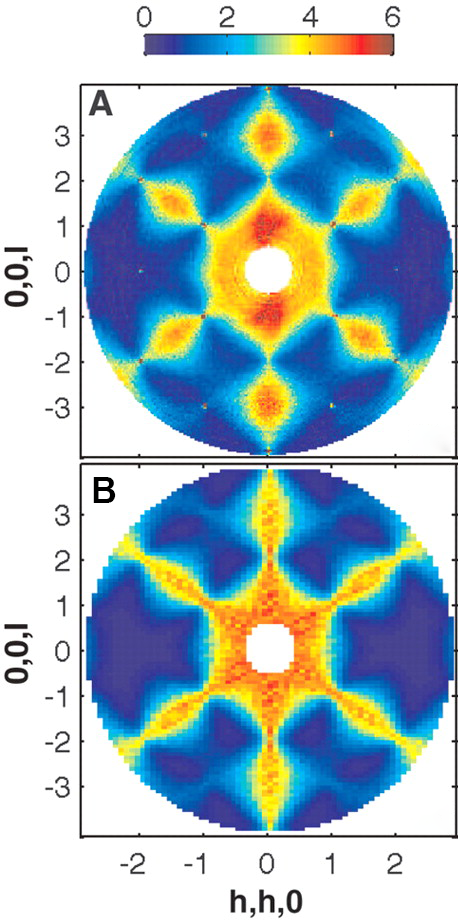}
	\caption{Pinch points in (a) the measured structure factor of \ch{Ho2Ti2O7} and (b) calculated for a near neighbour  model on the pyrochlore lattice. From \cite{Fennell}  reprinted with permission from AAAS.} \label{fig:PinchPoints}
\end{figure}

Remarkably, the gauge field has direct experimental consequences  \cite{CastelnovoARCMP,Henley,MoessnerPhilTrans}. The Fourier transform of \cref{eq:Guass} is $\bm q\cdot\bm B(\bm q)=0$. We denote the transverse part of the field, which satisfies this equation as $\bm B^\perp$. It  follows from  equipartition \cite{Henley,HenleyPRB} that the  structure factor, which is the Fourier transform of the correlation function discussed in \cref{sect:entag&correl}, is 
\begin{equation}
{\cal S}_{\mu\nu}(\bm q)=\left\langle B_\mu^\perp(-\bm q) B_\nu^\perp(\bm q) \right\rangle
= \frac{1}{K} \left( \delta_{\mu,\nu} - \frac{q_\mu q_\nu}{|{\bm q}|^2} \right).
\end{equation}
The spin structure factor can be measured via neutron scattering experiments. If the gauge field description is correct then one should see the same correlations in the spin structure factor because $\bm B$ is just a coarse-grained description of the spins. In particular, notice that the structure factor as, say, $\bm q\rightarrow (0,0,2\pi)$ depends on the direction from which we approach. That is, ${\cal S}_{\mu\nu}(0,0,2\pi)$ is singular (but not divergent). This gives rise to characteristic pinch points (also known as bow-ties). These are observed experimentally \cite{Henley,CastelnovoARCMP,MoessnerPhilTrans,Fennell}, providing a dramatic confirmation of the emergent gauge field theory, \cref{fig:PinchPoints}.

\subsection{Quantum  spin liquids}

In solids the atomic positions are so strongly correlated that they have long-range positional order. In a gas the correlations between atomic positions are weak. In many ways, the liquid  is  more interesting (and theoretically challenging): the correlations between atoms are strong enough that they cannot be neglected (as in the ideal gas), but not so strong that they dominate and produce long-range order (which, as  we saw in \cref{sect:SM}, there is a deep understanding of).

The same is true in magnetic materials. We have a deep understanding of phases with long-range magnetic order. Understanding the weakly interacting moments in classical paramagnets is straightforward \cite{Ashcroft}. But, if there are strong correlations between spins and no long-range order then more interesting and, to date, less well understood phenomena can arise. Such states are known as spin liquids.
While this idea is straightforward, even defining what a spin-liquid is remains contentious. But, as we are just going to discuss some examples, we do not need a rigorous definition. I will use the term `spin liquid' for any state that has well defined local moments (spins), does not have long-range magnetic order, and cannot be described by a model of weakly interacting spins.   
%

Long-range magnetic order is very common, so one question is: how we can suppress it? Low dimensionality  suppresses long-range order, and we have encountered an example of a 1D system without long-range order in \cref{sect:Haldane}
. This fits the hand-waving definition of a spin liquid given above. However, the reader should note that some people would prefer the definition to exclude 1D states from the definition of spin liquids. At low enough temperatures, crystalline materials are 3D, so we  expect long-range order to emerge in many chain systems at low-enough temperatures. However, low-enough can mean be significantly colder than any temperature ever achieved in the known universe \cite{Elise}.

Spin ice is an example of a classical spin liquid. 
Once quantum effects are included, there are many possible types of spin liquid that could emerge in more than one dimension. In fact, although many different types have been identified, it seems likely that the classification remains incomplete  (see  \cite{Savary} for a recent review). 

Spin liquids do not break any symmetry of the Hamiltonian, and cannot be described by a local order parameter. They are sufficiently strongly correlated that adiabatic continuity to a set of non-interacting spins does not provide us with a good description of their physics. Therefore, they are not described by the standard model of materials. The low-energy physics of many spin liquids is described by an emergent gauge field and fractionalised quasiparticles. We will consider some examples below that are reasonably tractable and illustrate the  rich physics found in quantum spin liquids.

\subsubsection{Dimer models and RVB states} 
 
One way to prevent long-range order is to look at geometrically frustrated materials, \cref{fig:tri}. Indeed the idea of a spin liquid was first proposed as a theory of the spin-1/2 Heisenberg model on the triangular lattice (we now know that this model  has long-range order). This idea started to be widely discussed when Anderson proposed that high temperature superconductivity in the cuprates is deeply connected to spin liquidity \cite{AndersonRVB}.

\begin{figure}
	\centering
	\includegraphics[width=\textwidth]{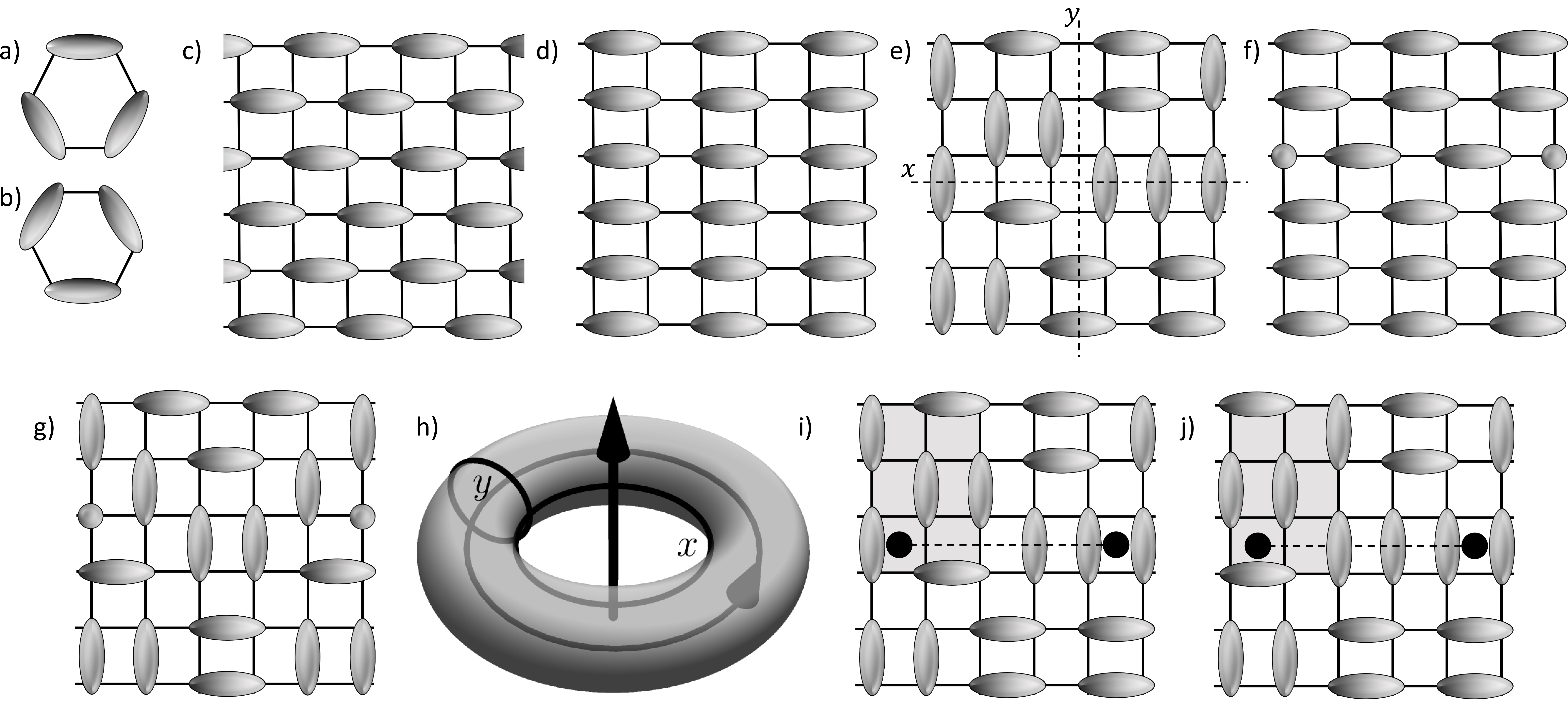}
	\caption{Valence bond states. Shaded ovals represent singlets or dimers. a,b) Two possible valence bond states on a single hexagon. These pictures can be interpreted as the $\pi$-bonds in benzene in the valence bond theory of chemistry. Their superposition explains the aromaticity of benzene \cite{valence}. The resonating valence bond theory of magnetic materials takes its name from this similar chemistry. 
	c) and d) show valence bond solids with broken translational symmetry: the staggered and columnar states respectively. 
	e) A possible dimer covering of the square lattice. This could be one state in a superposition making up an RVB state. The dashed lines $x$ and $y$ extend throughout the sample. If the lattice is on a torus then they are closed loops (panel h). 
	f) A pair on monomers (shaded circles) in an ordered phase leave a chain of unfavourably arranged dimers. The cost of this chain grows with its length, confining the dimers.
	g) In an RVB state there is no chain as the dimers are already disordered. Thus the monomers are deconfined.
	h) A torus with the two topologically distinct loops $x$ and $y$ marked, cf. panel (e). The arrows represent fluxes threading the hole and inside the torus.
	i) and j) show a pair of visons (black circles). Moving a monomer around on vison is equivalent to flipping the dimers on that path (here the perimeter of the shaded area). This changes the number of dimers crossing the (dashed) line, connecting the visons, by one.} \label{fig:RVB}
\end{figure}

The ground state of a pair of antiferromagnetically coupled spin-$1/2$s is a singlet, \cref{sect:QM+SBS} and \cref{eq:2estatesS}. A singlet is a maximally entangled state:  if a spin is in a singlet, it cannot be entangled with any other spins. Therefore, covering a lattice of spin-$1/2$s with a pattern of singlets completely specifies a quantum state.  The singlets are often called dimers in this context and one covering of the lattice is known as a valence bond state. But, of course, there are many possible coverings for any lattice, \cref{fig:RVB}. Anderson's idea was that a spin liquid can be built up from a quantum superposition of valence bond states;  known as a resonating valence bond (RVB) state, in analogy with the theory of aromaticity in chemistry (Fig. \ref{fig:RVB}a,b \cite{valence}).

Another motivation for looking at valence bond states is that they provide an alternative description of classical spin ices \cite{Gingras,MoessnerRaman}. So far we have have discussed 2-in/2-out ice rules. However, for   3-in/1-out and 1-in/3-out ice rules, we can derive a valence bond state working on the dual lattice and putting a dimer along any bond with the minority spin. Each ice state then maps to a dimer covering of the lattice.  Quantum dimer models are effective low-energy models of the quantum dynamics that allow the system to tunnel between ice states \cite{Gingras,MoessnerRaman}.\footnote{2-in/2-out ice states can be mapped to loop models, which are basically dimer models with two dimers on every site} Thus, they provide an important connection between classical and quantum spin liquids.

Quantum dimer models \cite{MoessnerRaman} are constructed to give valence bond states, and so provide insights into how such states might behave. These models do not describe the spins explicitly, but only include the dimers. On the square lattice\footnote{I will focus on the square lattice for simplicity. But, note that there is no extended spin liquid phase on the square lattice, only a special point with these properties. The model can be straightforwardly generalised to other lattices.} the Hamiltonian can be written as 
\begin{equation}
H = \sum_{\plaquette} \left[ - J \big(|\plaquettev\rangle\langle\plaquetteh|+|\plaquetteh\rangle\langle\plaquettev|\big)  
+ V\big(|\plaquettev\rangle\langle\plaquettev|+|\plaquetteh\rangle\langle\plaquetteh|\big)\right].
\end{equation}
The sum is over all `plaquettes', and the thicker lines represent the valence bonds. The $J$ term provides a `kinetic' energy for flipping the dimers on a plaquette, but the $V$ term adds a `potential' energy cost for having flippable plaquettes. 
We will limit our discussion to $J>0$. For $V/J\rightarrow\infty$, the system seeks to minimise the number of flippable plaquettes, which is achieved by the staggered phase, \cref{fig:RVB}c. In the limit $V/J\rightarrow-\infty$ the system seeks to maximise the number of flippable plaquettes, which is achieved by the columnar phase, \cref{fig:RVB}d.

In general the $J$- and $V$-terms compete, but at the so-called `Rokhsar-Kivelson' point, $V=J>0$, the model becomes
\begin{equation}
H_{RK} = \sum_{\plaquette} \left[  J \big(|\plaquettev\rangle - |\plaquetteh\rangle\big)  
\big(\langle\plaquettev|-\langle\plaquetteh|\big)\right].
\end{equation}
The Hamiltonian is now a projection operator, with eigenvalues $0$ or $J$. Therefore, any state that is annihilated by $H_{RK}$ will be a (zero energy) ground state. In general, states of dimer models have the form $\ket{\Psi} = \sum_c A_c \ket{c}$, where $\ket{c}$ is a covering of the lattice. Such a state will be annihilated by $H_{RK}$ if and only if the amplitudes, $A_c$, are the same for all coverings, $\ket{c}$, that differ only by one flipped plaquette. Therefore, if we denote by ${\sum_c}^\prime$ a sum over coverings that can be related by (repeatedly) flipping plaquettes, then any state of the form
\begin{equation}
\ket{\Psi_0} = {\sum_c}^\prime  \ket{c} \label{eq:RKwavefn}
\end{equation}
is a ground state.

An obvious question is: how many states are there with this form? Consider a reference line  spanning the lattice,  \cref{fig:RVB}e. Any flippable plaquette must cross this line an even number of times. Therefore, flipping a plaquette cannot change the parity of the state, which is defined as the number modulo two of dimers that cross this line. On a torus we can define a pair of winding numbers, $(F_x,F_y)$, that label different topological sectors. States in different topological sectors cannot be transformed into one another by flipping plaquettes. Changing between topological sectors requires a non-local move. While the values of $F_x$ and $F_y$ depend on which loops one picks, this does not change the topological sectors (only our labelling of them). On an arbitrary surface  the topological sector is labelled by $2g$ winding numbers, where $g$ is the genus of the surface: $g=0$ for the sphere, $g=1$ for a torus, and two objects with $g=2$ are shown in \cref{fig:vortex}a. As each winding number is a binary variable this leads to a ground state degeneracy of at least $2^{2g}$. To characterise the Rokhsar-Kivelson states, \cref{eq:RKwavefn}, we need to specialise to a particular lattice. On the square lattice there is believed to only be one such state in each topological sector. For example, the staggered state,  \cref{fig:RVB}c,  contain no flippable plaquettes and, therefore, is the only member of its topological sector and  has zero energy for any $J$, $V$. It is therefore trivially of the form of \cref{eq:RKwavefn}.

Another question one needs to address is: is the Rokhsar-Kivelson point part of a phase, with many other adiabatically connected states, or is it special? Isolated points are with unusual properties are far less interesting than extended phases as they are much harder to observe in any real material.  On the basis of several examples it is conjectured that  the  Rokhsar-Kivelson point is only part of an extended spin liquid phase in three or more dimensions or on frustrated 2D lattices \cite{MoessnerRaman}. In these cases, the Rokhsar-Kivelson point is the boundary between the, long-range ordered, staggered phase and the spin liquid.

We can  extend our Hilbert space to include two `monomers': spin-$1/2$s that are in a triplet state, also known as spinons. In an ordered phase, such as the staggered phase, moving the  monomers apart leaves a chain of unfavourably arranged dimers, \cref{fig:RVB}f. Therefore, the energy of the state grows linearly with the separation of the monomers and they are confined (it takes an infinite amount of energy to move them infinitely far apart). In a spin liquid phase there is no such chain and so one expects that  the spinons are deconfined, \cref{fig:RVB}g. This is in contrast to our expectation of spin-1 Nambu-Goldstone bosons in a magnetic materials with long range order.  Similarly, if we removed an electron, leaving a monomer and an empty site, we would expect spin-charge separation, where the charge and spin degrees of freedom behave as independent particles, in the spin liquid phase. This is reminiscent of the low-energy excitations in a Luttinger liquid \cite{Schofield}. 

\subsubsection{$Z_2$ spin liquids}

Dimer models also give rise to emergent gauge fields. Recall that in a classical gauge theory, such as electromagnetism, we replace the physical observables, $\bm B$ and $\bm E$, with new `gauge potentials', $\bm A$ and $\phi$. The gauge description is overcomplete: there are many choices of  $\bm A$ and $\phi$ that give rise to the same $\bm B$ and $\bm E$,  \cref{eq:EBgauge}. The gauge transformations, \cref{eq:gaugetrans}, take us between different choices of $\bm A$ and $\phi$ that give rise to the same $\bm B$ and $\bm E$. Therefore, a gauge transformation does not change the state of the system, it simply changes \textit{our} description of that state. This is why all physical properties of a system \textit{must} be gauge invariant. The nature of the universe does not depend on how we choose to label things. ``A rose by any other name would  smell as sweet''\footnote{Shakespeare was clearly aware of the difference between gauge and physical variables. Juliet laments that while Romeo is just a name, the fact that he is a Montague and she a Capulet is the source of their problems.}.  

Gauge descriptions of quantum matter arise in  the same way. We expand our Hilbert space and describe the system in terms of a new set of operators. These provide an overcomplete description of the system -- we have given multiple `names' to every state. Therefore, all physical states must be unchanged when we make a gauge transformation that merely changes the labels on the states. If you are wondering what the point is in such a purely formal manipulation, then you understand the process. The point is that, as in the case of classical electromagnetism, the new gauged description makes some calculations easier. For example, below we will use the a gauge field description to show that another type of particle exists in this model: a result that would be extremely challenging to derive directly.

I will not attempt to  derive the  $Z_2$ gauge field theory, which can arise in various contexts see \cite{MoessnerRaman,Wen,Savary}, and instead limit myself to a description of the theory. 
%
In quantum dimer model the dimers  live on the bonds between sites, so it is natural to introduce a variable that lives on the bonds too: $s_{ij}=s_{ji}=\pm1$. In some limits one can connect the two models by using $s_{ij}$ to label the presence or absence of a dimer \cite{MoessnerRaman}. We also define a gauge transformation,
\begin{equation}
	\tilde{s}_{ij}=W_i s_{ij} W_j^{-1},
	\label{eq:Z2gauge}
\end{equation}
where $W_i=\pm1$. The two possible values of $W_i$ are why this is called a $Z_2$ gauge theory. Now we need to count the size of our Hilbert space; to do so I will assume periodic boundary conditions, which is equivalent to putting our lattice on the surface of a torus. A single plaquette contains four sites, so there are $2^4=16$ different $Z_2$ gauge transformations on a plaquette. There are also four bonds on a plaquette so there are $2^4=16$ different $s_{ij}$ configurations on a plaquette. If the number of states were equal to the number of gauge transformations then there would only be one class of gauge-equivalent states and hence one physical state. But, we have been too hasty. Two of the gauge transformations do nothing: if all of the $W_i=1$ or all of the $W_i=-1$ then $\tilde{s}_{ij}=s_{ij}$. A do-nothing transformation is called the identity.\footnote{E.g., 1 is the identity for multiplication and 0 is the identity for addition.} We only have space for one identity in the gauge group \cite{ZeeGroup},  so we should not count them both. Similarly we should not count both of any pair of transformations that differ only by changing the signs of all $W_i$ as these must be equivalent. This decreases the number of gauge-equivalent states by a factor of two. Therefore, a single plaquette contains eight gauge equivalent states and $16/8=2$ physical states.

There are twice as many bonds as there as sites for a square lattice  on a torus. Therefore, for $N$ sites there are $2^{2N}$ possible $s_{ij}$ configurations. Na\"ively, there would appear to be $2^N$ different gauge transformations. But, again, we are double counting: there are $\frac{2^{2N}}{2^N/2}=2^{(N+1)}$ physical states.

We can represent these physical states as a $Z_2$ flux, $F_p$, threading the $p$th plaquette:
\begin{equation}
F_p = \prod_{i,j\in p} s_{ij} = \pm1.
\label{eq:Z2flux}
\end{equation}
For any two adjacent plaquettes the net flux is $F_{p+p'}=\prod_{i,j\in P_{p+p'}} s_{ij}$, where $P_{p+p'}$ is the perimeter of the rectangle $p+p'$. The internal bond on the shared edge does not contribute because it appears twice in the total product and $s_{ij}^2=1$. If we calculate the total flux through the holes on a torus, then every bond appears twice in the product, so $F_\text{total} = \sum_p F_p = \sum_{\langle ij\rangle} s_{ij}^2 =1.$ There are the same number of plaquettes as lattice sites on a torus. The constraint $F_\text{total} =1$ means that  on an $N$ site lattice, only $N-1$ fluxes can be chosen freely, and so there are $2^{(N-1)}$ possible arrangements of the $Z_2$ fluxes. 

Combining the results of the last two paragraphs, we see that there are $2^{(N+1)}/2^{(N-1)}=4$ physical states for every possible pattern of fluxes. This factor of four is related to the two distinct topologically non-trivial paths around the lattice,  \cref{fig:RVB}e,h. For a fixed pattern of fluxes there are four possible configurations of $s_{ij}$. These can be easily constructed. We simply state  one configuration, which we can call $s_{ij}^0$. We then define the function $X_{ij}$ to be $-1$ if the bond $ij$ crosses the $x$-line in   \cref{fig:RVB}e,h and 1 otherwise, with a similar definition for $Y_{ij}$. Now, the four related configurations are given by 
\begin{equation}
s_{ij}^{(F_x,F_y)}=X_{ij}^{F_x} Y_{ij}^{F_y} s_{ij}^0,
\label{eq:topoDegen}
\end{equation}
with $F_x$, $F_y=0$ or 1. So the winding numbers  that described the dimer model emerge naturally in the $Z_2$ gauge theory. In terms of the fluxes, one can elegantly view the winding numbers as a flux inserted either through the hole in the torus or inside it, \cref{fig:RVB}h. For higher genus surfaces, $2g$ topologically distinct loops spanning the lattice can be constructed, which leads to a topological degeneracy of $2^{2g}$.

Having discussed the Hilbert space in some detail, we are now ready to write down a Hamiltonian. To do this we need to identify the lowest order terms that are gauge invariant. $\sigma_{ij}^x$ and $F_p$ are gauge invariant, as is simple to check from  the definitions $\sigma_{ij}^x s_{ij} = -s_{ij}$, and \cref{eq:Z2gauge,eq:Z2flux}. Therefore, we expect the low energy physics to be described by 
\begin{equation}
H_{Z_2} = -v \sum_p F_p - t \sum_{\langle ij \rangle} \sigma_{ij}^x.
\end{equation}

For $v\gg|t|$ the ground state has all $F_p=1$. The lowest energy excitations involve changing the signs of a pair of fluxes (one cannot change the sign of a single flux because of the constraint $F_\text{total} =1$). These are topological excitations known as $Z_2$ vortices or visons. The $t$-term allows the visons to hop. Because the $\sigma_{ij}^x$ operators commute with one another  vison  exchange has bosonic statistics. 

The winding numbers $F_x$ and $F_y$ do not appear directly in the Hamiltonian. So it would appear that the energy does not depend on the winding numbers and we have a four-fold degeneracy on a torus. This is correct in the thermodynamic limit. But on a finite lattice of size $L\times L$ there is a small correction. We can change between the four states defined by \cref{eq:topoDegen} by acting $\sigma_{ij}^x$ on every $s_{ij}$ that crosses the $x$- and/or $y$-lines. This is equivalent to creating a pair of visons, moving them around the torus, and then annihilating them. Perturbation theory tells us that this will lift the four-fold degeneracy by $\sim t^L/v^{L-1} \sim e^{-cL} \rightarrow 0$ as $L\rightarrow\infty$, for some positive constant, $c$.
Note that in general, visons, like monomers, cannot be created individually by any local operator. Local operators always create pairs of visons with opposite winding numbers, as no local operator can change the total winding number of the system.

The visons and monomers have `mutual semionic statistics'.  We can wind a monomer around a vison by breaking a dimer in two, leaving one of the resulting monomers stationary and hopping the other around the vison, then annihilating the monomer to form a new dimer. This is equivalent to flipping all the dimers along some large flippable loop around the vison.  Because such a loop must necessarily cross a line connecting two visons an odd number of times, \cref{fig:RVB}i,j, this induces a $\pi$ phase shift in the wavefunction: $\ket\psi\rightarrow-\ket\psi$. The jargon used to describe this is that the vison and the monomer are mutual semions. Semions are a type of anyons.

In the absence of visons, monomers have bosonic statistics. This might seem counter-intuitive; they are spin-1/2 particles after all. However, Lorentz invariance is required for the spin statistics theorem, which states that integer spin particles are bosons and half-odd-integer particles are fermions. The lattice explicitly breaks  Lorentz invariance at low-energies and, therefore,  effective theories of strongly correlated matter do not have to respect the spin statistics theorem. However, the statistics of the visons can become fermionic if we attach a flux to them. In other words, the mutual statistics of a  bound vison-monomer pair with another  pair is fermionic. This is not what we would usually expect for the bound state of two bosons! and is another hallmark of the anyonic behaviour of the excitations.

For $t\gg|v|$ the model is equivalent to a lattice of uncoupled spins with a magnetic field in the $x$-direction. In the ground state every spin  points in the positive $x$-direction, i.e., $s_{ij}=(\ket{1}+\ket{-1})/\sqrt{2}$ for all $ij$. Therefore, the ground state is $\sum_{\{s_{ij}\}}s_{ij}$, where the sum runs over all possible configurations $\{s_{ij}\}$.

Therefore, we see that there are two phases of the $Z_2$ gauge theory. For $v\gg|t|$ the theory has topological order, the degeneracy of the ground state depends on the topology of the lattice, and the excitations are visons. This is known as the deconfined phase. But, for $t\gg|v|$ the theory is  topologically trivial and none of the characteristics of the $Z_2$ gauge theory are apparent from the low energy physics. This is known as the confined phase.
It is believed \cite{MoessnerRaman,Savary,Fradkin} that frustrated dimer models are in the deconfined phase.

$Z_2$ lattice gauge theories are of major interest to the quantum computing community. Kitaev devised a specific model, known as the toric code \cite{KitaevToric}, that is in the deconfined phase. In this context the monomers and visons are usually called e and m particles respectively, for reasons that will become clear in the next section. He proposed that the four degenerate ground states on a torus (or, equivalently, a plane with periodic boundary conditions) could be used as two qubits in a quantum computer. The topological order protects the qubits from noise, which is typically local  -- the qubits are said to be intrinsically decoherence free. For example, the probability of a vison spontaneously winding around the lattice, which would flip a qubit, becomes small as the size of the lattice becomes large. Potentially, any material described by a $Z_2$ lattice gauge theory could be used as a fault tolerant qubit. Two key design criteria would be large gaps to the monomer and vison excitations and the ability to manipulate the qubits to perform quantum gates. Savary and Balents \cite{Savary} recently gave a detailed review of the toric code from a condensed matter perspective.

\subsubsection{U(1) spin liquids and emergent electromagnetism}

Dimer models on bipartite lattices in three or more dimensions are described by  a compact $U(1)$ gauge theory \cite{MoessnerRaman,Savary,Fradkin,Wen}.\footnote{$U(1)$ is the name for the group of $1\times1$ unitary matrices, more simply called  phases, $e^{i\theta}$, with real $\theta$.} This is a discrete analogue  of theory of electromagnetism. This model contains two fields, $A_{ij}=-A_{ji}$ and $E_{ij}=-E_{ji}$, which again live on the bonds of the lattice (i.e., $E_{ij}=A_{ij}=0$ unless $i$ and $j$ are nearest neighbours). We choose these fields to be canonically conjugate: $[A_{ij},E_{ij}]=i$. 

We will require that our theory has a gauge invariance. $A_{ij}$ is the same state as $A_{ij}+2\pi$; \textit{i.e.}, $A_{ij}$ is a compact variable. This is different from electromagnetism, where the vector potential can be any real vector, and is not compact.
The commutation relation implies that $e^{-i\theta E_{ij}}A_{ij}e^{i\theta E_{ij}} = A_{ij} + \theta$. This is only  consistent with the required  gauge invariance if $e^{i2\pi E_{ij}}=1$, which can only true  if all the eigenvalues of $E_{ij}$ are integers.

We can make a gauge transformation by shifting $A_{ij}$ by the (discrete) gradient of an arbitrary scalar function, $\Lambda_i$. Specifically, $A_{ij} \rightarrow A_{ij} + \Lambda_i - \Lambda_j$. This is generated by a unitary operator, which the theory is named for, 
\begin{equation}
	U(\Lambda) = \exp\left[ i\sum_i \Lambda_i  (\nabla\cdot E)_i  \right],
\end{equation}
where we define the discrete divergence as $(\nabla\cdot E)_i \equiv \sum_j E_{ij}$. Comparing to  Maxwell's equations, it is natural to interpret a divergence as an electric charge, so we define $(\nabla\cdot E)_i\ket\psi=q_i\ket\psi$ for any state $\ket\psi$. Gauge invariance then requires that $U(\Lambda)\ket\psi=e^{i\sum_iq_i\Lambda_i}\ket\psi$, much like electromagnetism (cf. Eq. ref{eq:gaugetrans}a).

In order to write down a Hamiltonian we need to identify the lowest order terms consistent with gauge invariance. One such term is $B_p \equiv (\text{curl }  A)_p \equiv \sum_{i\in p} A_{i,i+1_p}$, where site $i+1_p$ is the site neighbouring $i$ in a clockwise direction around the plaquette, $p$. Because the gauge transformation at site $i$ acts  with opposite signs for $A_{i,i+1_p}$ and $A_{i-1_p,i}$ and a plaquette is a closed loop, $B_p$, is gauge invariant. However, the model also needs to respect the compactness of $A_{ij}$, so $B_p$ cannot appear directly in the Hamiltonian, but  $\cos B_q$ can. $E_{ij}$  commutes with $U(\Lambda)$ and so is also gauge invariant. Thus, our Hamiltonian is 
\begin{equation}
H = - K \sum_p \cos B_q + K'\sum_i q_i^2 + U \sum_{ij}E_{ij}^2.
\label{eq:H_U1}
\end{equation}

In 2+1 dimensions\footnote{That is for a 2D lattice plus time.} the theory is always confining \cite{Wen}. This explains why there are no spin liquid phases in dimer models on 2D bipartite lattices. But in 3+1 dimensions there is also a deconfined phase for small, but non-zero, $U$. In this limit the fluctuations in $B_p$ are small and we can Taylor expand the Hamiltonian (and ignore a constant, which cannot change the physics). If there are no charges present and we take the continuum limit, we find
\begin{equation}
H \simeq  \int d^3x \left[ \frac{\varepsilon^*}{2} |\bm E|^2 + \frac{1}{2\mu^*}|\bm B|^2 \right]. 
\end{equation}
This is  the Hamiltonian describing electromagnetism in the vacuum -- except, there is no reason to expect that the effective dielectric constant, $\varepsilon^*$, or permeability, $\mu^*$ take the the same values as they do in the vacuum. Nevertheless, a clear prediction of this model is that there should be a `photon' with the dispersion relation $\omega=c^*q$, where $c^*=\sqrt{\varepsilon^*\mu^*}$. Just as in ordinary electromagnetism, this model should have two transverse polarisations.

The appearance of a massless particle tells us immediately that the deconfined phase does not have topological order, as the $Z_2$ spin liquid does. States with topological order always have a gap between the ground states  (which has degeneracy dependent on the surface the lattice lives on) and the excited states. 

We have already seen that the theory also has electric charges. These are gapped excitations with energy $\sim K'$, cf. \cref{eq:H_U1}.
It also has magnetic monopoles. These are topological point defects in 3D. The existence of monopoles is connected to the compactness of $A_{ij}$. This means that $B_p\equiv (\text{curl }  A)_p$ is only defined modulo $2\pi$. A reasonable way to deal with this is to take the smallest possible value, $-\pi<B_p\leq\pi$. A topological defect is associated with a non-zero winding number, $m_q$, in $B_p$ around its core, \cref{fig:vortex}. This means that the surface integral (or rather its discrete analogue, the sum over plaquettes) around a topological defect is 2$\pi$ times the winding number, which by definition is a non-zero integer. As the monopoles arise from defects in the magnetic field we expect them to be gapped with energies $\sim K$, cf. \cref{eq:H_U1}.

The electric charges interact via a Coulomb potential $\propto q_iq_j/r_{ij}$, where $r_{ij}$ is the distance between them. Similarly, the magnetic monopoles experience a long-range long range force $\propto m_pm_{p'}/r_{pp'}$. Therefore, the deconfined phase of a compact $U(1)$ gauge theory is often called the Coulomb phase. The monopoles and electric charges cannot be anyons because those can only exist in 2D. However, neither particle can be created or destroyed by any local operator. Local operators can only create charge neutral pairs of  particles, as is familiar from quantum electrodynamics in the vacuum. 


\subsubsection{Spin-orbit coupling and the Kitaev honeycomb model}

\begin{figure}
	\centering
	\includegraphics[width=0.45\textwidth]{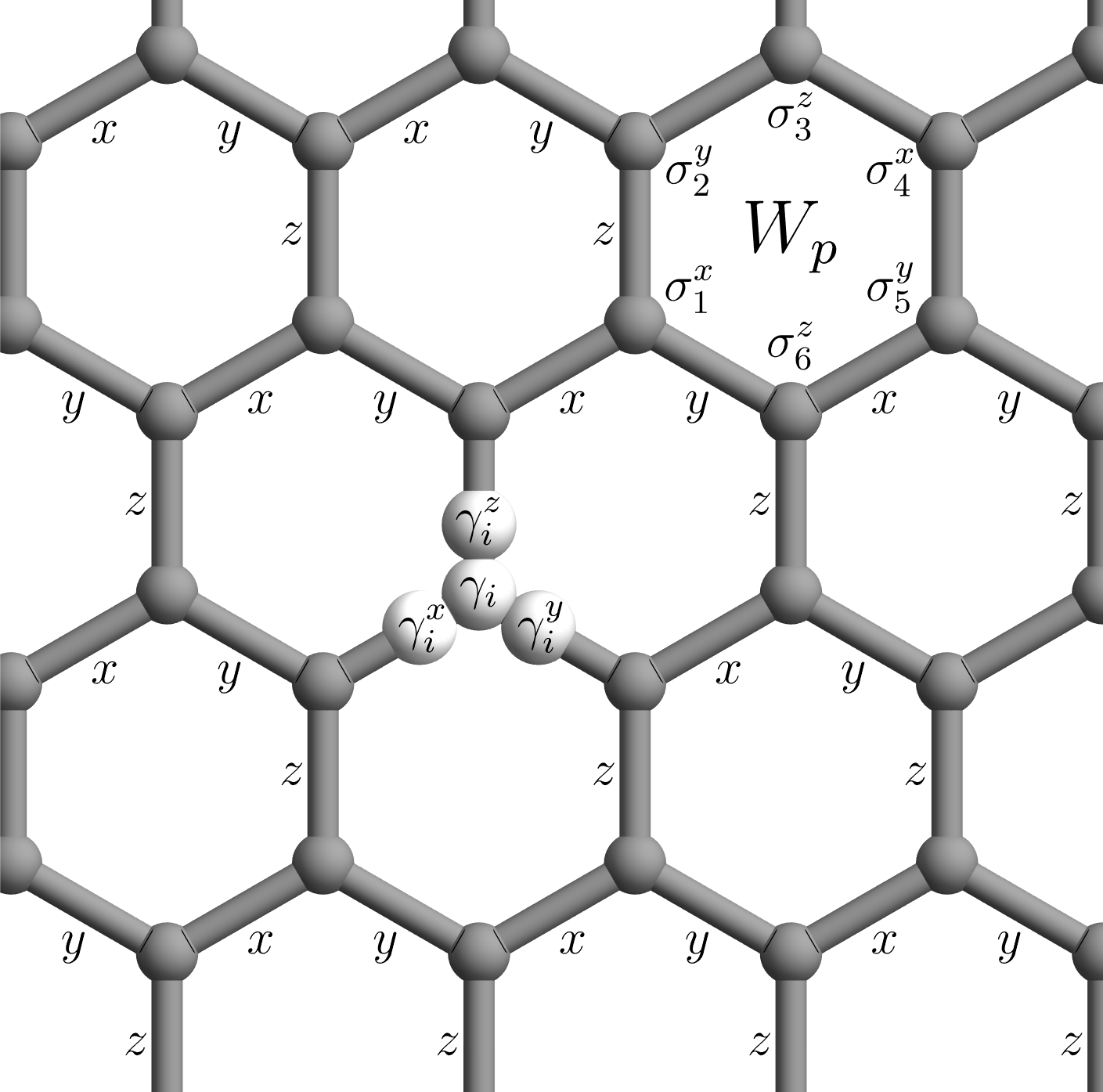}
	\caption{Illustration of Kitaev's honeycomb model. The  bonds are labelled $x$, $y$ or $z$. The plaquette flux operator $W_p$ is the product of the spin operators, $\sigma_i^\mu$, around the loop, as shown. The model is diagonalised by writing the spin operators in terms of four Majorana operators, $\hat\gamma_i$ and $\hat\gamma_i^\mu$, with $\mu\in\{x,y,z\}$, as sketched.} \label{fig:KitaevHoneycomb}
\end{figure}

A very different route to realise a spin liquid comes via systems with strong spin-orbit coupling \cite{Trebst}. The simplest model for this is the Kitaev honeycomb model  \cite{KitaevHoneycomb,KitaevHoneycombLecture},
\begin{equation}
	H_{Kh} = \sum_{\mu\in\{x,y,z\}}\sum_{\langle ij\rangle\in x}  J_\mu \sigma_i^\mu \sigma_j^\mu
\end{equation}
with a single spin-1/2 degree of freedom  on each vertex of the honeycomb lattice. Each site connects to exactly one $x$, $y$ and $z$ bond, \cref{fig:KitaevHoneycomb}. This model may seem very artificial, But, strong spin-orbit coupling can lead to exactly this kind of anisotropy in strongly correlated materials \cite{Trebst,Oshikawa,Khosla}. 

This model can be solved exactly by introducing four species of Majorana fermions; defined by $\sigma_j^\mu = i\hat\gamma_j \hat\gamma_j^\mu$, $\hat\gamma_j^\dagger = \hat\gamma_j$, and $(\hat\gamma_j^\mu)^\dagger = \hat\gamma_j^\mu$, with $\mu=x$, $y$ or $z$. In terms of the Majoranas 
\begin{equation}
	H_{Kh} = \frac{i}{2} \sum_{\mu\in\{x,y,z\}} \sum_{\langle ij\rangle\in \mu} J_\mu \hat u_{ij} \hat\gamma_i \hat\gamma_j,
\end{equation}
where $\hat u_{ij}= i\hat\gamma_i^\mu \hat\gamma_j^\mu$. However, in the Majorana representation we must constrain our Hilbert space to only include states that satisfy $\hat\gamma_i^x \hat\gamma_i^y \hat\gamma_i^x \hat\gamma_i \ket\psi = \ket\psi$ for all $i$. This can be achieved by introducing the projection operator $P=\frac12\prod(1+\hat\gamma_i^x \hat\gamma_i^y \hat\gamma_i^x \hat\gamma_i)$. This constraint introduces a $Z_2$ gauge structure to the Majorana picture. 

The Majoranas obey fermionic anticommutation relations, cf. \cref{eq:MajoranaCom}. Therefore, $[\hat u_{ij},\hat u_{kl}]=[H_{Kh},\hat u_{ij}]=0$ and the $u_{ij}=\pm1$ are good quantum numbers. It is therefore straightforward to diagonalise the Hamiltonian as it is quadratic in the $\hat\gamma_i$ operators. This yields the dispersion relation $\varepsilon_{\bm k}=\pm2|J_x e^{i (\sqrt{3}k_x,3k_y)/2} + J_y e^{i (-\sqrt{3}k_x,3k_y)/2} + J_z|$. This spectrum is gapless if every $J_\mu \leq \sum_{\nu\ne\mu} J_\nu$ and gapped otherwise.

As well as the Majoranas there are also topological defects. To see this it is helpful to introduce a plaquette flux operator, $\hat W_p =  \hat u_{12} \hat u_{23} \hat u_{34} \hat u_{45} \hat u_{56}  \hat u_{61} = \sigma_1^x \sigma_2^y \sigma_3^z \sigma_4^x \sigma_5^y \sigma_6^z$, where the sites are numbered clockwise around the plaquette, \cref{fig:KitaevHoneycomb}. The flux operator commutes with the Hamiltonian, $[H,\hat W_p]=0$, and has eigenvalues $W_p\pm1$. In the ground state $W_p=1$ for all plaquettes. This defines our vacuum and we therefore have a $\pi$-flux associated with any plaquette with $W_p=-1=e^{i\pi}$. The $\pi$-flux can hop to a neighbouring plaquette if we change the sign of $u_{ij}$ on the bond shared by the two plaquettes. This is achieved by  the $\sigma_i^\mu$ operator, which anticommutes with $\hat u_{ij}$. Therefore, if we have $\pi$-fluxes on neighbouring plaquettes we can move the first flux around the other by applying $\sigma_i^\mu$ successively to the six sites of the second plaquette. This is nothing but the operator $\hat W_p$ -- which must have eigenvalue $W_p=-1$ because of the $\pi$-flux on the second plaquette.
Therefore, moving one flux around another induces a $\pi$ phase shift  -- the same mutual semionic statistics as we found between the monomers (e) and visons (m particles) in the $Z_2$ spin liquid (toric code). We can arbitrarily identify alternating columns of plaquettes  with each type of particle.
The $c_i$ Majoranas obey fermionic statistics, and so we can identify these as equivalent to the bound state of a dimer and a monomer ($\varepsilon = \text{e m}$ in the language of the toric code). 
Therefore, this model also has emergent anyonic (semionic) quasiparticles. The gapped phases of the Kitaev honeycomb model are examples of  $Z_2$ spin liquids and therefore topologically ordered. 

In the gapless state the flux statistics are ill-defined: adiabatic transport is not possible as moving the fluxes affects the Majorana fermions. However, introducing a time reversal symmetry breaking perturbation opens a gap. In this phase a Majorana binds to each flux. It can be shown explicitly that the resulting particles are non-Abelian anyons \cite{KitaevHoneycomb,KitaevHoneycombLecture}. A key signature of this phase is that robust chiral edge modes emerge. For neutral particles there is no Hall effect, but there is a thermal Hall effect, where the voltage in \cref{fig:Hall} is replaced by a heat gradient, and, in the presence of a magnetic field, a perpendicular heat gradient develops. The thermal Hall conductivity, $\kappa_{xy}$, is given by 
\begin{equation}
\frac{\kappa_{xy}}{T} = q \frac{\pi k_B^2}{6\hbar}.
\end{equation}
For conventional fermions $q=\nu$, an integer known as the Chern number. But for Majoranas $q=\nu/2$ as two Majoranas are required to build a fermion, see \cref{sect:Majoranas,fig:Majoranas}. Therefore, the  observation of  quantisation of the thermal Hall conductance with $q=1/2$ in $\alpha$-\ch{RuCl3} \cite{Kasahara}, combined with earlier experimental characterisation of the spin excitations \cite{Trebst,Nejc}, suggests that a Kitaev spin liquid with non-Abelian anyons may be realised in $\alpha$-\ch{RuCl3}  in a magnetic field.

Overall, the experimental evidence for quantum spin liquids, and fractionalisation and emergent gauge fields, is not incredibly strong. But a number of candidate materials, that do not order magnetically down to the lowest temperatures have been identified. Key examples are the kagome lattice material herbertsmithite and triangular lattice organic materials \cite{Savary,RPP}.

\subsection{The fractional quantum Hall (FQH) effect}

\begin{figure}
	\centering
	\includegraphics[width=0.4\textwidth]{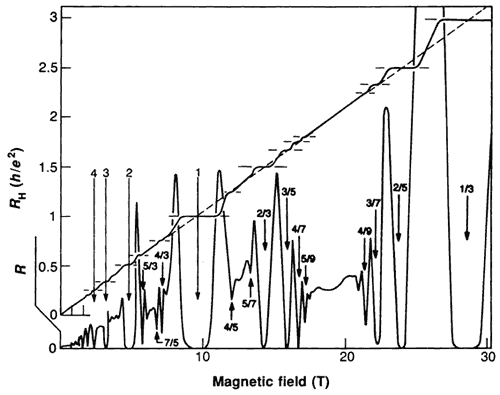}
	\caption{The Hall resistance, $R_H$ and the longitudinal resistance, $R$, both show signatures characteristic of the fractional quantum Hall effect. From \cite{Eisenstein}. Reprinted with permission from AAAS. 
	} \label{fig:FQHE}
\end{figure}

2DEGs that are even cleaner than those which give the IQHE, \cref{sect:IQHE},  show additional plateaus, \cref{fig:FQHE}. All of the additional plateaus occur at rational filling fractions, $\nu=p/q$. This is the FQH effect. Most fractions have odd $q$, but there are exceptions, most notably the prominent plateau at $\nu=5/2$. 

The FQH effect was the first context where fractionalisation and topological order were discovered.  FQH systems provide the best experimental evidence for these phenomena  in a quantum material in more than one dimension. Furthermore, the fractionalised particles in FQH systems show collective phenomena that are interesting in their own right.

The many-body wavefunction for the  IQH state is just the antisymmetrised product of the Landau level wavefunctions, \cref{eq:LandauLevelWF}. For example, the $\nu=1$ wavefunction is
\begin{equation}
\Psi_1 = \left| 
		\begin{array}{ccc}
		1 & 1 & \dots \\
		z_1 & z_2 & \dots \\
		z_1^2 & z_2^2 & \dots \\
		\vdots & \vdots & \ddots \\
		\end{array}
 \right| e^{\sum_i |z_i|^2/4\ell_B^2}
 = \prod_{i<j} (z_i-z_j) e^{\sum_i |z_i|^2/4\ell_B^2}
\end{equation}
Laughlin generalised this  wavefunction and thereby explained the plateaus at $\nu=1/(2p+1)$, where $p$ is an integer:
\begin{equation}
\Psi_{(2p+1)} = \prod_{i<j} (z_i-z_j)^{(2p+1)} e^{\sum_i |z_i|^2/4\ell_B^2}.
\end{equation}
Each factor $(z_i-z_j)$ describes a vortex threaded a by flux quantum. Therefore the Laughlin wavefunction describes the $\nu=1$ state for  composite fermions formed from an electron and $2p$ flux quanta \cite{Jain}. In general the $n$th integer quantum Hall state of composite fermions leads to  plateaus at 
$\nu = \frac{n}{2pn + 1} $.
Similarly, composite fermions can form from  holes in Landau levels. This leads to plateaus at 
$\nu = \frac{n}{2pn - 1}$.
Note that all of these fractional filling factors have odd denominators, which is a consequence of the even number of flux quanta attached to form the composite fermions.

We expect to accumulate an  Aharonov-Bohm phase, $\theta=\pi/(2p+1)$, when we exchange composite fermions, as they are bound states of a charged particle and a flux, cf. \cref{sect:anyons}. This implies that the excitations of the Laughlin states are Abelian anyons. Furthermore, the excitations of the composite fermion fluid are dressed. A simple counting argument shows that this produces quasiparticles with a fractional charge $e/(2np+1)$ \cite{Goldhaber,Jain}.

There is strong experimental evidence supporting these conclusions. Current noise measurements show that the quasiparticles in the $\nu=1/3$ state carry a charge of $e/3$ \cite{Saminadayar,Picciotto,Heiblum}.  There is even  evidence for anyonic statistics. When bosons scatter  they tend to bunch together. Fermions avoid one another; this is known as antibunching. Anyons are expected to do something in between. It was recently reported that in the $\nu=1/3$ state the quasiparticles show the level of bunching expected for anyons with $\theta=\pi/3$, consistent with  predictions for the Laughlin state \cite{Bartolomei}. Even more recently evidence for the braiding of the anyons in the $\nu=1/3$ state has been reported \cite{Nakamura}.

Although there is an infinite family of Laughlin states, there are fractional plateaus observed experimentally that are not predicted for the Laughlin states. The emergence of new particles is not done. We can build Laughlin states from the composite fermions. These are known as hierarchical FQH states, which explains most of the other plateaus that have been observed experimentally.  

The plateau at $\nu=5/2$ is surprising as none of the Laughlin or hierarchical states have even denominators and we do not expect there to be a gap for the composite fermions at this filling. One proposal for how a gap opens is that  residual weak attractive interactions between composite fermions  lead to  spinless $p$-wave superconductivity \cite{Read52}.  This suggests, cf. \cref{sect:anyons}, that the quasiparticle excitations from this state should be non-Abelian anyons \cite{Read52,Moore}. 
In the $\nu=5/2$ state the thermal Hall conductivity is quantised at $2.5$ times the value expected for electrons \cite{Banerjee}. This half-odd-integer value is consistent with non-Abelian anyons \cite{Heiblum}.

The existence of fractionalised and anyonic quasiparticles is highly suggestive of topological order. Indeed, this is where the idea was first explored. It is predicted that the degeneracy of the ground states of the FQH liquid depends on the topology of the manifold it lives on \cite{Wen}. 

The FQH effect can also be explained in terms of an emergent Chern-Simons gauge field theory \cite{Tong,Wen}. This is a topological field theory that can only occur in 2+1 dimensions.

\section{Conclusions: the materials multiverse}

I have reviewed just a small selection of the particles and fields that can emerge in strongly correlated materials. In the standard model of materials the emergent particles can be understood in terms of two key concepts: spontaneously broken symmetry and adiabatic continuity between interacting and non-interacting systems. Adiabatic continuity explains why we see electron-like quasiparticles in metals and semiconductors, and why we can understand many of the properties of these materials without resorting to quantum many-body theory. Spontaneously broken symmetry explains the existence of Nambu-Goldstone modes, the Anderson-Higgs mechanism and topological defects in the order parameter. 

We have seen that the range and variety of particles and fields increases dramatically when we consider  correlated matter. Again adiabatic continuity provides  understanding of emergent matter, but now from exact solutions at special points in parameter space. Topological matter provides a unifying theme beyond the standard model of materials: SPT states when  entanglement is absent or short-range and topological order when  entanglement is long-range.  Two new phenomena emerge beyond the standard model of materials: fractionalised quasiparticles, which carry only part of a `charge' that is quantised in the underlying building blocks of the system; and emergent gauge fields, which arise as an accounting trick. But, accounting is important if you want to balance your books. And descriptions of materials in terms of emergent gauge fields allow us to predict new phenomena and particles, such as the photon in the quantum Coulomb phase.

The hierarchy of  FQH states and the $\nu=5/2$ state show that emergent particles in FQH liquids can form new emergent states of matter,  their own FQH states or superconductors respectively. One might ask how  general such collective phenomena are for emergent quasiparticles. Will we discover other hierarchies of emergent quantum matter where the remaining correlations between emergent quasiparticles lead to even more exotic physics? It would  surprise me if we do not. As Swift  \cite{Swift} put it,
\begin{quote}
	So, Nat'ralists observe, a Flea \\
	Hath smaller Fleas that on him prey,\\
	And these have smaller yet to bite 'em,\\
	And so proceed ad infinitum.
\end{quote}

Another exciting development is the effort to control and manipulate fractionalised quasiparticles. This has a strong overlap with efforts in quantum information processing. Topological matter is attractive for such applications because of its intrinsic ability to resist computational errors \cite{Lahtinen,Lutchyn}.\footnote{Because the system can only be moved between states in different topological sectors by operations that act on the entire system, whereas errors typically occur from local `noise'.} So far most progress has been made with weakly correlated topological matter. But, the greater range of behaviours in strongly correlated matter may ensure that this is harnessed in time. It may also lead to efforts to fabricate strongly correlated metamaterials. In fact, one could view a general purpose quantum computer as  a metamaterial that can be prepared in an arbitrary strongly correlated state (even this is not enough for a quantum computer, one must be able to manipulate and read out the state as well).

It is now clear that  almost any kind of particle/field can emerge as a collective excitation in a sufficiently correlated system. This has led researchers in both condensed matter and high energy physics to ask whether  the  standard model of particle physics and gravity might actually be emergent \cite{Wen,Padmanabhan,Verlinde}. This question remains  open. But,  we do know that that the emergent, low-energy excitations of strongly interacting quantum matter are often described by gauge fields and matter fields. These are precisely the building blocks of the standard model of particle physics. 

A related question about the standard model of particle physics is: why does it contain fermions? Gauge fields are a natural consequence of conservation laws. Bosons appear naturally when we quantise a classical theory. But, fermions have to be put into a theory by hand and necessitate complicated mathematical structures with anticommuting algebras. On the other hand, the  Pauli exclusion principle is essential for chemistry, and hence life, as we know it. If fermions emerged as fractionalised excitations of some more fundamental theory that was not built from fermions, that would provide a elegant resolution of these aesthetic concerns.

What is already clear, however, is that strongly correlated materials can form their own universes, each with its own set of particles and fields emerging at low-energies. It is clear that materials are the multiverse.

\section*{Acknowledgements}

I thank Sean Barrett, Oliver Bellwood, Janani Chander, Jarad Cole, Jace Cruddas, Andrew Doherty, Anthony Jacko, Elise Kenny, Ian McCulloch, Ross McKenzie, Jaime Merino, Henry Nourse, Sayed Saadatmand, Nic Shannon and Tom Stace for helpful conversations;   Jace Crudas and Gordon Powell for help preparing the figures; and Henry Nourse and Miriam Ohlrich for  feedback on a draft  manuscript.

\section*{Disclosure statement}

No potential conflict of interest was reported by the author.

\section*{Funding}

This work was supported by the Australian Research Council through grants  DP181006201 and DP200100305.

\section*{Notes on contributor}

Ben Powell is a Professor of Physics at the University of Queensland. His research focuses on emergent phenomena in molecular materials and other chemically complex systems.

\includegraphics[width=0.2\textwidth]{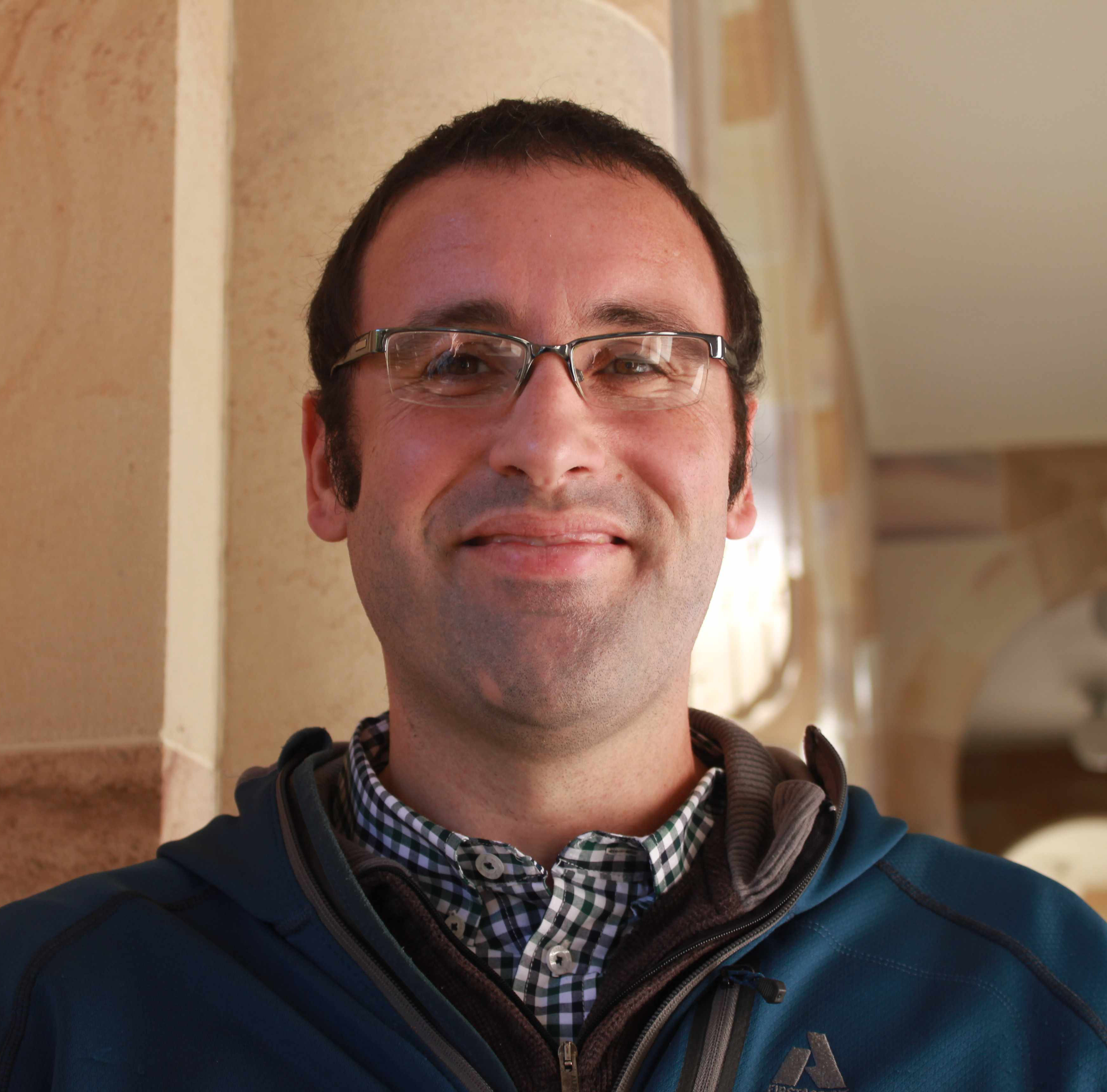}

\bibliographystyle{tfnlm}
\bibliography{qp}

\end{document}